\documentclass[namedreferences]{solarphysics}

\def\referee#1{{#1}}
\def\editor#1{{#1}}

\usepackage[optionalrh,solaromanenum,linksfromyear]{spr-sola-addons} 

\usepackage{graphicx}                    
\usepackage[usenames]{color}                       
\usepackage{url}                         

\usepackage{style_wei_symbols}
\usepackage{style_wei}

\usepackage{solar_physics_additional_label_def}

\usepackage{solaheader}

\begin{document}

\begin{article}

\begin{opening}

\title{Advances in Observing Various Coronal EUV Waves in the SDO Era and Their Seismological Applications
\editor{(Invited Review)}}


\author{Wei~\surname{Liu}$^{1, \, 2}$\sep
        Leon~\surname{Ofman}$^{3, \, 4}$      
       }

\runningauthor{W.~Liu, L.~Ofman}
\runningtitle{EUV Wave Observations by SDO/AIA}

\institute{
 $^{1}$ W.W.~Hansen Experimental Physics Laboratory, Stanford University, Stanford, CA 94305, USA \\
 $^{2}$ Lockheed Martin Solar and Astrophysics Laboratory, Bldg.~252, 3251 Hanover Street, Palo Alto, CA 94304, USA \\
   email: \url{weiliu@lmsal.com} \\ 
 $^{3}$ Catholic University of America and NASA Goddard Space Flight Center, Code 671, Greenbelt, MD 20771, USA \\
 $^{4}$ Visiting Associate Professor, Department of Geophysics and Planetary Sciences, Tel Aviv University, Tel Aviv 69978, Israel  \\
   email: \url{Leon.Ofman@nasa.gov} \\
}

\begin{abstract}	

Global extreme ultraviolet (EUV) waves are spectacular traveling disturbances in the solar corona
associated with energetic eruptions such as coronal mass ejections (CMEs) and flares. 
\editor{Over the past 15 years,
observations from three generations of space-borne EUV telescopes}
have shaped our understanding of this phenomenon and at the same time 
led to controversy about its physical nature. 
Since its launch in 2010, the {\it Atmospheric Imaging Assembly} (AIA) onboard the 
{\it Solar Dynamics Observatory} (SDO) has observed more than 210 global EUV waves in exquisite detail, 
thanks to its high spatio--temporal resolution and full-disk, wide-temperature coverage. A combination of
statistical analysis of this large sample, more than 30 detailed case studies, and data-driven MHD modeling,
has been leading their physical interpretations to a convergence, 	
favoring a bimodal composition of an outer, fast-mode magnetosonic wave component 
and an inner, non-wave CME component. 
Adding to this multifaceted picture, AIA has also discovered new EUV wave and wave-like phenomena
associated with various eruptions, including quasi-periodic fast propagating (QFP) wave trains, 
magnetic Kelvin--Helmholtz instabilities (KHI) in the corona and associated nonlinear waves, 
and a variety of mini-EUV waves.
Seismological applications using such waves are now being actively pursued, especially
for the global corona.
We review such advances in EUV wave research focusing on recent SDO/AIA observations, 
their seismological applications, related data-analysis techniques,
and numerical and analytical models.

\end{abstract}


\end{opening}

\section{Introduction}
\label{sect_intro}


The dynamic, magnetized solar corona hosts a variety of waves and wave-like phenomena
that are believed to play important roles in many fundamental, yet enigmatic processes,
such as 	
corona heating ({\it e.g.} \opencite{IonsonJ.Alfven.wave.heat.corona.1978ApJ...226..650I},
 \opencite{HeyvaertsPriest.Alfven-wave-heat-corona.1983A&A...117..220H})
and solar-wind acceleration (see reviews by \opencite{Ofman.solarwind-review.2010LRSP....7....4O}
and \opencite{CranmerS.solar.wind.model.review.2012SSRv..172..145C}).
Such waves also carry critical information that can be used to decipher otherwise elusive 	
physical parameters of the corona, such as the magnetic-field strength,
via a technique called {\it coronal seismology}
\cite{Uchida.coronal-seismology.1970PASJ...22..341U,Roberts.coronal-seismology.1984ApJ...279..857R,%
Nakariakov.TRACE-loop-oscil.1999Sci...285..862N,NakariakovOfman.B-from-oscil.2001A&A...372L..53N,%
Nakariakov.wave-review.2005LRSP....2....3N}.

Space-borne extreme ultraviolet (EUV) imagers have been the prime instruments observing traveling coronal disturbances
for decades, thanks to a wide range of EUV emission produced by ions at various coronal temperatures.	
The most spectacular examples are EUV disturbances
expanding across a fraction of the solar disk, often in annular shapes and commonly
associated with coronal mass ejections (CMEs) and flares.
They were discovered by the {\it Extreme-ultraviolet Imaging Telescope} (EIT: \opencite{SOHO.EIT.1995SoPh..162..291D})
onboard the {\it Solar and Heliospheric Observatory} (SOHO) 
and rekindled broad interest in large-scale coronal (shock) waves
\cite{Moses.EIT-wave.1997SoPh..175..571M,DereK.EIT.1st-result.1997SoPh..175..601D,%
ThompsonB.EIT-wave-discover.1998GeoRL..25.2465T}.
They are thus often called ``EIT waves", 
as well as ``(global) EUV waves" \cite{Patsourakos.Vourlidas.EIT-wave-review.2012SoPh..281..187P}, 
``coronal bright fronts" \cite{GallagherP.LongD.EIT.wave.review.2011SSRv..158..365G},
or ``large-scale coronal propagating fronts" \cite{NittaN.AIA.wave.stat.2013ApJ...776...58N}.
Here we adopt the most commonly used term ``EIT waves", while we
reserve ``EUV waves" for propagating EUV disturbances in general -- the subject of the present review.

Over the past decade and a half, three generations of EUV telescopes, 
notably SOHO/EIT, the {\it Extreme UltraViolet Imager} (EUVI: \opencite{WuelserJ.STEREO-EUVI.2004SPIE.5171..111W}) onboard
the {\it Solar TErrestrial RElations Observatory} (STEREO: \opencite{Kaiser.STEREO-mission.2008SSRv..136....5K}), and
the {\it Atmospheric Imaging Assembly} (AIA: \opencite{LemenJ.AIA.instrum.2012SoPh..275...17L}) 
onboard the {\it Solar Dynamics Observatory} (SDO: \opencite{PesnellD.SDO.mission.2012SoPh..275....3P}),
have each contributed to shaping our evolving understanding of EIT waves in specific and EUV waves in general. 
SDO/AIA in particular, as the most advanced solar EUV imager to date, has led to breakthroughs
in coronal-wave research. During its 3.5 years of operation, 
AIA has not only been bringing the long-standing debate on the nature of EIT waves 
to a closure by establishing a bimodal composition with
both wave and non-wave components \cite{Patsourakos.Vourlidas.EIT-wave-review.2012SoPh..281..187P}, 
but also discovered new types of EUV waves, 
especially quasi-periodic fast propagating (QFP) wave trains \cite{LiuW.FastWave.2011ApJ...736L..13L}
and nonlinear waves associated with magnetic Kelvin--Helmholtz instabilities
(\opencite{Ofman.Thompson.AIA.KH.instab.2010AGUFMSH14A..02O},
\citeyear{Ofman.Thompson.AIA.KH.instab.2011ApJ...734L..11O};
\opencite{Foullon.AIA.KH.instab.2011ApJ...729L...8F}),
adding to the multitude of aspects of this complex phenomenon.
Global coronal seismology utilizing these large-scale EUV waves
is becoming a reality.

As an active research subject, observations and models of EUV waves have been
reviewed extensively in the past, each with a somewhat different focus.
Interested readers are referred to 
\inlinecite{WarmuthA.EIT-wave-review.2007LNP...725..107W} for a review
based mainly on SOHO and multiwavelength observations,
to \inlinecite{Wills-Davey.EIT-wave-review.2009SSRv..149..325W},
\inlinecite{GallagherP.LongD.EIT.wave.review.2011SSRv..158..365G},
and \inlinecite{ZhukovA.EIT.wave.review.STEREO.2011JASTP..73.1096Z} 
for updates with early STEREO observations,
to \inlinecite{Patsourakos.Vourlidas.EIT-wave-review.2012SoPh..281..187P}
for a synthesized view from SOHO, STEREO, {\it Hinode}, and SDO in its first year of operation,
and to \inlinecite{VrsnakCliver.corona-shock-review.2008SoPh..253..215V} 
and \inlinecite{ChenPF.CME.review.2011LRSP....8....1C} for related subjects of coronal shocks and CMEs, respectively.

The aim of this review is to summarize the current knowledge of EUV waves,
focusing on the unique and revolutionary contributions made by SDO/AIA
to observations of three types of waves generally associated with eruptions,
{\it i.e.} EIT waves, QFP wave trains, and small-scale waves including mini-EUV waves 
and magnetic Kelvin--Helmholtz instability nonlinear waves.
We strive to make our review complementary to the existing literature
with different perspectives and minimal overlap yet without sacrificing completeness. 
\editor{We review here mainly published material, together with some new results
such as structural oscillations of wide-ranging periods triggered by EIT waves,
long periodicities of EIT waves themselves, and new clues to the relationship
between quasi-periodic wave trains inside and outside CME bubbles 
(see \figs{908_oscil-fits.eps}, \ref{0908_long-fft.eps}, and \ref{QFP-2trains.eps}).
}	

Other types of waves that can be seen in EUV and are generally associated with traditional local coronal seismology
are not covered in this review. This is partly because these waves have been extensively studied in the last decade and a half,
especially with SOHO and TRACE, while SDO/AIA has not yet made significant advances in their observations.
Such waves include standing (oscillations) or propagating magnetosonic waves of slow modes
\cite{Ofman.UVCS-polar-hole-wave.1997ApJ...491L.111O,DeMoortel.TRACE-slow-mode-discover.2000A&A...355L..23D},
fast kink modes \cite{Aschwanden.1st-TRACE-wave.1999ApJ...520..880A,Nakariakov.TRACE-loop-oscil.1999Sci...285..862N,%
NakariakovOfman.B-from-oscil.2001A&A...372L..53N},	
and fast sausage modes \cite{Nakariakov.sausage-mode-Nobeyama.2003A&A...412L...7N},
as well as \Alfven waves \cite{Tomczyk.Alfven-wave.2007Sci...317.1192T,Jess.Alfven-wave.2009Sci...323.1582J}.
Interested readers are referred to relevant reviews
\cite{AschwandenM2004psci.book.....A,Nakariakov.wave-review.2005LRSP....2....3N,%
Banerjee.Magnetoseismology-review.2007SoPh..246....3B,RobertsB.coronal.seismology.progress.2008IAUS..247....3R,%
Ofman.oscil.model-review.2009SSRv..149..153O}
and recent AIA observations ({\it e.g.} \opencite{Aschwanden.Schrijver.AIA.loop.oscil.2011ApJ...736..102A};
\opencite{McIntoshS.Alfven.wave.heat.corona.solar.wind.2011Natur.475..477M};
\opencite{WangTJ.AIA.transverse.oscil.2012ApJ...751L..27W};
\opencite{GosainS.FoullonC.2010-09-08_EUV.wave.flmt.oscil.2012ApJ...761..103G};
\opencite{WhiteRS.Verwichte.AIA.loop.oscil.2012A&A...537A..49W};
\opencite{SrivastavaAK.kink.oscil.2011Aug9.X69.flare.2013ApJ...777...17S};
\opencite{ThrelfallJ.cmpr.wave.CoMP.AIA.2013A&A...556A.124T}).


We organize this article as follows: After a brief description of relevant EUV telescopes
in Section~\ref{sect_instr}, we review in Sections~\ref{sect_EIT}\,--\,\ref{sect_mini-wave}
observations of the three types of EUV waves mentioned above.
We then review their coronal seismological applications in Section~\ref{sect_seism}
and related data-analysis techniques and numerical and analytical models in Section~\ref{sect_method},
followed by conclusions and future prospects in Section~\ref{sect_conclude}.

\section{Instruments Observing EUV Waves}
\label{sect_instr}

\begin{figure}[tbhp]      
 \begin{center}
 \includegraphics[width=1.0\textwidth]{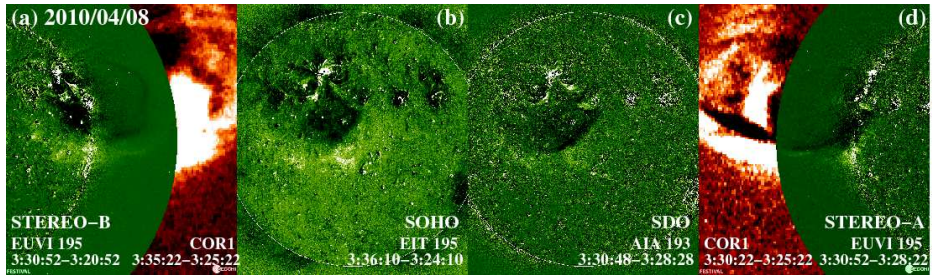}	
 \end{center}
 \caption[]{\footnotesize    
 Running-difference images of the 08 April 2010 EIT wave
 observed by three generations of EUV imagers, SOHO/EIT, STEREO/EUVI (and COR1), and SDO/AIA, 
 at the nearest times from three different viewing angles in near quadrature.
  STEREO-A and -B were $68 \degree$ and $71 \degree$ ahead of and behind the Earth, respectively.
  The AIA cadence is downgraded to match that of STEREO-A/EUVI.
 \referee{The different image contrasts are mainly due to different cadences of running difference 
 and integration of emission along different lines of sight.} 
 Panels~c and d are from \inlinecite{LiuW.AIA-1st-EITwave.2010ApJ...723L..53L}.
 } \label{3mission.eps}
 \end{figure}

The evolution of our understanding of EUV waves in the last 15 years
has been primarily driven by technological advances of EUV imagers in three generations.	
Here we briefly review their capabilities, summarized in \tab{table_instr},
as well as their outstanding contributions and limitations. 
\fig{3mission.eps} shows an example of an EIT wave observed simultaneously 
by three representative instruments.
%
\begin{table}[tbhp]	
\caption{\small Capabilities of the recent three generations of EUV imagers.}
\tabcolsep 0.04in	
\tiny 
\begin{tabular}{llllllll}
\hline
 Instruments & Operation 
 & Caden.\tabnote{The highest typical operational cadence. Even higher cadences are generally available (for example, 66~seconds for EIT, 12~seconds for TRACE, and 10~seconds for AIA), but have rarely been used.} 
 & Expos.\tabnote{Typical exposure.} &  FOV   & CCD   
 & Resol.,\tabnote{Angular resolution, pixel size. All instruments, except AIA, have Nyquist-limited resolution at twice the pixel size.}  & EUV Channels \\
             & Period    & [seconds]  &  [seconds]  &             &   &    pixel             & [\AA] \\
\hline
 SOHO/EIT    & 1995--     & 600         & $\ge$1.5  &  $45 \arcmin ^2$   & $1024^2$ &  $5.2\arcsec$, $2.6\arcsec$   & 171, 195, 284, 304 \\	
 TRACE       & 1998\,--\,2010 & 20\,--\,30      & $\ge$15   &  $8.5 \arcmin ^2$  & $1024^2$ &  $1.0\arcsec$, $0.5\arcsec$    & 171, 195, 284 \\
\hline
 STEREO/EUVI & 2006--     & 75\,--\,150     & 4\,--\,8      &  $54 \arcmin ^2$   & $2048^2$ &  $3.2\arcsec$, $1.6\arcsec$   & 171, 195, 284, 304  \\	
 PROBA2/SWAP & 2010--     & 60\,--\,120     & 10        &  $54 \arcmin ^2$   & $1024^2$ &  $6.4\arcsec$, $3.2\arcsec$   & 174  \\	
\hline
 SDO/AIA     & 2010--     & 12	        & 2\,--\,3    &  $41 \arcmin ^2$   & $4096^2$ &  $1.5\arcsec$, $0.6\arcsec$  & 94, 131, 171, 193, \\
            &&&&&&& 211, 304, 335 \\


\hline  \end{tabular}
\label{table_instr} 
\end{table}
%

SOHO/EIT has been the primary first-generation EUV wave imager for over a solar cycle. 
Unfortunately, its operational 12\,--\,18~minute cadence significantly 
under-samples	
the typically hour-long lifetimes of EIT waves and 
results in large uncertainties in their kinematics measurements.
Such observations led to under-constrained models and considerable controversy.	
TRACE \cite{HandyB.trace.1999SoPh..187..229H} had much higher cadences of 20\,--\,30~seconds 
and the best spatial resolution [$1\arcsec$] to date, but its small field of view (FOV)
made it incapable of tracking global EUV waves	
and its long exposures can smear rapidly evolving features.
As such, TRACE detected only a handful EIT waves during its 12~year mission
({\it e.g.} \opencite{Wills-DaveyThompson.TRACE-EUV-wave.1999SoPh..190..467W}).	

STEREO/EUVI, representing the second-generation EUV wave imagers, 
has alleviated some of the above limitations. It images the full-Sun corona with the largest
FOV to date ($54 \arcmin \times 54 \arcmin$, up to $r=1.7 \Rsun$) from two vantage points, 
thus offering crucial clues to the 3D geometry ({\it e.g.} \opencite{TemmerM.3D.stereo.EUVwv.2011SoPh..273..421T}),
especially the height range, of EIT waves within the limitations of optically thin coronal emission. 
The coronagraphs within the same {\it Sun Earth Connection Coronal and Heliospheric Investigation} 
(SECCHI: \opencite{HowardR.STEREO-SECCHI.2008SSRv..136...67H}) package 
are critical for establishing the CME--EIT wave relationship
({\it e.g.} 	
\opencite{Patsourakos.EUVI-fast-mode-wave.2009ApJ...700L.182P}).
However, its common cadence of 150~seconds (occasionally 75~seconds) is inadequate for 
capturing high-speed waves like QFPs.
The {\it Sun Watcher using Active Pixel detectors and Image Processing}
(SWAP: \opencite{SeatonD.PROBA2.SWAP.instr.2013SoPh..286...43S}) instrument
onboard the {\it PRoject for On-Board Autonomy 2} (PROBA2: \opencite{SantandreaS.PROBA2.mission.2013SoPh..286....5S})
mission is a derivative of EIT with an improved cadence and FOV similar to EUVI.
Its single 174~\AA\ channel provides limited temperature coverage
for EUV waves ({\it e.g.} \opencite{Kienreich.STA-PROBA2.EUV.wave.reflect.2013SoPh..286..201K}).
%
%

SDO/AIA came with revolutionary improvements over previous generations.
Its typical cadence of 12~seconds, one to two orders of magnitude higher than those of 
EIT and EUVI, short exposure of two to three~seconds, and high resolution of $1.5\arcsec$ 
with $0.6\arcsec$ pixels are all crucial for detecting fast propagating EUV waves.
Its seven EUV channels (including previously underused 94, 131, 211, and 335 \AA)
cover a wide temperature range, offering unparalleled thermal diagnostic power
\cite{ODwyer.AIA-T-response.2010A&A...521A..21O}.
Its high sensitivity (signal-to-noise ratio) and dynamic range
\cite{Paul.Boerner.initial.AIA.calib.2012SoPh..275...41B} allow for
detecting faint waves in the presence of bright emission ({\it e.g.} flares).
 \editor{In addition, AIA has a high duty cycle of $\ge$95\,\% with only interruptions due to
 eclipses lasting for up to 90 minutes a day for a period of about two weeks once every six months.
 This enables monitoring the full-Sun corona virtually uninterruptedly
 and thus significantly increases the detection rate of transients like EUV waves.
 }  
On the other hand, its $41 \arcmin \times 41 \arcmin$ FOV is 
the smallest among the full-disk EUV imagers in the recent three generations
and can miss high-corona features above the limb especially near the Equator.
Overall, the combination of unprecedented capabilities makes 
AIA the best ever instrument for observing EUV waves. 
By the time of writing, 3.5 years in operation, AIA has detected 	
more than 210 EIT waves 	
  \editor{(see the list compiled by N.~Nitta	
  at http://www.lmsal.com/nitta/movies/AIA\_Waves)},
over 15 QFP wave trains, and various small-scale EUV waves,
represented in more than 40 publications.	

Other than imagers, EUV spectrometers can provide valuable plasma diagnostics for EUV waves
such as the density, temperature, and Doppler and nonthermal broadening velocities
(for a review, see Section~8 of \opencite{Patsourakos.Vourlidas.EIT-wave-review.2012SoPh..281..187P}).
Due to the low chance of capturing a transient EUV wave in a narrow slit,
{\it e.g.} with the SOHO/{\it Coronal Diagnostic Spectrometer} (CDS: \opencite{Harrison.CDS.instr.1995SoPh..162..233H})
and the {\it Hinode}/{\it EUV Imaging Spectrometer} (EIS: \opencite{Culhane.EIS.instr.2007SoPh..243...19C}),
such studies have been performed for only a few events 	
\cite{HarraSterling.fast.leading.edge.EIT-wave.2003ApJ...587..429H,%
ChenFeng.DingMD.EIS-evapor.0.6MK-downflow.2010ApJ...724..640C,West.EIT-wave-seismology.2011ApJ...730..122W,%
Harra.Sterling.EIS.2010Feb16.EITwv.2011ApJ...737L...4H,ChenF.Ding.ChenPF.EIS.EITwv.nonWave.2011ApJ...740..116C,%
Veronig.EIS.2010Feb16.EITwv.2011ApJ...743L..10V,YangLH.AIA.EIS.EUV-wave.2013ApJ...775...39Y,%
LongD.EIT.wave.seismology.2013SoPh..288..567L}.

\section{Global EUV (EIT) Waves}
\label{sect_EIT}

\subsection{General Properties from Pre-AIA Observations}
\label{subsect_gen}


Early SOHO/EIT observations indicated that the majority (93\,\%) of EIT waves are diffuse pulses 
on the order of 100~Mm wide, while the rest are arc-like, sharp fronts called ``S-waves" 
that eventually become diffuse \cite{Biesecker.EIT-wave-association.2002ApJ...569.1009B}. 
EIT waves have a broad speed distribution within 50\,--\,$700 \kmps$
\cite{ThompsonB.EIT-wave-catalog.2009ApJS..183..225T}.		
Their height range was determined by 
STEREO/EUVI to be one to two gravitational scale heights ($\approx$100~Mm)
above the solar surface \cite{Patsourakos.EUVI-wave.2009SoPh..259...49P,Kienreich.2009ApJ...703L.118K}.
Their counterparts in the high corona up to $r=3 \Rsun$ were recently imaged in
white light by STEREO/COR1 \cite{Kwon_STEREO-EUV-wave-high-corona.2013ApJ...766...55K}.

EIT waves are associated with a variety of solar activity,
including CMEs, Type-II radio bursts, EUV dimmings, stationary brightenings, and remote filament oscillations
(see the review by 
\opencite{GallagherP.LongD.EIT.wave.review.2011SSRv..158..365G}
 and references therein).
They appear to stop at coronal hole boundaries and avoid active regions
({\it e.g.} \opencite{ThompsonB.EIT-Moreton-wave.1999ApJ...517L.151T}).	
Some EIT waves, especially S-waves, are associated with 
{\it chromospheric} \ion{He}{1}~10830~\AA\ waves
\cite{Vrsnak.HeI.wave.2002A&A...394..299V,
GilbertH.HeI.EIT.wave.cospatial.2004ApJ...607..540G}	
and \Ha Moreton waves \cite{MoretonRamsey.1960PASP...72..357M} 	
which are interpreted as ``sweeping skirts" of	
{\it coronal} fast-mode (shock) waves \cite{Uchida.Moreton-wave-sweeping-skirt.1968SoPh....4...30U}.
Counterparts of EIT waves at other wavelengths have been imaged in
soft X-rays \cite{Narukage.SXT-EIT-wave.2002ApJ...572L.109N,HudsonH.SXT-wave.2003SoPh..212..121H}
and microwave and metric radio bands
\cite{WarmuthA.multiw-EIT-wave.2004A&A...418.1101W,%
White.Thompson.Nobeyama-EIT-wave.2005ApJ...620L..63W,Vrsnak2005.radio.EIT.wave.ApJ...625L..67V}.

\subsection{EIT Wave Models and New Constraints from AIA Observations}
\label{subsect_model}


A number of models were proposed for EIT waves, 
falling into three categories: wave, non-wave, and hybrid models,
as described below and summarized in \tab{table_models}.
Over the last 15 years, there has been an intense debate between wave and non-wave
interpretations, each with its own pros and cons under-constrained by early EIT observations. 
In recent years, STEREO/EUVI and particularly SDO/AIA have provided new constraints
leading to a converged view that includes both wave and non-wave components.
\begin{table}[bthp]	
\scriptsize		
\caption{\small Proposed models for EIT waves and their supporting evidence.}
\tabcolsep 0.04in	
\begin{tabular}{llll}
\hline
Categories &  MHD Wave (Shock)   &  Non-wave                   &  Hybrid/Bimodal \\
\hline
Models   & fast mode              & field-line stretching      &  wave $+$ non-wave \\
         & slow mode (soliton)    & current shell              &   \\
         &                        & reconnection front         &   \\
\hline
Evidence
         & $v_{\rm EIT} \approx v_{\rm f}$, deceleration & $v_{\rm EIT} < c_{\rm s} < v_{\rm f}$, erratic  & CME--wave decoupling \\
         & reflection, transmission     &  stationary brightenings\tabnote{Stationary brightenings favor non-wave models, but can also be explained by wave models (see Section~3.6).}  & multiple components   \\    
         & periodicity, compressibility   & long-term dimmings   &  {\it plus evidence for wave}   \\
         & sequential oscillations        &                      &  {\it and non-wave models} \\
         & Moreton, Type~II, SEP assoc.   &  lack of such assoc. &     \\
%
%
\hline \end{tabular}
\label{table_models} \end{table}
	%

{\bf Wave models} interpret EIT waves as true waves,
most likely {\it fast-mode magnetosonic waves}
\cite{ThompsonB.EIT-Moreton-wave.1999ApJ...517L.151T,%
WangYM.EIT-fastMHDwave.2000ApJ...543L..89W,
Wu.EIT-fast-MHD-wave.2001JGR...10625089W,WarmuthA.EIT-Moreton-wave-fast-mode.2001ApJ...560L.105W,%
OfmanThompson.EIT-wave-fast-mode.2002ApJ...574..440O}.
This explains their quasi-circular shapes,
because fast modes are the only MHD waves that can propagate perpendicular 
to magnetic fields with a weak direction dependence
in a low-$\beta$ plasma such as the solar corona.
The measured typical EIT wave speeds of 200\,--\,$400 \kmps$ 	
\cite{ThompsonB.EIT-wave-catalog.2009ApJS..183..225T} and 
especially the $>$\,$600 \kmps$ median speed revealed by AIA
\cite{NittaN.AIA.wave.stat.2013ApJ...776...58N} are within the expected
range of coronal fast-magnetosonic speeds [$v_{\rm f}$].
Other supporting evidence established or confirmed by AIA 	
includes quasi-periodic wave trains within broad EIT wave pulses and sequential structural oscillations
({\it e.g.} \opencite{LiuW.cavity-oscil.2012ApJ...753...52L}),
reflections, transmissions, and refractions at structural boundaries
({\it e.g.} 	
\opencite{OlmedoO.2011-2-15_X2.AIA.EUV.wave.2012ApJ...756..143O}),
compressional heating and cooling cycles ({\it e.g.} \opencite{DownsC.MHD.2010-06-13-AIA-wave.2012ApJ...750..134D}),
cospatial Moreton waves ({\it e.g.} \opencite{AsaiA.2010AugX6.9.Moreton.AIA.wave.2012ApJ...745L..18A}),
and deceleration accompanied by pulse broadening and dispersion 
({\it e.g.} \opencite{LongD.AIA.EUV-wave.2011ApJ...741L..21L}).
EIT waves were also interpreted as {\it slow-mode shocks} and velocity vortices
surrounding CMEs \cite{WangHongjuan.EIT-slow-mode-wave.2009ApJ...700.1716W}
or {\it slow-mode solitons} formed by a balance between dispersive decay and
nonlinear steepening \cite{Wills-Davey.EIT-wave-soliton.2007ApJ...664..556W}.

{\bf Non-wave models} interpret EIT waves not as true waves but as signatures of
CME-caused reconfiguration of the corona, including plasma compression due to
successive {stretching of magnetic-field lines} 	
\cite{ChenPF.EIT-wave-MHD.2002ApJ...572L..99C,ChenPF.EIT-wave.2005ApJ...622.1202C},
Joule heating in {current shells} surrounding CMEs  
\cite{Delannee.EIT-pseudo.wave.2000ApJ...545..512D,Delannee.EIT-wave-current-shell.2008SoPh..247..123D},
{sequential magnetic reconnection} between CME flanks and the ambient corona
with favorably-oriented field lines \cite{Attrill.EIT-wave-CME-Footprint.2007ApJ...656L.101A}, 
or simply line-of-sight (LOS) {projection of CME bubbles} themselves \cite{AschwandenM.4D-CME-dimming.2009AnGeo..27.3275A}.
These alternative models were proposed to resolve difficulties facing the fast-mode wave model,	
such as erratic kinematics, stationary brightenings,	
long-lasting dimmings,	 
and some EIT waves being slower than expected fast-mode speeds $v_{\rm f}$ and even sound speeds $c_{\rm s}$.
Meanwhile, non-wave models have their own difficulties. For example, 
the inferred magnetic topology of the quiet Sun does not support 
a coherently propagating reconnection front
\cite{DelanneeC.B-topology.no.reconn.front.2009A&A...495..571D}.


%
\begin{table}[bthp]	
\scriptsize			
\caption{\small AIA observed EIT waves and their characteristics.}
\tabcolsep 0.012in	
\begin{tabular}{lllllllllrlllll}
\hline 
\multicolumn{3}{c}{Flare} && 
\multicolumn{5}{c}{Characteristics\tabnote{Abbreviations: B: Bimodal, W: Wave, N: Non-wave; 
H: Heating, C: Compression; P: Periodicity (of EIT waves), O: Oscillations (triggered by EIT waves); 
R: Reflection, T: Transmission (or Refraction);
according to the original interpretation in the reference.}} &  Notes/References   \\
\cline{1-3}                \cline{5-9}
Date  & Start & GOES    && Speed\tabnote{Maximum speed measured at 193 or 211~\AA.}
                                    &  Bimod. & Heat. & Period. &  Refl.     \\
{\tiny ddmmmyy} 
        & Time  & Class && {\tiny [$\kmps$]}  &         & Comp. & Oscil.  &  Tran.            \\
\hline 
08Apr10 & 02:30 & B3.7 && 240 & B  & H & &  & QFPs \cite{LiuW.AIA-1st-EITwave.2010ApJ...723L..53L}   \\
12Jun10 & 00:53 & M2.0 && 360 & {\hskip 0.4cm} N & &    \multicolumn{3}{r}{reduced blueshifts \cite{ChenF.Ding.ChenPF.EIS.EITwv.nonWave.2011ApJ...740..116C}}  \\
         &       &      && 1280 & & &   \multicolumn{3}{r}{dome, SEPs \cite{Kozarev.AIA.EUVwv.2010Jun12.2011ApJ...733L..25K}}  \\

13Jun10 & 05:30 & M1.0 && 730 & & {\hskip 0.13cm} C &  \multicolumn{3}{r}{dome, SEPs \cite{Kozarev.AIA.EUVwv.2010Jun12.2011ApJ...733L..25K}}  \\
         &       &      && 600 & {\hskip 0.1cm} W & HC & \multicolumn{3}{r}{shock, $M_{\rm A}$=1.35 \cite{MaSL.AIA.shock.2011ApJ...738..160M}}  \\
         &       &      && 640 & & {\hskip 0.13cm} C & \multicolumn{3}{r}{$B$$\approx$1.4~G \cite{GopalswamyN.2010Jun13.AIA.shock.Bseismology.2012ApJ...744...72G}}  \\	
         &       &      && 600 & B & HC & & & \cite{DownsC.MHD.2010-06-13-AIA-wave.2012ApJ...750..134D}  \\
         &       &      &&     &   & {\hskip 0.13cm} C &   \multicolumn{3}{r}{Type~II \cite{KouloumvakosA.typeII.in.sheath.CME.EUV.wave.2014SoPh..289.2123K}}  \\	

27Jul10 & 08:46 & A6.0 && 560 & B & & &  & \cite{ChenPF.Wu.Coexis-fast-mode.2011ApJ...732L..20C} \\
14Aug10 & 09:38 & C4.4 && 420 & {\hskip 0.1cm} W &  & & & dispersion (Long et al.~\citeyear{LongD.AIA.EUV-wave.2011ApJ...741L..21L}) \\
08Sep10 & 23:05 & C3.3 && 830 & B & HC & PO & {\hskip 0.1cm} T & QFPs \cite{LiuW.cavity-oscil.2012ApJ...753...52L}  \\
         &       &      &&     & B &    & {\hskip 0.1cm} O  &   & \cite{GosainS.FoullonC.2010-09-08_EUV.wave.flmt.oscil.2012ApJ...761..103G} \\

16Oct10 & 19:07 & M2.9 && 1390& B  &  & {\hskip 0.1cm} O & & \cite{KumarP.AIA.EUV.wave.oscil.2013SoPh..282..523K}  \\
         &       &      &&     &    &  & {\hskip 0.1cm} O & \multicolumn{2}{r}{ \cite{Aschwanden.Schrijver.AIA.loop.oscil.2011ApJ...736..102A}}  \\

13Nov10 & 17:04 & B5.7 && 350 & {\hskip 0.1cm} W & &  \multicolumn{3}{r}{surge \cite{ZhengRS.EUV.wave.surge.2013ApJ...764...70Z}}  \\	
27Jan11 & 11:53 & C1.2 && 550 & B & H & {\hskip 0.1cm} O &  & \cite{DaiY.EUVI.AIA.wave.bimodal.2012ApJ...759...55D}  \\
14Feb11 & 17:20 & M2.2 &&     &   &   &   &  \multicolumn{2}{r}{Moreton wave (White~et~al.\citeyear{WhiteS.2011Feb.Moreton.AIA.waves.2011SPD....42.1307W})} \\
15Feb11 & 01:44 & X2.2 && 730 & {\hskip 0.4cm} N & HC & {\hskip 0.1cm} O &  \multicolumn{2}{r}{curr.~shell \cite{Schrijver.Aulanier.2011Feb15.X2.AIAwv.2011ApJ...738..167S}} \\
         &       &      && 780 & B & & & RT & \cite{OlmedoO.2011-2-15_X2.AIA.EUV.wave.2012ApJ...756..143O}  \\

16Feb11 & 14:19 & M1.6 && 500 & B & {\hskip 0.13cm} C & & \multicolumn{2}{r}{red/blueshifts \cite{Harra.Sterling.EIS.2010Feb16.EITwv.2011ApJ...737L...4H}}  \\
         &       &      && 590 & {\hskip 0.1cm} W & {\hskip 0.13cm} C & &  \multicolumn{2}{r}{\cite{Veronig.EIS.2010Feb16.EITwv.2011ApJ...743L..10V}}  \\

24Mar11 & 12:01 & M1.0 && 950 & B & & & {\hskip 0.1cm} T & \cite{XueZK.EUV.wave.deform.decel.2013A&A...556A.152X}  \\	
25Mar11 & 23:08 & M1.0 && 1020& B  &  & {\hskip 0.1cm} O & RT & \multicolumn{1}{r}{\cite{KumarP.AIA.plasma.blob.EUV.wave.2013A&A...553A.109K} } \\
07Jun11 & 06:16 & M2.5 && 960 & B & & &  & \cite{ChengX.ZhangJie.EUV-wave-CME-decouple.2012ApJ...745L...5C}  \\
         &       &      && 780 & B & H & & RT & dome \cite{LiTing.2010Jun07.AIAwave.reflect.2012ApJ...746...13L}  \\

04Aug11 & 03:41 & M9.3 && 910 & {\hskip 0.1cm} W & H & PO & RT & redshifts \cite{YangLH.AIA.EIS.EUV-wave.2013ApJ...775...39Y}  \\
09Aug11 & 07:48 & X6.9 && 760 & B & & {\hskip 0.1cm} O & \multicolumn{2}{r}{Moreton wave \cite{AsaiA.2010AugX6.9.Moreton.AIA.wave.2012ApJ...745L..18A}}  \\
         &       &      && 1000& B & & \multicolumn{3}{r}{photosph.~\cite{ShenYD.LiuY.Moreton.wave.photosphere2012ApJ...752L..23S}}  \\
         &       &      &&     &   & & {\hskip 0.1cm} O & \multicolumn{2}{r}{\cite{SrivastavaAK.kink.oscil.2011Aug9.X69.flare.2013ApJ...777...17S}} \\	

22Sep11 & 10:29 & X1.4 && 480 &   & H &  \multicolumn{3}{r}{$M_{\rm A}$=2.4, SEPs \cite{CarleyE.AIA.EUV.wave.shock.e-acc.2011-09-22X1.4.2013NatPh...9..811C}} \\
24Sep11 & 09:32 & X1.9 && 690 & {\hskip 0.1cm} W & & {\hskip 0.1cm} O  &  R  & {$M_{\rm f}$=1.4 \cite{ShenYD.LiuY.true.wave.2012ApJ...754....7S}}  \\
30Sep11 & 02:46 & C1.0 && 1100& {\hskip 0.1cm} W & & & \multicolumn{2}{r}{jet driven \cite{ZhengRS.1000km/s.EUV.wave.mini.filament.2012ApJ...753..112Z}} \\

23Apr12 & 17:38 & C2.0 && 960 & B & & & RT & \cite{ShenYD.LiuYu.AIA-wave.diffr.refr.refl.2013ApJ...773L..33S} \\
	

\hline
Range   &       & A\,--\,X  && \multicolumn{6}{l}{240\,--\,1390; speed mean: 740, median: 730, stand.~dev: 280  $\kmps$} \\
\multicolumn{5}{l}{Occurrences\tabnote{Each event is counted only once in each category.}} 
                               & 13B    &  8H  &  2P  &  6R  &  2 Moreton waves    \\
\multicolumn{5}{l}{(Total: 21 events, 33 articles)}           & 7W/2N  &  4C  &  8O  &  7T      \\

\hline
\multicolumn{2}{l}{Not included:} &
    \multicolumn{8}{l}{(1) The rest of the 171 AIA events in \inlinecite{NittaN.AIA.wave.stat.2013ApJ...776...58N}} \\
 && \multicolumn{8}{l}{(2) The six~Type-II associated events in \inlinecite{GopalswamyN.shock.formation.height.2013AdSpR..51.1981G}}  \\
 && \multicolumn{8}{l}{(3) The 12~SEP associated events in \inlinecite{ParkJ.SEP.STEREO.AIA.EUV.wave.2013ApJ...779..184P}} \\
  
\hline \end{tabular}
\label{table_EITwv-events} \end{table}

%
{\bf Hybrid models} {\it including both wave and non-wave components}
were suggested to reconcile the above controversy  \cite{Zhukov.EIT-wave2004A&A...427..705Z}. 
In this {\it bimodal} picture, a faster, but often weaker, outer component 
of a fast-mode wave travels ahead of a generally slower, but stronger 
inner component of CME-caused reconfiguration. This shift of paradigm has been
backed up by increasing evidence from numerical simulations 
({\it e.g.} \opencite{DownsC.MHD.2010-06-13-AIA-wave.2012ApJ...750..134D})
and observations especially those from AIA ({\it e.g.} \opencite{LiuW.cavity-oscil.2012ApJ...753...52L}). 
The hybrid concept has been widely recognized and
incorporated in a coherent picture explaining a broad range of observations
(\opencite{Patsourakos.Vourlidas.EIT-wave-review.2012SoPh..281..187P}, see their Section~11).
Another similar attempt to resolve the wave {\it vs.} non-wave controversy was
the recent classification of EIT waves 
(\opencite{Warmuth.Mann.3class.EIT.wave.2011A&A...532A.151W}; see \sect{subsect_kinem}).


In the following subsections, we elaborate on the above aspects to which AIA has made key contributions
and that are categorized in \tab{table_EITwv-events} for an exhaustive list of AIA detected EIT waves.
\tab{table_EITwv_prop} summarizes AIA updates of EIT-wave properties.
%
\begin{table}[bthp]	
\scriptsize		
\caption{EIT-wave properties updated with AIA observations.
  See Gallagher and Long~(2011) and Patsourakos and Vourlidas~(2012) for early values
  and more complete lists. 
  }
\tabcolsep 0.05in	
\begin{tabular}{lllllllllll}
 \hline
 Properties      &  Initial Speed     &  Acceleration     &  Period     &  Energy Content  \\ 
 \hline
 Typical Values  &  600\,--\,$700\kmps$ &  $(-400)$\,--\,$(+300)$ $\m \pss$ &  2\,--\,70~minutes &	$10^{28}$\,--\,$10^{29} \erg$  \\
\hline  \end{tabular}

\label{table_EITwv_prop} \end{table} 
%


\subsection{Wave/Non-wave Bimodality and Multiple Fronts}
\label{subsect_bimod}	



%
 \begin{figure}[thbp]      
 \begin{center}
 \includegraphics[height=4.2cm, bb=188 1 378 212, clip=]{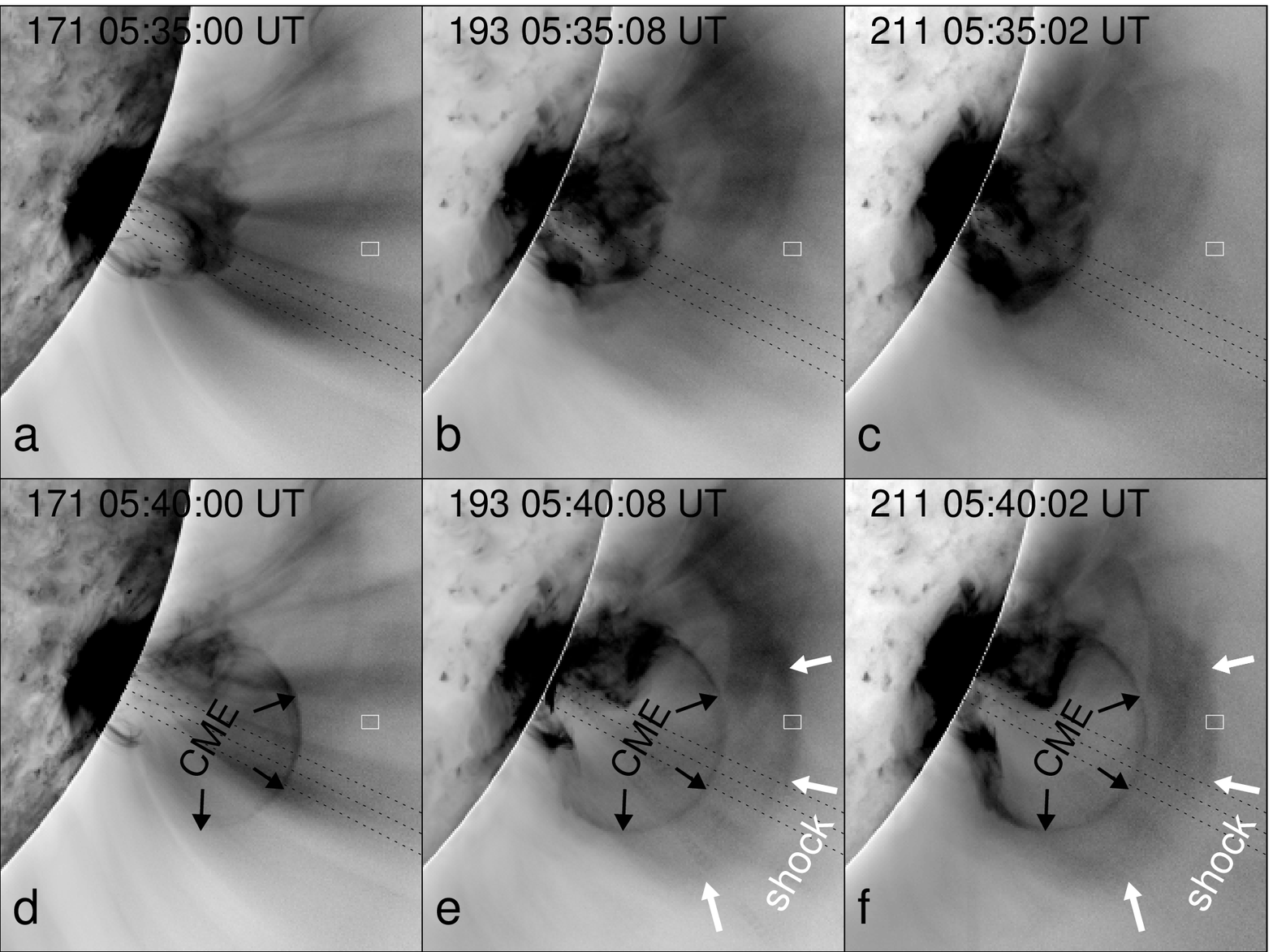}	
 \includegraphics[height=4.2cm]{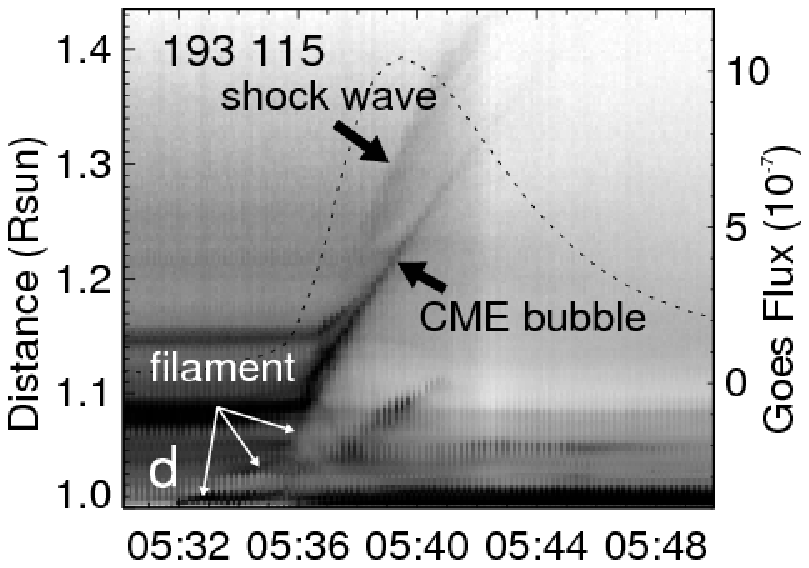}
 \includegraphics[height=4.2cm]{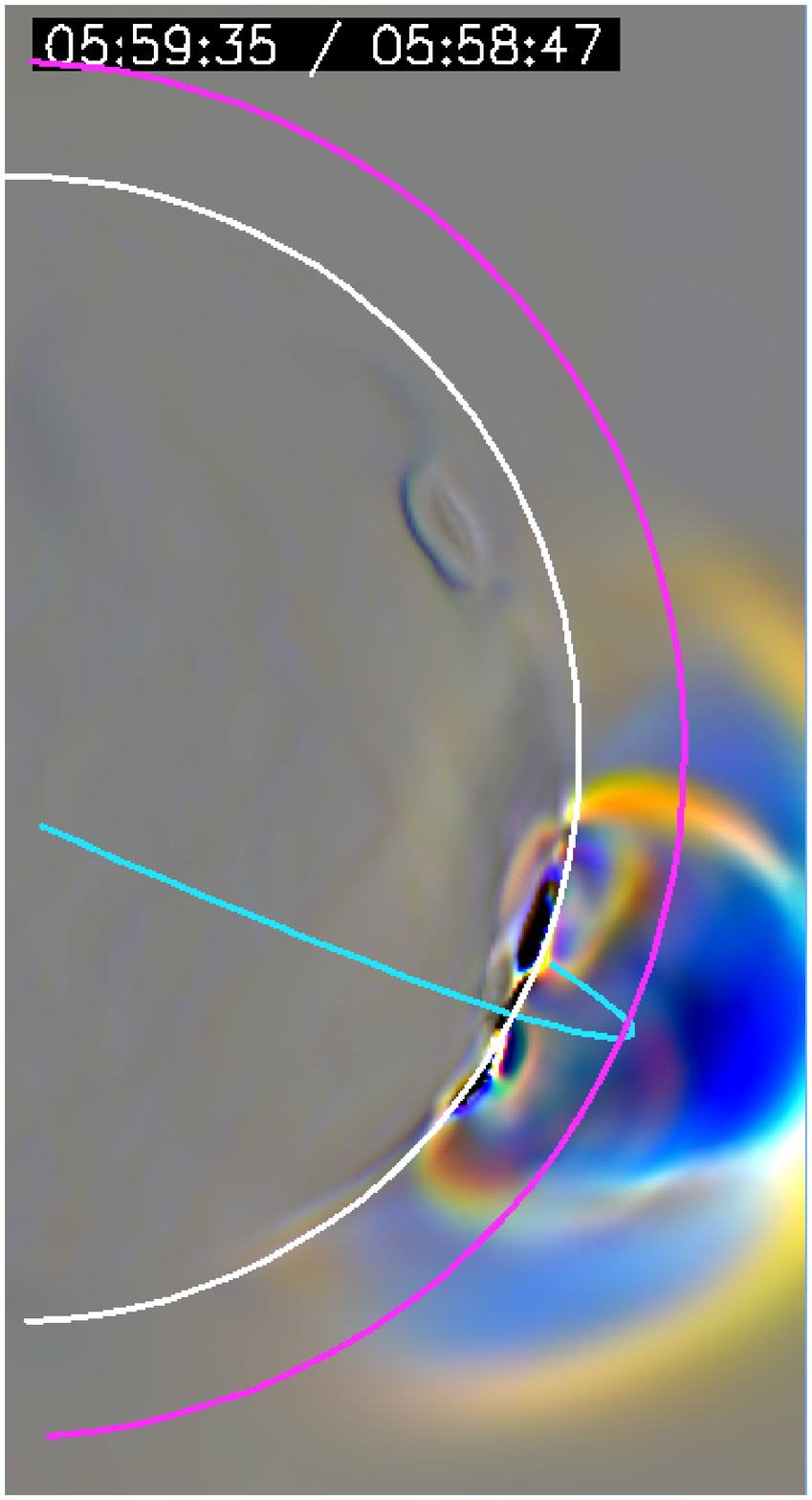}
 \end{center}
 \caption[]{	
 Wave/non-wave bimodality evidenced in a shock running ahead of its CME driver on 13 June 2010
 in an AIA 193~\AA\ image ({left}) and space--time plot from a nearly vertical cut ({middle})
 (from \opencite{MaSL.AIA.shock.2011ApJ...738..160M}).
  {Right}: Simulated composite image (red: 211 \AA, green: 193 \AA, and blue: 171 \AA) 
 for the same event showing a heated outer wave front in yellow followed by a cooling region in blue
 immediately ahead of an inner CME front (from \opencite{DownsC.MHD.2010-06-13-AIA-wave.2012ApJ...750..134D}).
 } \label{bimod_MaSL.eps}
 \end{figure}
%
Although multiple fronts are not uncommon in
Moreton waves ({\it e.g.} \opencite{Narukage.3-Moreton-wave-fronts.2008ApJ...684L..45N})
and \ion{He}{1}~10830~\AA\ waves ({\it e.g.} \opencite{Gilbert.HeI.wave.2004ApJ...610..572G}),
EIT waves were originally considered as a single-pulse phenomenon
\cite{Wills-Davey.EIT-wave-soliton.2007ApJ...664..556W},
This view has recently changed.		
The first indication of a {\it bimodal composition} was found in a 2D MHD model  
\cite{ChenPF.EIT-wave-MHD.2002ApJ...572L..99C},	
in which a series of fast-mode waves	
travel ahead of a CME-driven compression.	
Chen~{\it et al.}~called the former ``coronal Moreton waves", ascribing their non-detection to
EIT's low cadence, and instead interpreted the latter as an EIT wave. 
Recent 3D global MHD simulations 	
(\opencite{CohenO.EITwave.non-wave.both.2009ApJ...705..587C};
\opencite{DownsC.MHD.EUVIwave.2011ApJ...728....2D}, \citeyear{DownsC.MHD.2010-06-13-AIA-wave.2012ApJ...750..134D})
confirmed the composition of an outer fast-mode front
and an inner CME compression front. Additional, but non-critical
contributions to the inner EUV emission may come from a current shell or magnetic reconnection.

Arguably the first observational evidence of the {wave/non-wave bimodality} was provided
by TRACE and SOHO/CDS \cite{HarraSterling.fast.leading.edge.EIT-wave.2003ApJ...587..429H},
while other analyses of the same event reached an opposite conclusion favoring a single pulse 
\cite{Wills-DaveyThompson.TRACE-EUV-wave.1999SoPh..190..467W,Delannee.EIT-pseudo.wave.2000ApJ...545..512D}.
More convincing evidence came from quadrature STEREO observations 	
of an EIT wave front decoupling from a CME flank as the latter slows down
\cite{Patsourakos.EUVI-fast-mode-wave.2009ApJ...700L.182P}. 
Recent AIA observations have established the bimodality as a general trend, 	
as seen in 13 out of 21 EIT waves (see \tab{table_EITwv-events}).
One clear example, as shown in \fig{bimod_MaSL.eps}, is a coronal shock propagating at $600 \kmps$ 
ahead of a slower CME front at $410 \kmps$ \cite{MaSL.AIA.shock.2011ApJ...738..160M}. 	
In another example \cite{ChenPF.Wu.Coexis-fast-mode.2011ApJ...732L..20C}, 	
the slower front decelerates and stops at a separatrix surface, as predicted for	
field-line stretching \cite{ChenPF.EIT-wave.2005ApJ...622.1202C}.	
In addition to kinematic differences, the two components can exhibit distinct thermal properties, 
as discussed in \sect{subsect_therm}.
 \begin{figure}[thbp]      
 \begin{center}
 \includegraphics[height=6.7cm]{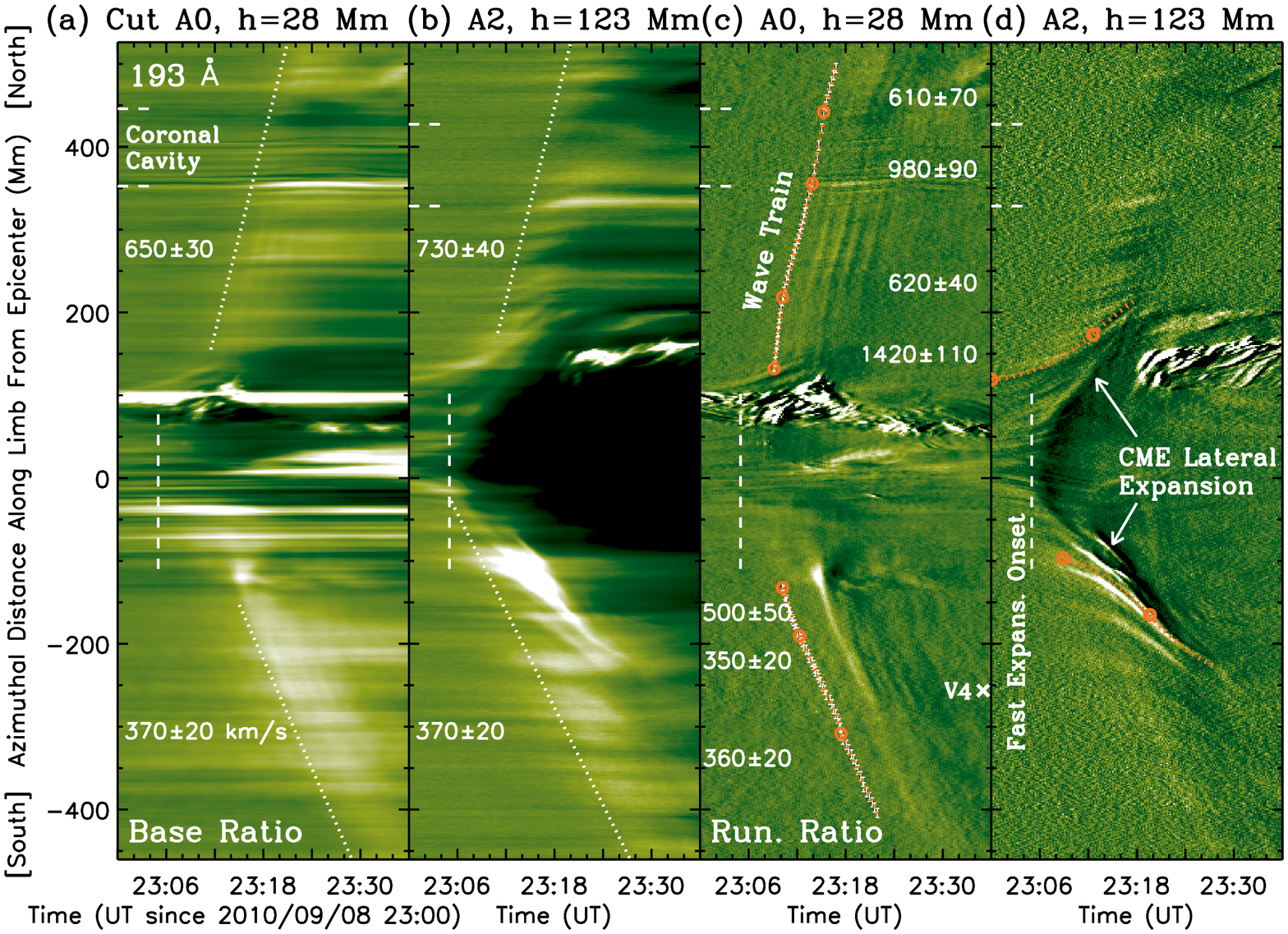}
 \includegraphics[height=6.7cm]{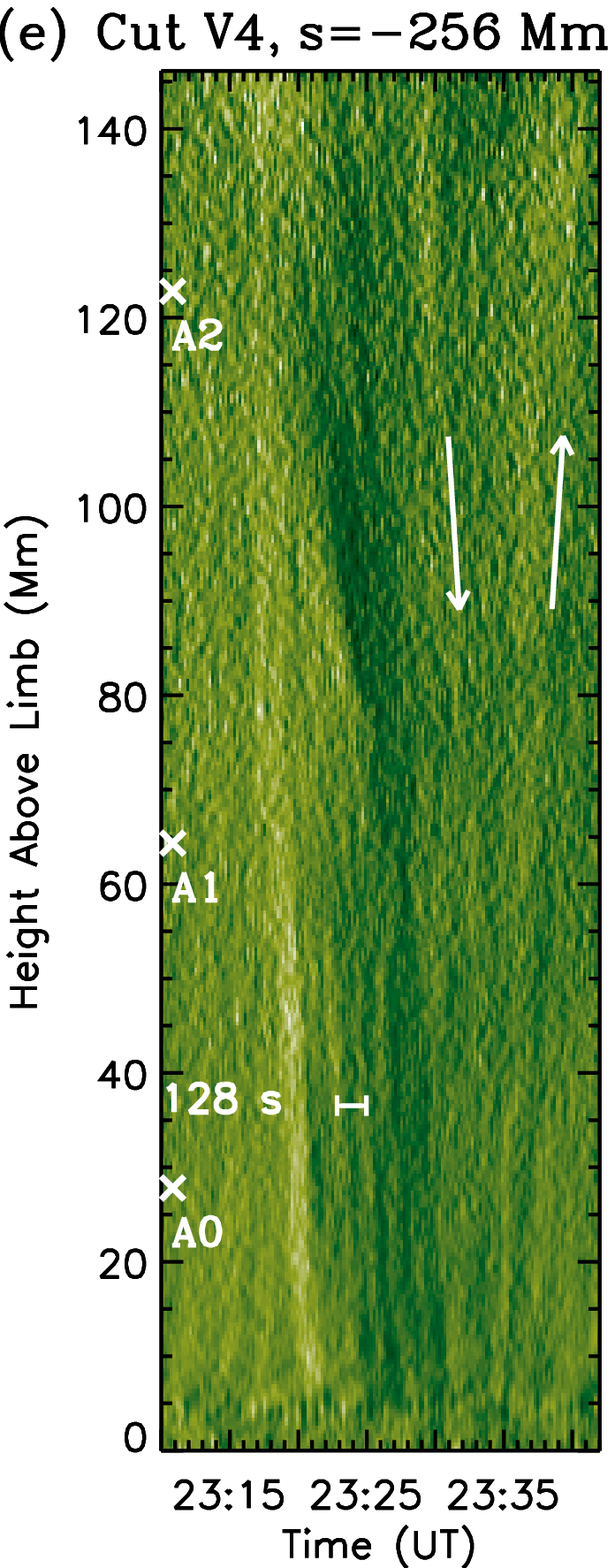}
 \end{center}
 \caption[]{
 Wave/non-wave bimodality exemplified in a low-corona ($h \lesssim 100 \Mm$) 
 quasi-periodic wave train within a broad EIT wave pulse that 
 decouples from the high-corona CME lateral expansion on 08 September 2010
 (from \opencite{LiuW.cavity-oscil.2012ApJ...753...52L}). 
 This is evident in AIA 193~\AA\ base-ratio ((a) and (b)) and running-ratio ((c) and (d)) 
 space--time plots from off-limb cuts at two constant heights,
 and a running-ratio plot (e) from a vertical cut at 256~Mm South of the epicenter.
 The slanted stripes in (e) indicate delayed arrivals of the wave train at lower heights and
 thus wave fronts forwardly inclined toward the solar surface.
 The change of slope near 23:35 from negative to positive is suggestive of an echo or reflection 
 from the underlying chromosphere, as revealed in an MHD simulation 
 \cite{WangHongjuan.EIT-slow-mode-wave.2009ApJ...700.1716W}.
 } \label{0908_global-train.eps}
 \end{figure}
	%

A critical addition by AIA to this bimodal picture is not only one, but {\it multiple,
quasi-periodic fronts} within the outer wave component itself {\it ahead} of a CME flank, 
manifesting its true wave nature \cite{LiuW.cavity-oscil.2012ApJ...753...52L}.
As schematically shown in \fig{QFP-2trains.eps}a, such a {quasi-periodic wave train} 
comprises low-corona wave fronts forwardly inclined toward the solar surface 
and travels along it to distances $\gtrsim\,$$\Rsun/2$.  	
In the example shown in \fig{0908_global-train.eps},
there is a dominant two~minute period and high initial speeds up to $1400 \kmps$.
Such wave trains
were potentially detected by EUVI at a lower 75-second cadence
(\opencite{Patsourakos.early-CME_EIT-wave-train.2010A&A...522A.100P}, see their Figure 11)
and resemble simulated coronal Moreton waves \cite{ChenPF.EIT-wave-MHD.2002ApJ...572L..99C}.
Their relationship with QFP wave trains in coronal funnels {\it behind} CME fronts
will be discussed in \sect{subsect_2trains}.

Likewise, the inner non-wave component can contain multiple fronts as well.
In the first EIT wave detected by AIA on 08 April 2010
\cite{LiuW.AIA-1st-EITwave.2010ApJ...723L..53L},
{\it multiple, ripple-like sharp fronts} travel in a wide speed range of 40\,--\,$240 \kmps$
behind a diffuse front at uniform speeds of 200\,--\,$240 \kmps$.
Some sharp fronts undergo acceleration, suggestive of
being driven by lateral CME expansion.
Some faster sharp fronts even overtake slower ones, which, with a hindsight, is possibly
due to LOS projection of loops expanding at different heights.	

\subsection{Kinematics}		
\label{subsect_kinem}

%


Kinematics of EIT waves, usually measured at their leading fronts, 	
has a strong bearing on their physical nature.
According to a catalog of 176 EIT waves from January 1997 to June 1998
on the early rising phase of solar cycle 23
\cite{ThompsonB.EIT-wave-catalog.2009ApJS..183..225T},
their average speeds are distributed in a wide range of 50\,--\,$700 \kmps$ 
with typical values of 200\,--\,$400 \kmps$.
One cycle later, a new catalog of 171 events detected by SDO/AIA  
from April 2010 to January 2013 was compiled \cite{NittaN.AIA.wave.stat.2013ApJ...776...58N}.
As shown in \fig{Nitta_stat_hist.eps}, a subset of 138 on-disk events has 
an even broader speed distribution of 200\,--\,$1500 \kmps$ with a {\it much higher 
mean} of $644 \kmps$, a median of $607 \kmps$, and a standard deviation of $244 \kmps$.	
The large lower bound of $200 \kmps$ is partly due to the sample selection threshold
(angular width $\geq\,$$45 \degree$, travel distance $\geq\,$200~Mm from the epicenter).

\begin{figure}[thbp]      
 \begin{center}
 \includegraphics[width=0.44\textwidth]{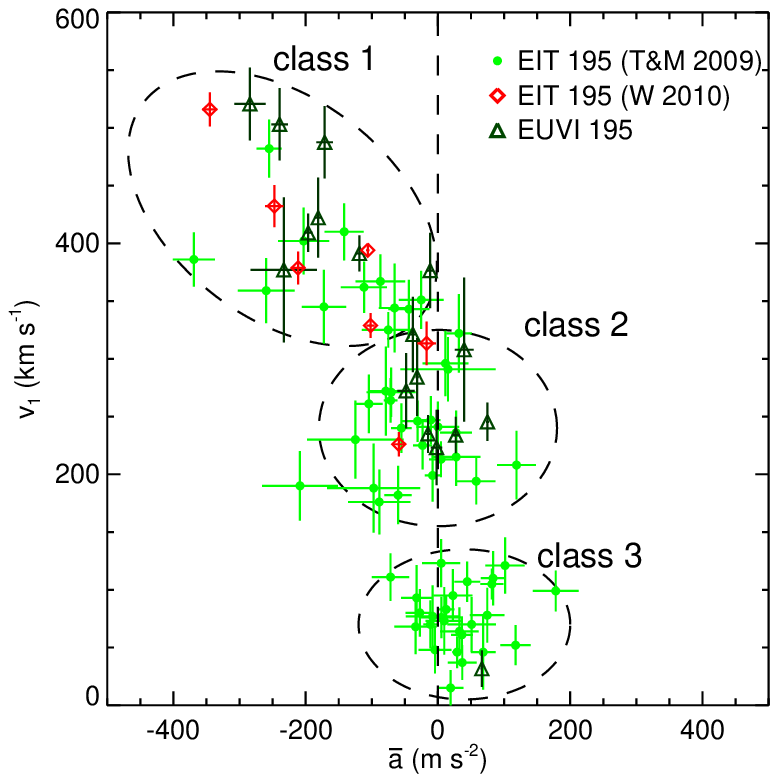}
 \includegraphics[width=0.545\textwidth]{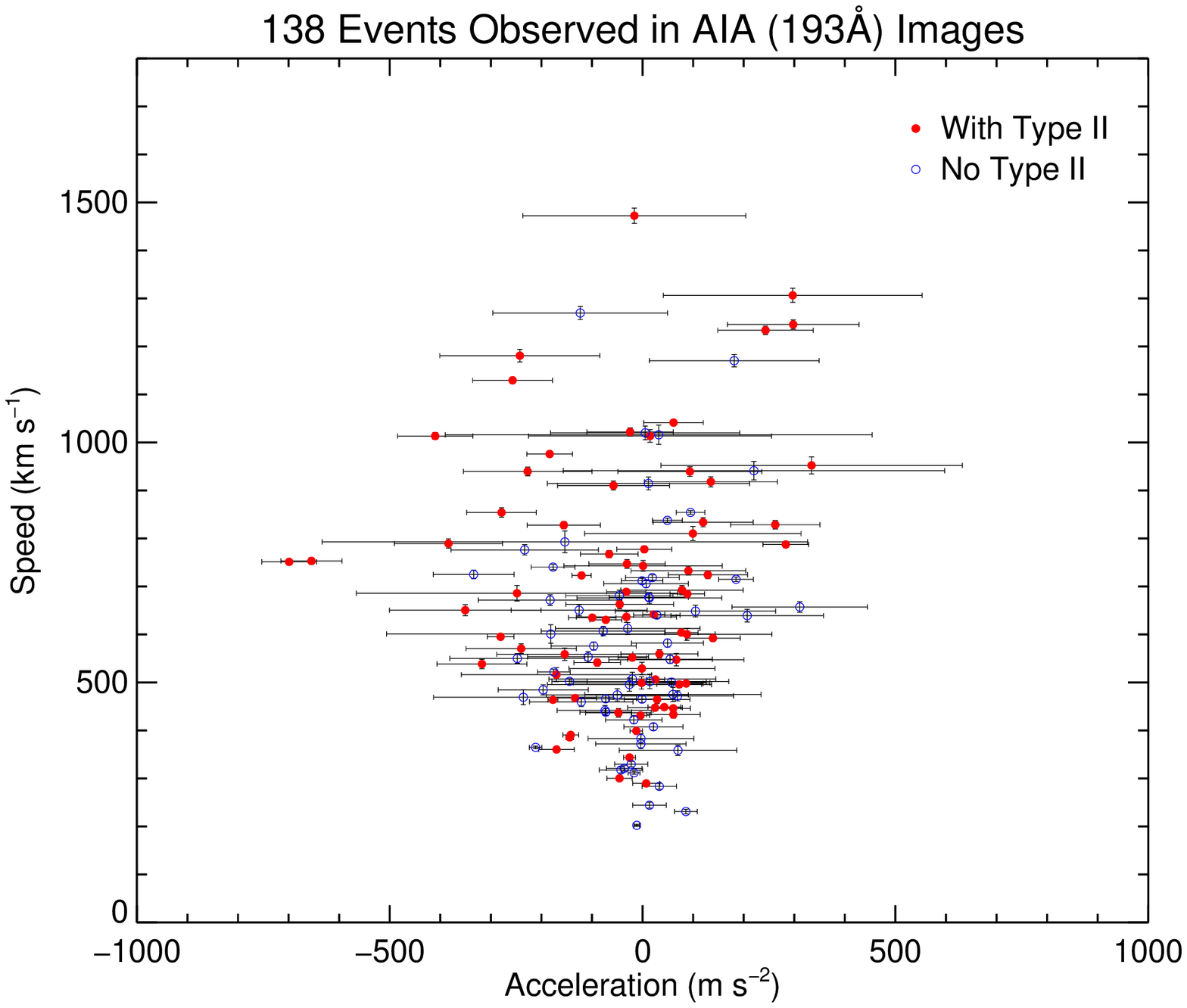}
 \end{center}
 \caption[]{	
 Distributions of EIT wave speeds {\it vs.}~accelerations obtained from 
 EIT and EUVI showing three distinct classes 
 (left; from \opencite{Warmuth.Mann.3class.EIT.wave.2011A&A...532A.151W})	
 and from AIA appearing more continuously but with some similar trends
 (right; from \opencite{NittaN.AIA.wave.stat.2013ApJ...776...58N}).
 } \label{Nitta_stat_v-a.eps}
 \end{figure}

Observational case studies 	
with three generations of EUV imagers 	
have found many EIT waves of nearly {\it constant speeds} ({\it e.g.} 	
\opencite{MaSL.EUVI-wave.2009ApJ...707..503M}; \opencite{LiuW.AIA-1st-EITwave.2010ApJ...723L..53L})	
and many others with {\it decelerations} 	
({\it e.g.} \opencite{WarmuthA.multiw-EIT-wave.2004A&A...418.1101W}; \opencite{LongD.EUVI-wave.2008ApJ...680L..81L};
\opencite{ZhaoXH.WuST.EIT-wave-2010-01-17.2011ApJ...742..131Z};
\opencite{XueZK.EUV.wave.deform.decel.2013A&A...556A.152X}).	
A smaller number of events exhibit acceleration 	
and even erratic acceleration and deceleration episodes
({\it e.g.} \opencite{Zhukov.EIT-wave-acc-decel.2009SoPh..259...73Z}). 
Such accelerations often occur at low speeds $\lesssim$\,$200 \kmps$, 
suggestive of a physical connection with CME expansion or filament eruptions.
This general trend was confirmed by the new AIA catalog \cite{NittaN.AIA.wave.stat.2013ApJ...776...58N}
in which more events (57\,\% {\it vs.}~43\,\%) experience deceleration than acceleration.
The accelerations, as shown in \fig{Nitta_stat_v-a.eps} (right), are compactly distributed around 
and concentrated below zero, with a mean of $-37 \mpss$, a median of $-12 \mpss$, 
and a standard deviation of $162 \mpss$,
indicating nearly uniform speeds for a large fraction of the events.
%

A statistical study including 61 events observed by EIT and 17 by EUVI
revealed distinct types of EIT waves of possibly different physical origins
\cite{Warmuth.Mann.3class.EIT.wave.2011A&A...532A.151W}.
As shown in \fig{Nitta_stat_v-a.eps} (left), the distribution of 
the initial speed [$v_1$] {\it vs.}~the average acceleration [$\bar{a}$] appears to cluster in {\it three classes}.
  Class~1 includes fast waves of $\geq\,$$320 \kmps$ with 
stronger decelerations at higher speeds, interpreted as
{\it nonlinear fast-mode} waves or shocks decelerating due to amplitude decay
resulting from dissipation and geometric expansion.
  Class~2 corresponds to waves of moderate and nearly uniform speeds of 170\,--\,$320 \kmps$
with or without slight decelerations or accelerations,
interpreted as {\it linear fast-mode} waves. The final speeds of these two classes 
fall in a compact range of 200\,--\,$300 \kmps$, similar to the 180\,--\,$380 \kmps$ range
found by \inlinecite{Patsourakos.Vourlidas.EIT-wave-review.2012SoPh..281..187P}
from published EUVI and AIA observations. Such final speeds are comparable to 
the quiet-Sun fast-magnetosonic speeds, further suggesting closely related origins.
  Class~3 includes very slow waves 	
$\leq\,$$130 \kmps$ showing constant speeds or slight accelerations, which were interpreted
as non-wave, CME-related {\it coronal reconfiguration}, 
although slow-mode waves were also suggested 
({\it e.g.} \opencite{Podladchikova.EUVI.mini.wave.2010ApJ...709..369P}).

These distinct classes cannot be readily identified	
in the recent AIA ensemble (\opencite{NittaN.AIA.wave.stat.2013ApJ...776...58N}; see \fig{Nitta_stat_v-a.eps}, right),
which appears in a more continuous distribution.	
However, within 450\,--\,$800 \kmps$, there is a similar, but weak positive correlation
between decelerations and speeds, corresponding to the above Class~1.
Events in the 200\,--\,$450 \kmps$ range with small absolute accelerations
correspond to Class~2,
while slow events of Class~3 are left out, 	
partly because of the sample selection thresholds mentioned earlier.
We stress that distinct class boundaries
are less important than general trends and may not be well-defined at all
because of the diversity in physical conditions among different events.
The above classification is not incompatible with the hybrid models.
The general presence of both wave and non-wave components can have different observational manifestations
that may fall in one of the three classes, depending on specific circumstances.	
For example, if the stand-off distance between a fast-mode shock and its driving CME is too small to be
resolved by an EUV imager, only one component would be detected and identified
as a nonlinear shock (Class~1).

The statistical analysis discussed here has its limitations. 
Each EIT wave is represented by only two numbers, its speed and acceleration,
{\it e.g.} measured in the direction of the highest speed, 
while other kinematics determining factors, such as the direction (anisotropy), 
height ({\it e.g.} speeds increasing with height; \opencite{LiuW.cavity-oscil.2012ApJ...753...52L}; 
\opencite{Kwon_STEREO-EUV-wave-high-corona.2013ApJ...766...55K}),
and observing passband or temperature \cite{LongD.AIA.EUV-wave.2011ApJ...741L..21L},
are ignored and should be included in future studies.



\subsection{True Wave Behaviors: Reflections, Transmissions, and Refractions}
\label{subsect_optic}	



The propagation of any true wave can be altered by the inhomogeneity of its medium and exhibit
such behaviors as reflections, transmissions, and refractions at interfaces with
strong gradients of the characteristic wave speed.	
Active regions and coronal holes are regions of high \Alfven and fast-magnetosonic speeds 
where anomalies of EIT wave propagation take place, as predicted in numerical models
\cite{WangYM.EIT-fastMHDwave.2000ApJ...543L..89W,OfmanThompson.EIT-wave-fast-mode.2002ApJ...574..440O}.
Early low-cadence observations indicated that EIT waves avoid such regions
(\opencite{ThompsonB.EIT-wave-discover.1998GeoRL..25.2465T}, \citeyear{ThompsonB.EIT-Moreton-wave.1999ApJ...517L.151T}).
EUVI discovered reflections from a coronal hole 
\cite{Gopalswamy.EIT-wave-reflect-CH-illusion.2009ApJ...691L.123G},
while AIA provided more definitive evidence of such behaviors.	
%
 \begin{figure}[thbp]      
 \begin{center}
 \includegraphics[height=7cm]{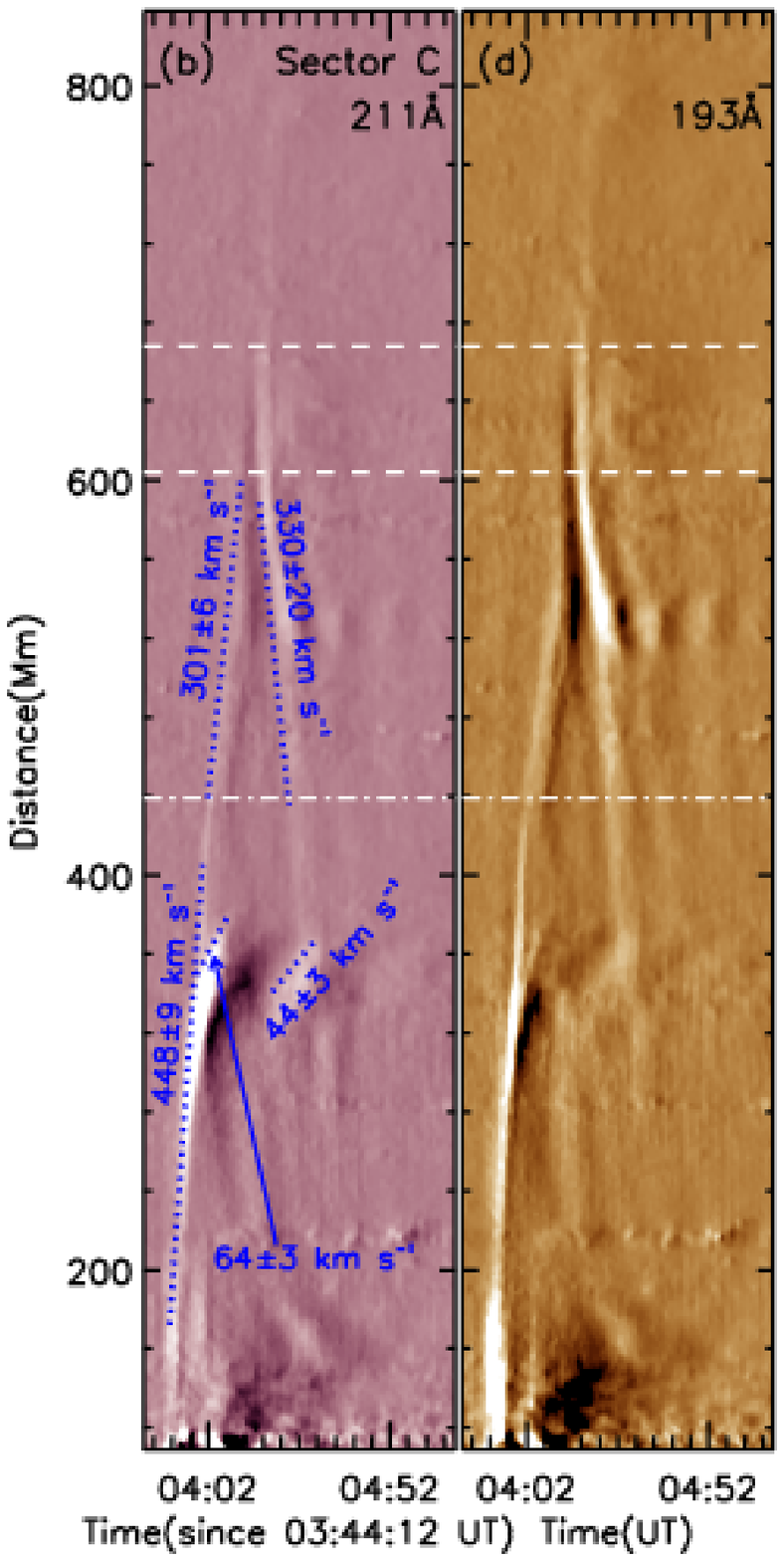}	
 \includegraphics[height=7cm]{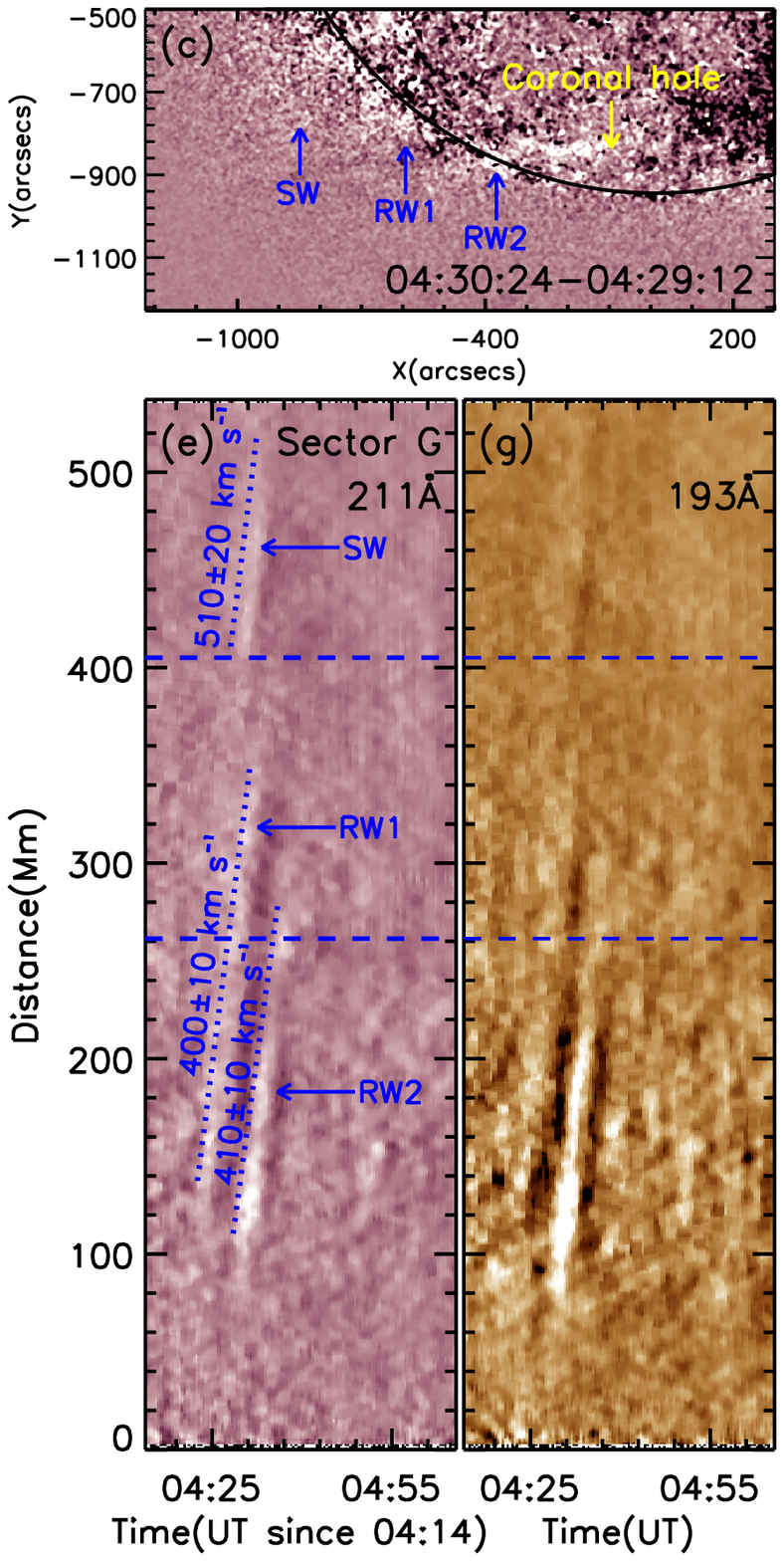}
 \includegraphics[height=7cm]{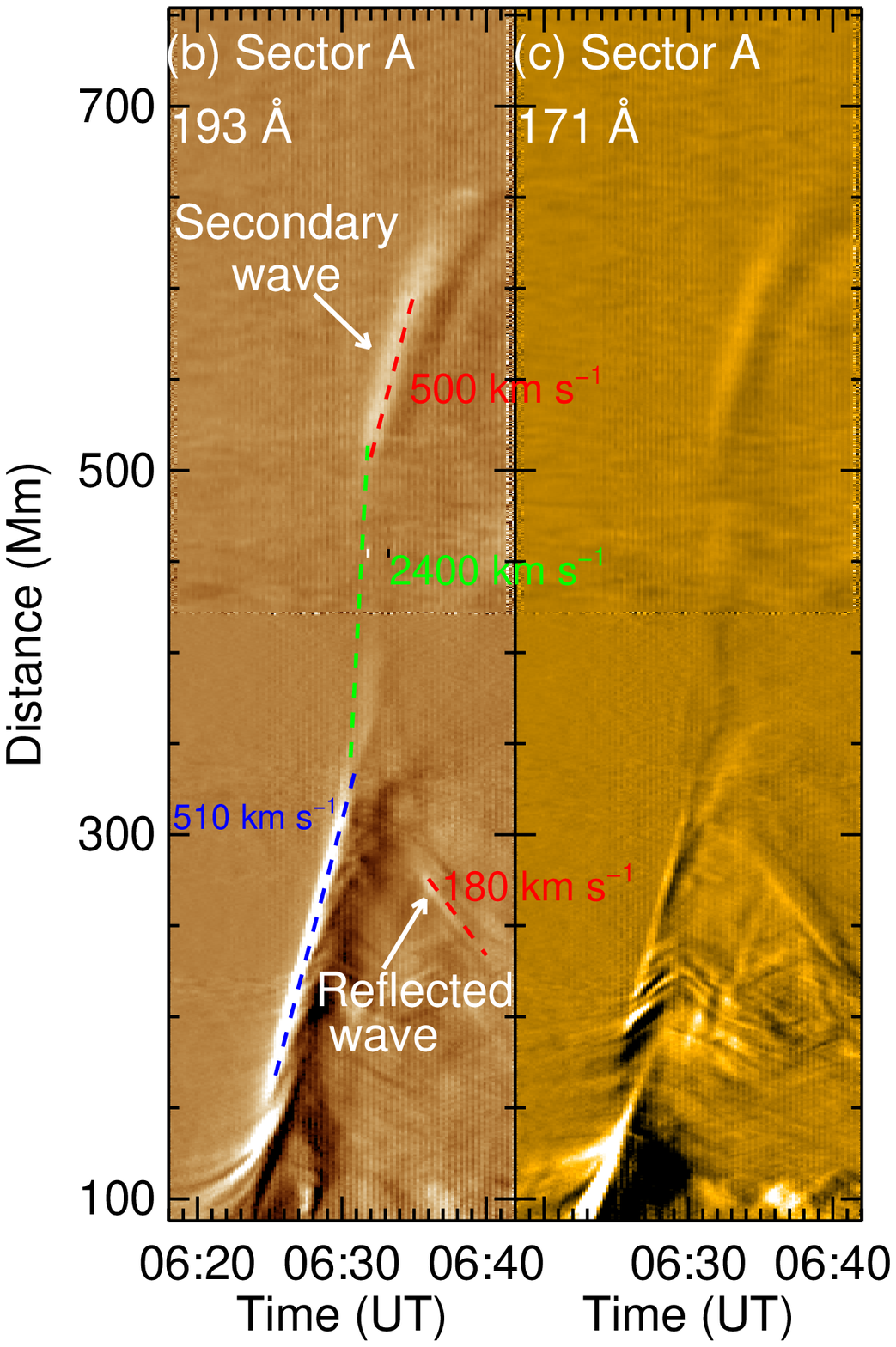}
 \end{center}
 \caption[]{	
 Secondary waves shown in AIA space--time plots
 on 04 August 2011 (left and middle; from \opencite{YangLH.AIA.EIS.EUV-wave.2013ApJ...775...39Y}) 
 and 07 June 2011 (right; from \opencite{LiTing.2010Jun07.AIAwave.reflect.2012ApJ...746...13L}).
  {Left}: Reflection from a coronal bright point, stationary intensity oscillations
 (at $\approx$500~Mm), and loop displacement (at $\approx$300~Mm).
  {Middle}: Double-pulsed reflections (RW1,2) from a polar coronal hole shown in the images on the top
 and their own secondary wave (SW).	
  {Right}: Apparent discontinuity or disappearance of a wave 
 inside an active region, implying a high transmission speed of $2400 \kmps$.
 } \label{reflect.eps}
 \end{figure}
%

{\bf Reflections} of EIT waves from various coronal structures, including coronal holes, active regions,
and quiet-Sun bright points, have been detected by AIA 
\cite{LiTing.2010Jun07.AIAwave.reflect.2012ApJ...746...13L,OlmedoO.2011-2-15_X2.AIA.EUV.wave.2012ApJ...756..143O,%
ShenYD.LiuYu.AIA-wave.diffr.refr.refl.2013ApJ...773L..33S,YangLH.AIA.EIS.EUV-wave.2013ApJ...775...39Y}.
The reported primary waves travel at typically 300\,--\,$800 \kmps$ while the reflected waves travel at similar or 
fractionally lower speeds in the range of 100\,--\,$500 \kmps$ ({\it e.g.} \fig{reflect.eps}, left). 
Primary and reflected waves share similar thermal
properties, such as appearing as brightening at 193~\AA\ but darkening at 171~\AA\
\cite{LiTing.2010Jun07.AIAwave.reflect.2012ApJ...746...13L,YangLH.AIA.EIS.EUV-wave.2013ApJ...775...39Y},
suggestive of a common physical nature. 
Strong EIT waves 	
can produce cascades of secondary waves by multiple encounters with local structures.	
For example, the EIT wave associated with the 04 August 2011 M9.3 flare		
travels more than $1 \Rsun$ to reach a polar coronal hole,		
where double reflected waves emerge and again produce their own secondary wave when encountering
large-scale loops (see \fig{reflect.eps}, middle; \opencite{YangLH.AIA.EIS.EUV-wave.2013ApJ...775...39Y}).
Multiple-pulsed reflections, as modeled numerically \cite{Schmidt.Ofman.3D-MHD-eitwv.2010ApJ...713.1008S},
are possible signatures of resonance of the coronal hole in response to the primary wave impact.

A puzzle about reflected EIT waves is that many of them are
{\it fractionally slower} (by up to a factor of four) than the incident waves in the same medium.	
One possibility is that the incident wave is shocked with a fast-magnetosonic Mach number
$M_{\rm f}>1$ while the reflected wave is a linear fast-mode. 
In addition, the passage of a shock or CME 
({\it e.g.} \opencite{GopalswamyN.SEP.twin-CME.2004JGRA..10912105G})	
can modify the magnetic and plasma conditions and thus the fast-magnetosonic speed [$v_{\rm f}$]
at which a reflected wave will travel.
However, for the extreme cases of $1/4$ speed ratio, this would imply a high $M_{\rm f}$ up to four, 
indicating a strong shock that would not survive dissipation.	
We suggest that LOS projection of reflected waves traveling upward at large angles from the solar surface,
when viewed from above, could \referee{reduce apparent wave speeds and} contribute to this discrepancy.	
 \referee{\inlinecite{Kienreich.STA-PROBA2.EUV.wave.reflect.2013SoPh..286..201K}
 suggested that in the presence of a bulk plasma flow behind a primary wave traveling in
 the same direction ({\it e.g.} the downstream flow of a shock), 
 the reflected wave would propagate through the oppositely directed flow and thus appear at a reduced speed seen in the rest frame.
 }

{\bf Transmissions} of EIT waves are somewhat more subtle than reflections and have been found in
flux-rope coronal cavities \cite{LiuW.cavity-oscil.2012ApJ...753...52L},
active regions \cite{ShenYD.LiuYu.AIA-wave.diffr.refr.refl.2013ApJ...773L..33S},
and corona holes (\opencite{OlmedoO.2011-2-15_X2.AIA.EUV.wave.2012ApJ...756..143O};
{\it cf.} Moreton-wave transmission, \opencite{VeronigA.EIT.Moreton.wave.in.coronal.hole.2006ApJ...647.1466V}).
As expected for higher fast-magnetosonic speeds in these regions, the speeds of the transmitted
waves are higher (by $\approx$10\,--\,60\,\%) or comparable to those of the incident waves.
For example, as shown in \fig{0908_global-train.eps}, the quasi-period wave train has an elevated
speed of $\approx$$1000 \kmps$ within the coronal cavity but a lower speed of $\approx$$600 \kmps$
before and after \cite{LiuW.cavity-oscil.2012ApJ...753...52L}.
Transmissions through topological separatrix surfaces around such structures
are strong evidence of a true fast-mode wave component, because a non-wave component
related to field line stretching is supposed to stop there \cite{ChenPF.EIT-wave.2005ApJ...622.1202C}.


Another phenomenon that we call ``apparent discontinuity" is related to transmission
and could be its special case. In the example reported by \inlinecite{LiTing.2010Jun07.AIAwave.reflect.2012ApJ...746...13L},
as shown in \fig{reflect.eps} (right), when an EIT wave reaches 	
an active region, the primary wave ``disappears" or exhibits marginal signals,
while a new wave front reemerges from its far side.	
The authors offered two open interpretations:
i)~One is a fast-mode wave with a weak signal often below detection transmitting through the active region,
as numerically modeled \cite{OfmanThompson.EIT-wave-fast-mode.2002ApJ...574..440O},
at a high speed up to $2400 \kmps$, five times the original wave speed.	
The low signal and high speed are expected due to the strong magnetic field
and magnetic pressure/tension of an active region, making it very rigid, which can result in less density or 
temperature perturbations and thus EUV intensity variations.	
This can also be explained by {\it conservation of wave-energy flux},
a product of energy density and group velocity.
 \editor{In general, the relative amplitudes and energy densities of the incident,
 reflected, and transmitted waves depend on their speeds in the respective media and the incident angle
 among other quantities.	
 In addition to energy loss due to reflection and damping,}
a higher transmission wave speed can lead to an even lower transmission energy density and thus amplitude
than those of the incident wave, as manifested in \fig{0908_global-train.eps}c for a flux-rope coronal cavity.
ii)~Another possibility for ``apparent discontinuity" is sequential stretching of neighboring high-altitude magnetic-field lines
whose footpoints straddle the active region, causing a jump of a non-wave signal from one side of it to the other
\cite{ChenPF.EIT-wave.2005ApJ...622.1202C}.
iii)~\inlinecite{ShenYD.LiuYu.AIA-wave.diffr.refr.refl.2013ApJ...773L..33S} proposed a third interpretation
as wave diffraction around the active region, which, in our view,
cannot explain the unusually high speed of ``transmission", because it would take a much longer time
if the wave were to travel around the active region at the original wave speed on the quiet Sun.

One of the reasons leading to these open possibilities is the ambiguity due to LOS projection onto the solar disk
where the height-dependent information is lost and where most of the reported secondary waves were detected.
An alternative limb view is needed to recover {\it height-dependent propagation}, as reported by 
\inlinecite{LiuW.Ofman.transmit.reflect.2014}. They found within an active region
a continuous change of the wave-front orientation
from forward to backward inclination toward the solar surface, 
meaning larger wave speeds at lower heights.	
This is consistent with the expected fall-off of \Alfven and fast-magnetosonic speeds with height
in an active region and indicates that the EIT wave indeed transmits through it.


{\bf Refraction} of EIT waves can be referred to as a subset of transmissions that	
involve gradual deflections of the wave toward regions of progressively lower wave speeds
of an inhomogeneous medium. This is predicted for fast-mode waves by geometric acoustics 
(\opencite{Uchida.Moreton-wave-sweeping-skirt.1968SoPh....4...30U}; 
\opencite{Afanasyev.Uralov.shock-EUV-wave-I.2011SoPh..273..479A})
and 3D MHD simulations \cite{OfmanThompson.EIT-wave-fast-mode.2002ApJ...574..440O}.
Observational evidence includes anisotropic lateral propagation avoiding active regions 
or coronal holes ({\it e.g.} \opencite{ThompsonB.EIT-Moreton-wave.2000SoPh..193..161T}, their Figure~5), 
the change in vertical inclination of the wave front during its traverse through
an active region noted above \cite{LiuW.Ofman.transmit.reflect.2014},
and forward inclinations of the low-corona wave fronts toward the solar surface
away from the high corona of greater fast-mode speed [$v_{\rm f}$]
({\it e.g.} \opencite{LiuW.cavity-oscil.2012ApJ...753...52L}; 
\opencite{Kwon_STEREO-EUV-wave-high-corona.2013ApJ...766...55K};
especially during the late stage when the effect of the downward compression of the wave driver diminishes,
{\it cf.} \sect{subsect_gener}).
Near a null point where the magnetic field and \Alfven speed [$v_A$] approach zero
and $v_{\rm f}$ drops to the sound speed [$c_{\rm s}$], refraction can cause a fast-mode wave to wrap around 
and converge toward it ({\it e.g.} Figure~2 of \opencite{McLaughlinJ.HoodA.MHD.wave.2D.null.2004A&A...420.1129M}),
but such observations of EIT waves have not been reported.
At times, refractions and reflections may be observationally indistinguishable.
For example, 	
an EIT wave can be refracted at a large-angle away from an active region, 
depending on the incident angle and the spatial gradient of $v_{\rm f}$, and seemingly appear as a reflection. 

Similar to the ``apparent discontinuity" in {\it large} active regions discussed above,
a secondary wave can appear ahead of a primary EIT wave 
that encounters {\it small} coronal bright points or loops.
The former can travel faster than the latter and both co-exist for some time,
forming a ``bifurcation" shape in space--time plots.
Examples have been reported by \citeauthor{LiTing.2010Jun07.AIAwave.reflect.2012ApJ...746...13L}
(\citeyear{LiTing.2010Jun07.AIAwave.reflect.2012ApJ...746...13L}, see their Figures~4 and 5),
who interpreted this as evidence of LOS projection of multiple components of refraction
at different heights. We suggest another possibility that dispersion due to
the increased inhomogeneity of the medium at such locations can lead to multiple fronts 
of different speeds \cite{Wills-Davey.EIT-wave-soliton.2007ApJ...664..556W}.
Such open interpretations need to be validated, {\it e.g.} by numerical simulations.

\subsection{Perturbations to Local Structures: Sequential Deflections/Oscillations, Stationary Brightenings,
and Sympathetic Eruptions}
\label{subsect_oscil}




%
 \begin{figure}[thbp]      
 \begin{center}
 \includegraphics[height=9cm]{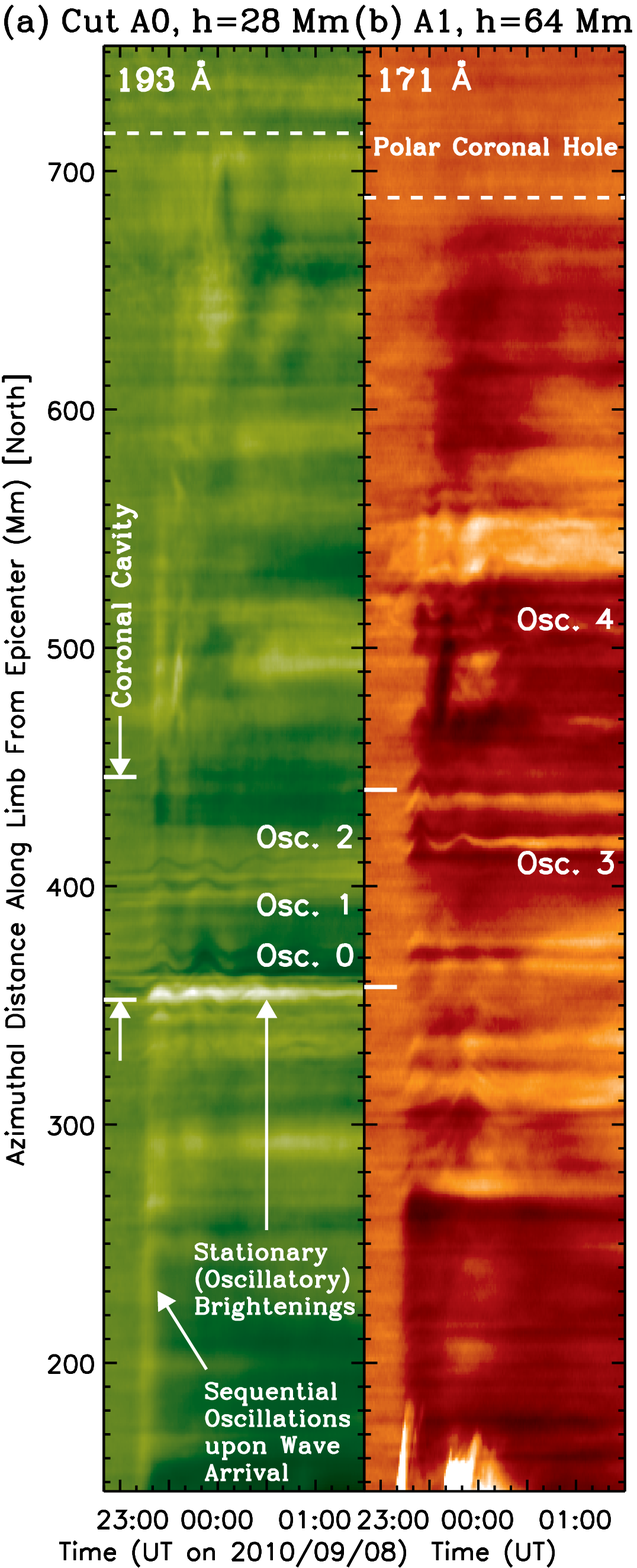}	
 \includegraphics[height=9cm]{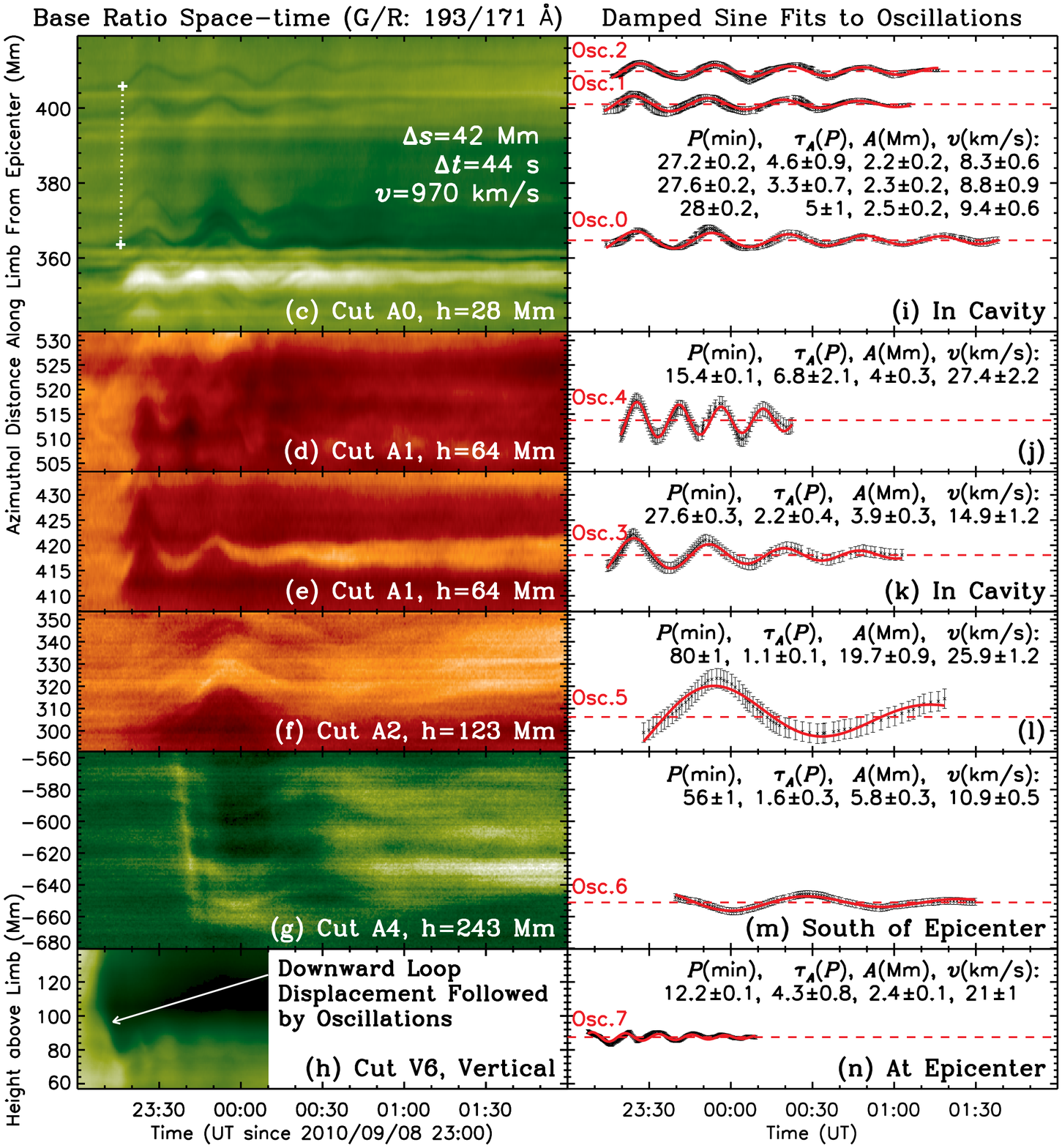}	
 \end{center}
 \caption[]{
 Sequential transverse oscillations set off by the arrivals of an EIT wave at increasing
 distances on 08 September 2010, shown in base-ratio space--time plots from off-limb cuts at
 constant heights for 193~\AA\ (a) and 171~\AA\ (b).
 Note the stationary and oscillatory brightenings at the coronal-cavity boundary.
 Selected oscillations in enlarged views and their damped sine fits are shown on the right.
 The fitted period [$P$], e-folding damping time [$\tau_A$],	
 and the initial spatial and velocity amplitudes [$A$ and $v$] are listed.	
 Panels a, c and i are from \inlinecite{LiuW.cavity-oscil.2012ApJ...753...52L}.
 } \label{908_oscil-fits.eps}
 \end{figure}
AIA's high cadence allows for detection of an {unbroken chain sequence}
of displacements, including {\bf deflections and transverse oscillations},
of local coronal loops set off by an EIT wave ({\it e.g.} \opencite{LiuW.cavity-oscil.2012ApJ...753...52L}).
The delayed onsets of displacements at increasing distances 	
agree with the wave travel times.
These displacements, signaling the {\it first} response to the arrival of a traveling disturbance, 
are suggestive of its fast-mode wave nature, {\it i.e.}
propagation at the {\it highest} speed (in the linear regime) supported by the medium.
The subsequent oscillations are sustained by a restoring force
(dominated by the Lorentz force in a low-$\beta$ plasma)
after the passage of a {\it transient} perturbation,
most like a true wave, rather than a CME expansion or reconnection front 
that would otherwise cause {\it permanent} deflections 
(see Figure~8 of \opencite{Patsourakos.Vourlidas.EIT-wave-review.2012SoPh..281..187P}).

The example in \fig{908_oscil-fits.eps}
evidently shows concurrent arrivals of an EIT wave and
onsets of oscillations of various local structures.
These oscillations occur in a {\it broad range of periods} --		
12\,--\,15~minutes in low-altitude (short) loops, 28~minutes in a filament-hosting coronal cavity, 
and 56\,--\,80~minutes in high-altitude (long) loops --
which are positively correlated with loop lengths and thus suggestive of fast kink modes.
They have initial amplitudes of 2\,--\,20~Mm, velocity amplitudes of 8\,--\,$27 \kmps$,
and e-folding damping times of 1\,--\,7 periods.
In another example shown in \fig{reflect.eps} (left, near $300 \Mm$), the initial large-amplitude loop displacement 
appears as a short-lived secondary intensity front that bifurcates from the primary wave and lags {\it behind} at $\lesssim 1/7$ of its speed.	
Such fronts differ from the secondary waves at coronal bright points that emerge {\it ahead} of the primary waves at greater speeds,
as mentioned above in \sect{subsect_optic}.

Oscillations triggered by EIT waves are common in AIA observations,
as manifested in 8 out of 21 events listed in \tab{table_EITwv-events}.
Other instruments with lower cadences also detected similar but often spatially isolated cases,
such as off-limb loop oscillations ({\it e.g.} \opencite{Patsourakos.EUVI-fast-mode-wave.2009ApJ...700L.182P}),
filament oscillations ({\it e.g.} 
\opencite{HershawJ.promin.oscil.2011A&A...531A..53H};
{\it cf.} those triggered by Moreton waves: 	
	\opencite{LiuR.1990-05-24-Moreton.wave.2013ApJ...773..166L}),
and streamer deflections ({\it e.g.} \opencite{TripathiD.EITwave.deflect.streamer.2007A&A...473..951T};
{\it cf.} those triggered by CMEs: \opencite{ChenYao.streamer.wave.2010ApJ...714..644C}).
{\bf Stationary brightenings} occur upon arrival or passage of EIT waves at  
structural boundaries, {\it e.g.} of active regions, coronal holes, coronal cavities, and quiet-Sun bright points,
which can be cospatial with topological separatrices
({\it e.g.} \opencite{DelanneeAulanier.EITwave-current-shell.1999SoPh..190..107D}).
If no wave signal is detected beyond such a boundary, a stationary brightening 
or a decelerating secondary front noted above can appear as if an EIT wave came to a stop.
This could be the case for low-cadence EIT observations,
which may miss (especially faint) reflections or transmissions now clearly seen with AIA.
Stationary brightenings can result from persistent heating and/or compression
produced by wave or non-wave mechanisms, such as localized energy release triggered by MHD waves
\cite{OfmanThompson.EIT-wave-fast-mode.2002ApJ...574..440O,Terradas.Ofman.MHD.wave.compress.2004ApJ...610..523T},
opening of magnetic-field lines \cite{Delannee.EIT-pseudo.wave.2000ApJ...545..512D,ChenPF.EIT-wave.2005ApJ...622.1202C},
Joule heating in current sheets \cite{Delannee.EITwv.stationary.2007A&A...465..603D},
and continual magnetic reconnection 
\cite{Attrill.EIT-wave-CME-Footprint.2007ApJ...656L.101A}.	
In addition, high-cadence AIA observations revealed {\it periodic stationary brightenings and darkenings},
indicative of MHD oscillations and waves ({\it e.g.} \figs{reflect.eps} and \ref{908_oscil-fits.eps}, left;
\opencite{LiuW.cavity-oscil.2012ApJ...753...52L}, their Figure~9d).
In fact, trapped fast-kink modes can be responsible for stationary brightenings 
at streamer footpoints \cite{Kwon_STEREO-EUV-wave-high-corona.2013ApJ...766...55K}.

{\bf Sympathetic flares} and eruptions in general
\cite{BeckerU.sympathetic.flare.1958ZA.....44..243B,PearceHarrison.sympathetic-flare.1990A&A...228..513P}
have been long thought to be triggered by some waves that carry a disturbance and its energy
from one eruption to another. This has been demonstrated in MHD models 
({\it e.g.} \opencite{SakaiJ.magnetosonic.wave.trigger.flare.1982ApJ...258..823S};
\opencite{OfmanThompson.EIT-wave-fast-mode.2002ApJ...574..440O}).
The continuous full-Sun coverage of AIA sparked
renewed interest in this topic and revealed profound global connections among large-scale
eruptions 	
\cite{Schrijver.Title.2010Aug01-long-range-couple.2011JGRA..116.4108S,SchrijverC.global-coupling-pathway.2013ApJ...773...93S}.
Thus far, very few EIT waves have been identified as triggers of sympathetic eruptions
({\it e.g.} \opencite{KhanJ.HudsonH.sympathetic.transeq.loop.disappear.2000GeoRL..27.1083K}).
Among recent examples observed by AIA are a minor flare 	
induced by an M2 flare across the Equator 
\cite{LiuW.Ofman.counter.wave.2014} and high-speed (200\,--\,$400 \kmps$), low-corona mass ejecta
in a wide angular range 	
\cite{LiuW.Ofman.transmit.reflect.2014}.
A statistical survey is yet to be carried out to determine how
common such sympathy is and what role EIT waves play.

\subsection{EIT Wave Periodicity}
\label{subsect_period}


\referee{
As EIT waves were largely considered as a single-pulse phenomenon in the past
\cite{Wills-Davey.EIT-wave-soliton.2007ApJ...664..556W}, 
very little attention was paid to their possible periodicities. The only prior detection,	
to the best of our knowledge, was a traveling 195~\AA\ intensity oscillation 
with an average period of about seven~minutes 	%
found by \inlinecite{Ballai.EIT-wave-trigger-loop-oscil.2005ApJ...633L.145B}
in the extensively studied 1998 June 13 TRACE event 
({\it e.g.} \opencite{Wills-Davey.auto-detect-EIT-wave.2006ApJ...645..757W}).
}  
Now, AIA has revealed a variety of periodicities
pertinent to the true wave components of EIT waves,
reflecting the characteristics of their {\it drivers} and {\it media}.	
 \begin{figure}[thbp]      
 \begin{center}
 \includegraphics[width=1.0\textwidth]{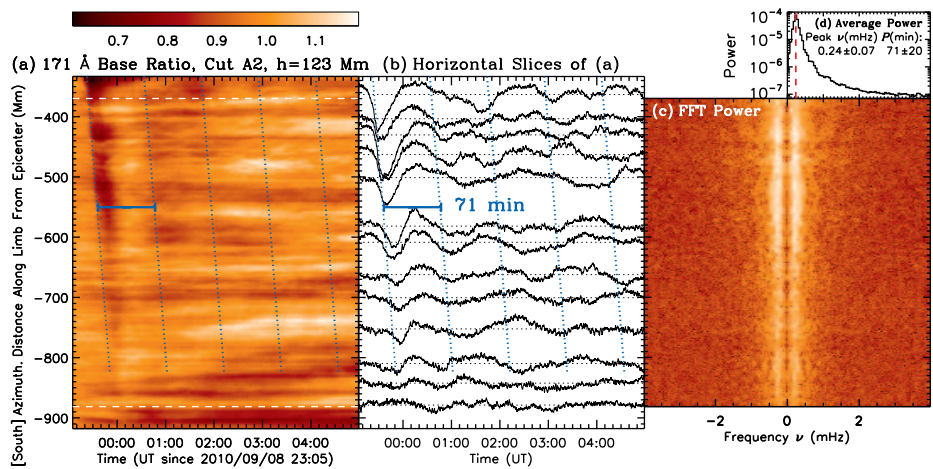}
 \end{center}
 \caption[]{
 Multiple-pulsed, long-period traveling intensity oscillations of the
 EIT wave on 08 September 2010 \cite{LiuW.cavity-oscil.2012ApJ...753...52L}.
  (a) Base-ratio space--time plot at 171~\AA\ from 
 azimuthal off-limb cut~A2		
 to the South of the eruption where the quiet Sun is relatively homogeneous.		
  (b) Temporal profiles from horizontal slices of (a) at selected distances
 marked by the black-dotted line. The slanted lines indicate propagating pulses
 at a dominant period of $71 \pm 20$~minutes.
  (c) Fourier power of the temporal profiles 	
 at each distance in (a) within the range marked by the two white dashed lines.
  (d) Average Fourier power 	
 ($\nu \ge 0 $ only) by collapsing (c) in distance.
 } \label{0908_long-fft.eps}
 \end{figure}
%

For example, 	
the EIT wave trains shown in \fig{0908_global-train.eps}
have a broad period range of 36\,--\,212~seconds \cite{LiuW.cavity-oscil.2012ApJ...753...52L}. 
The dominant two~minute period matches X-ray flare pulsations,
but is well below the 12\,--\,80~minute periods of local oscillations on the
wave path (see \fig{908_oscil-fits.eps}). 
This suggests a {\it periodic driver} related to flare pulsations		
that determines the wave periods.
This and other potential periodicity contributors, including a dispersive medium,
will be discussed in \sect{subsect_freq}.

The {\it medium} through which EIT waves propagate, namely the solar corona, 
can further modify their periods if they represent
propagating eigenmode oscillations due to MHD resonances. At first glance, this seems unlikely, 
because the corona is inhomogeneous and threaded by magnetic structures of 
a wide range of characteristic periods, as shown in \fig{908_oscil-fits.eps}.	
However, this may happen under favorable conditions.

One example was found from our additional analysis of the
event presented in \inlinecite{LiuW.cavity-oscil.2012ApJ...753...52L}.
As shown in \fig{0908_long-fft.eps}, the passage of the EIT wave produces
an initial darkening at 171~\AA\ followed by 	
multiple cycles of intensity oscillations that generally decrease with time and distance in amplitude.
The periods are on the order of one to two~hours and some oscillations exhibit more than one period, 
which could result from LOS superpositions of multiple off-limb structures.
One can identify a series of coherent wave pulses traveling over $500 \Mm$
with a dominant period of $71 \pm 20$~minutes that translates into a wavelength of 
$1400 \Mm$ or $2 \Rsun$, considering the measured wave speed of $330 \kmps$.
Such traveling oscillations last for six~hours, meaning that the global corona 
is left undulating for a long time after being swept by the initial wave.
This further points to the global nature of EIT waves and solar eruptions 	
({\it e.g.} \opencite{ZhukovA.global.CME.2007ApJ...664L.131Z}; 
\opencite{SchrijverC.global-coupling-pathway.2013ApJ...773...93S}).

As to the origin of such {\it coherent long periods}, one possibility is 	
collective kink oscillations of large-scale loops of similar lengths and thus periods,
such as vertical kink modes in trans-equatorial loops 
\cite{ZaqarashviliT.radio.seismology.2013A&A...555A..55Z}
and those trapped in streamers \cite{Kwon_STEREO-EUV-wave-high-corona.2013ApJ...766...55K}.
Another, yet less likely, possibility is fast magnetosonic--gravity surface waves	
\cite{BallaiI.magneto-gravity-wave-EIT.2011A&A...527A..12B},
provided a vertical density jump lies in an everywhere-vertical magnetic field,	
for which no observational evidence has been found.

\subsection{Thermal Properties}
\label{subsect_therm}



Thermal properties of EIT waves are inferred from observations with multiple passbands 
each covering a specific temperature range of the emitting plasma. 
The seven EUV channels of AIA have significantly increased the temperature coverage,
which now ranges from $\approx$0.08~MK (304~\AA) in the transition region
to $\gtrsim\,$10~MK (131 and 193~\AA) in flares
\cite{ODwyer.AIA-T-response.2010A&A...521A..21O}.
In general, EIT waves are best seen as intensity enhancements in the 193 and 211~\AA\ channels
that have peak temperature responses at 1.6 and 2.0~MK, respectively, 
but as intensity reductions in the 171~\AA\ channel
that peaks at 0.8~MK. This anti-correlation indicates {\it plasma heating} 
taking place between 0.8 and 1.6\,--\,2.0~MK, consistent with similar trends seen
by TRACE \cite{Wills-DaveyThompson.TRACE-EUV-wave.1999SoPh..190..467W}
and EIT \cite{Gopalswamy.Thompson.early.life.CME.2000JASTP..62.1457G}.
In other channels, intensity enhancements become progressively weaker from
335 to 94, 131, and 304~\AA.

There are exceptions to this general trend. First, not all EIT waves are 171~\AA\ dark.
As shown in \fig{Nitta_stat_hist.eps}d, among 138 events observed by AIA,
17\,\% appear as 171~\AA\ bright, 41\,\% as dark, and the rest with no clear signal.
171~\AA\ brightening indicates a {\it low temperature} [$\lesssim$\,0.8~MK] of the pre-event plasma 
or a {\it strong density enhancement} ({\it e.g.} at a CME front) dominant over temperature effects.
In fact, in the high speed regime $\gtrsim$\,$900 \kmps$, 
there are relatively more 171~\AA\ bright waves,	
which are likely shocks with strong compression.
An opposite trend was found in 34 EIT waves observed by EUVI
\cite{NittaN.flare.CME.solar.min.2014SoPh..289.1257N}
with 59\,\% bright and only 18\,\% dark.		
One possible explanation is that these EUVI events occurring during the extended solar minimum
could have different environments, {\it e.g.} lower temperatures or weaker magnetic fields, 
compared with those AIA events during the rise phase of the present solar cycle.

Another exception is that the 193~\AA\ wave signal is not always bright,
although this trend persists in all of the AIA events reported by \inlinecite{NittaN.AIA.wave.stat.2013ApJ...776...58N}.
For example, two recent homologous EIT waves 	
both appear dark at 171 and 193~\AA, but bright at 211 and 335~\AA, 
suggesting a {\it high initial temperature} 	
in the 1.6\,--\,2.0~MK range \cite{LiuW.Ofman.counter.wave.2014}.
Among other examples are 195~\AA-dark wave fronts detected by EIT (\opencite{Zhukov.EIT-wave2004A&A...427..705Z}; their Figure~2)
and EUVI (\opencite{NittaN.flare.CME.solar.min.2014SoPh..289.1257N}; their Table~2).

%
 \begin{figure}[thbp]      
 \begin{center}
 \includegraphics[height=4.3cm]{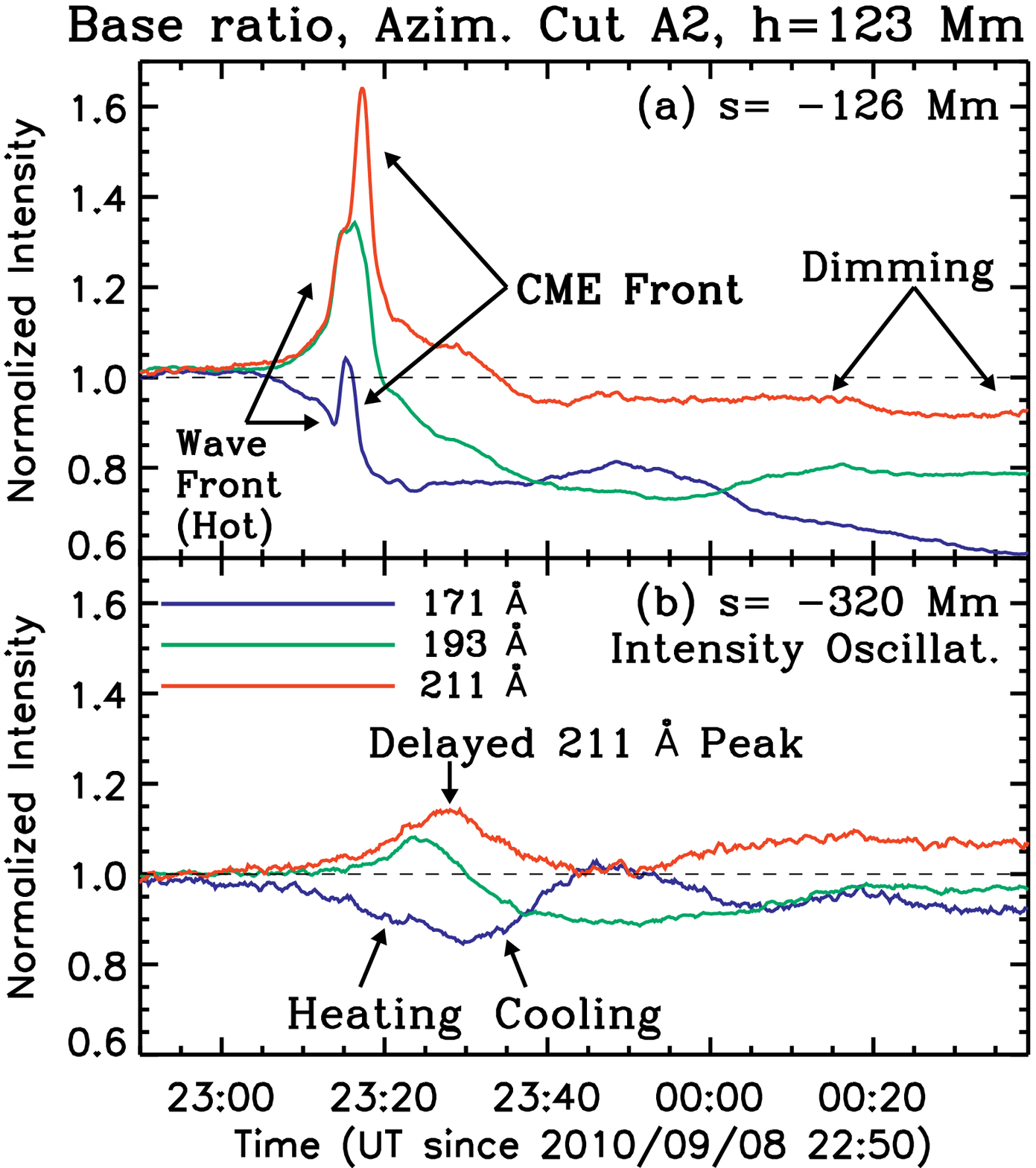}
 \includegraphics[height=4.3cm]{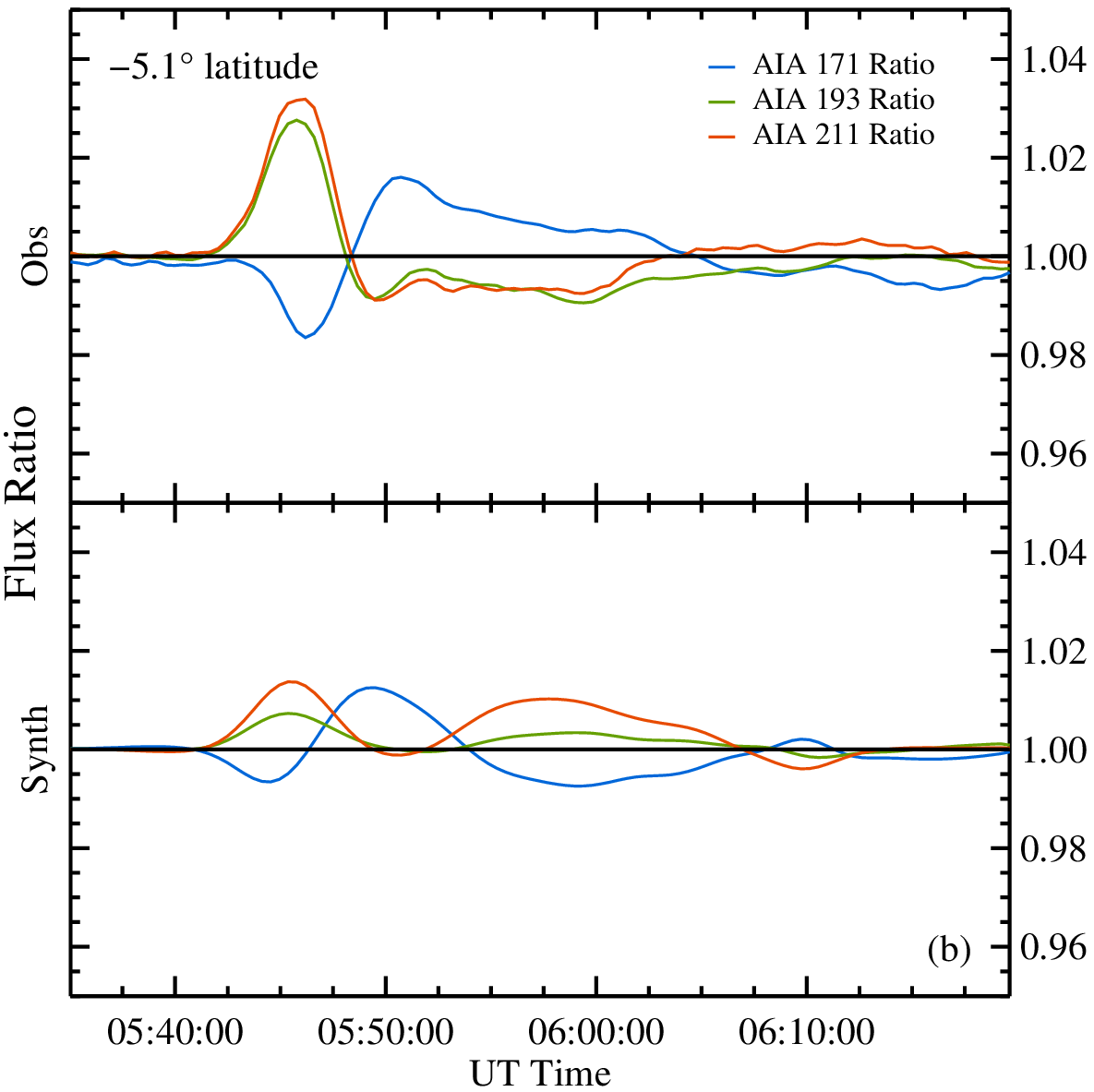}		
 \includegraphics[height=4.3cm]{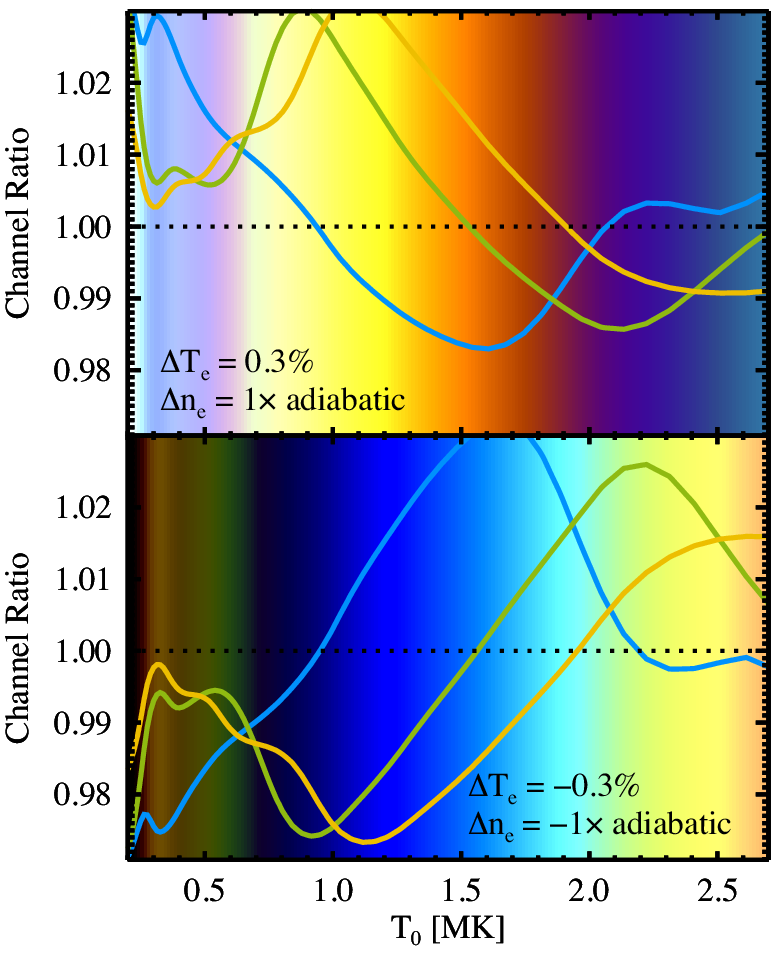}	
 \end{center}
 \caption[]{	
 Thermal properties of EIT waves.
  {Left}: AIA intensity variations on 08 September 2010 at two locations on the EIT wave path
 showing distinct thermal characters of the wave and non-wave components (top)
 and delay across passbands and oscillations (bottom) 
 (from \opencite{LiuW.cavity-oscil.2012ApJ...753...52L}).
  {Middle}: Observed (top) and simulated (bottom) AIA intensity variations on 13 June 2010
 showing multiple heating and cooling cycles.
 The left and middle panels share the same color scheme for passbands.
  {Right}: Predicted fractional intensity changes of AIA channels (blue: 171~\AA, green: 193~\AA, orange: 211~\AA)
 for adiabatic heating (top) and cooling (bottom) as a function of the initial temperature [$T_0$].
 The background is the combined tri-color signal at each $T_0$.
 The middle and right panels are from \inlinecite{DownsC.MHD.2010-06-13-AIA-wave.2012ApJ...750..134D}.
 } \label{thermal.eps}
 \end{figure}
%
As shown in \fig{thermal.eps} (middle),
 after the general initial heating phase with brightening at 193 and 211~\AA\ and darkening at 171~\AA,
a cooling phase follows with reversed intensity variations. 
This can be explained by adiabatic heating due to compression 
followed by cooling with subsequent expansion/rarefaction driven by a restoring pressure gradient force.	
Mild adiabatic compression was estimated to cause $\approx$10\,\% increases in both
density and temperature for the EIT wave associated with the 15 February 2011 X2.2 flare 
\cite{Schrijver.Aulanier.2011Feb15.X2.AIAwv.2011ApJ...738..167S}. 
Such compression-expansion is consistent with 	
the red-blueshift sequence detected by {\it Hinode}/EIS in coronal lines
\cite{Harra.Sterling.EIS.2010Feb16.EITwv.2011ApJ...737L...4H,Veronig.EIS.2010Feb16.EITwv.2011ApJ...743L..10V},
similar to the well-known down-up swings in chromospheric \Ha Moreton waves.
Post-EIT wave cooling can continue even after the perturbed plasma returns to the initial temperature,	
unexpected for conductive cooling. The observed cooling timescales on the order of ten~minutes
are significantly shorter than radiative cooling times in the quiet Sun, which are typically orders of hours
({\it e.g.} 	
\opencite{LiuW.Berger.Low.flmt-condense.2012ApJ...745L..21L}).
Such behaviors, however, agree with
{\it multiple cycles of adiabatic compression and expansion} 	
of a compressible, fast-magnetosonic wave. Wave dissipation may also contribute to the initial heating.

AIA's rapid cadence allows for tracking detailed temporal evolution of thermal structures.
For example, as shown in \fig{thermal.eps} (upper left), a thin layer with the general
anti-correlation between 193/211~\AA\ and 171~\AA, indicative of heating,
is followed by a sharp brightening and then prolonged dimming in all three channels,
indicative of density enhancement and depletion, respectively. 
This suggests a heated and compressed sheath of a (shocked) wave component preceding
a non-wave component of CME expansion,	
each with {\it distinct thermal characters}.
There is a delay of the maximum intensity variations
at 211 and 171~\AA\ from that at 193~\AA\ (\fig{thermal.eps}, lower left), which could be related to 
passband-dependent wave kinematics	
\cite{LongD.AIA.EUV-wave.2011ApJ...741L..21L}.
In a coronal shock imaged by AIA, \inlinecite{MaSL.AIA.shock.2011ApJ...738..160M} 
found similar {\it delays indicative of progressive heating} and ionization 	
to higher charge states. They inferred 		
a temperature of 2.8~MK in qualitative agreement with the result from differential-emission-measure analysis
of the same event \cite{Kozarev.AIA.EUVwv.2010Jun12.2011ApJ...733L..25K}.

\subsection{Pulse Evolution: Amplitude and Width}	
\label{subsect_pulse}



The evolution of the pulse shape, or so-called perturbation profile,
can give additional clues to the nature of EIT waves. 
The pulse shape is usually measured with the EUV intensity profile
as a function of distance or time. It often appears as a hump that can be 
fitted with a Gaussian and characterized with an amplitude [$A$] and full-width half maximum [FWHM].

\editor{	
 Early observations of a few EIT waves by TRACE within its limited FOV	
 suggested a dispersionless behavior, {\it i.e.} the wave pulse maintains its coherence without increase in FWHM \cite{Wills-Davey2003PhDT.........8W}.
 This was interpreted as evidence of solitons \cite{Wills-Davey.EIT-wave-soliton.2007ApJ...664..556W}.
 }  
However, EIT observations indicated a more general trend of {\it amplitude decay and pulse broadening} 	
often accompanied by wave deceleration,
which is consistent with a non-linear fast-magnetosonic wave or shock degenerating into a linear wave 
due to dispersion, dissipation, and geometric expansion
({\it e.g.}	
\opencite{Warmuth.EIT.wave.debate.2010AdSpR..45..527W}).
EUVI added that the amplitude sometimes grows for a few minutes prior to its decay and
the FWHM often increases from $\approx$50 to $200 \Mm$ over the course of the event
\cite{Veronig.dome-wave.2010ApJ...716L..57V,MuhrN.STEREO.EUVwave.2011ApJ...739...89M,%
LongD.STEREO.EUV-wave.Dispersion.2011A&A...531A..42L,LiTing.EUVwave.AIA.STEREO.2012RAA....12..104L}.

AIA brought more details, some unexpected, to this general picture.
\inlinecite{LiuW.AIA-1st-EITwave.2010ApJ...723L..53L} found in certain directions 
an increase followed by a decrease in pulse amplitude that is anti-correlated with the FWHM,
implying pulse steepening follow by broadening.
In other directions, the pulse amplitude decreases linearly with distance $r$,
which is faster than a surface wave ($A \propto r^{-1}$) but slower than a spherical wave ($A \propto r^{-2}$)
due to geometric expansion alone.
This is compared to earlier dependencies of $r^{-1}$ found by TRACE \cite{Wills-Davey2003PhDT.........8W}
and $r^{-2.5 \pm 0.3}$ by EUVI \cite{Veronig.dome-wave.2010ApJ...716L..57V}.
These results imply {\it complex interplay between dispersive decay and nonlinear steepening}.
\inlinecite{LongD.AIA.EUV-wave.2011ApJ...741L..21L} found 	
passband-dependent intensity enhancements (amplitudes) ranging from 10\,\% at 304~\AA\ to 90\,\% at 211~\AA.
By fitting perturbation profiles with sinusoidal waves within a Gaussian envelope,
they obtained pulse expansion rates of $\approx$$200 \kmps$ and dispersion rates of
$\mathrm{d}^2 \omega / \mathrm{d} k^2 \approx 10 \Mm^2 \ps$.

The common practice using {\it intensity-perturbation profiles} must be applied with care
for a number of reasons: Because of the inhomogeneity of the corona 
and LOS integration of optically thin EUV emission, each image pixel can contain 
contributions from unrelated structures, {\it e.g.} each with its own periodicity
(see \sect{subsect_oscil}), which can lead to complex temporal variations. 
This can be further complicated by multiple (wave and non-wave) components, wave periodicities,
and 	
thermal effects (see \sects{subsect_bimod}, \ref{subsect_period}, and \ref{subsect_therm}).
Some refinements have been adopted to mitigate such impacts, for example,
by excluding the trailing side of a perturbation profile 
from Gaussian fits to avoid the non-wave contribution such as dimming
\cite{MuhrN.STEREO.EUVwave.2011ApJ...739...89M}.

\subsection{Generation of EIT Waves}
\label{subsect_gener}




It has been established that it is CMEs (large-scale), not flares (localized), that generate EIT waves traveling across the solar disk
({\it e.g.} see \sect{subsect_relat_flare-CME}; \opencite{WarmuthA.EIT-wave-review.2007LNP...725..107W};
\opencite{VrsnakCliver.corona-shock-review.2008SoPh..253..215V}; 
\opencite{ZhukovA.EIT.wave.review.STEREO.2011JASTP..73.1096Z}).
Specifically, the {\bf lateral expansion} of the flank of a CME	
was recently recognized to play an important role 
(for a review, see Section~10 of \opencite{Patsourakos.Vourlidas.EIT-wave-review.2012SoPh..281..187P}).
There has been strong supporting evidence from EUVI and AIA observations
({\it e.g.} 	
\opencite{Patsourakos.CME-genesis.AIA.2010ApJ...724L.188P};
\opencite{TemmerM.dome.over-expansion.2010Jan17.2013SoPh..287..441T})	
and numerical or analytical models ({\it e.g.} \opencite{PomoellJ.MHD-EIT-wave-lateral-expansion.2008SoPh..253..249P};
\opencite{TemmerM.analytic-model-Moreton-wave.2009ApJ...702.1343T}).
A lateral inflation or over-expansion is characterized by a faster growth of a CME bubble 
in the lateral than in the radial direction.	
The latter is critical to generating an upward propagating (shock) wave in the high corona, 	
but less relevant to an on-disk EIT wave.
When a CME expands rapidly enough in all directions, a dome-shaped EIT wave
can be detected ({\it e.g.} \opencite{Veronig.dome-wave.2010ApJ...716L..57V};
\opencite{LiTing.2010Jun07.AIAwave.reflect.2012ApJ...746...13L}).

A {\bf downward compression}, in addition to the lateral CME expansion, 
can also be important.	
As schematically shown in \fig{QFP-2trains.eps}a, 
\inlinecite{LiuW.cavity-oscil.2012ApJ...753...52L} found the onset
of an EIT wave being associated with the downward expansion of a growing CME bubble 
originally centered at an elevated height of $\approx$100~Mm.
This	
can happen when a CME self-expands faster than the rise of its center.
It pushes the low-corona plasma against the underlying chromosphere,
resulting in {\it enhanced compression efficiency}.
A narrow angle formed between 	
the forwardly inclined lower portion of a laterally expanding CME bubble and the chromosphere underneath it has a similar effect.
It may contribute to such EIT wave characteristics as
i) initial redshifts followed by blueshifts
\cite{Veronig.EIS.2010Feb16.EITwv.2011ApJ...743L..10V},
ii) the forward inclination of the early, low-corona wave front toward the solar surface,
and iii) the dominance of wave signals at lower heights $\lesssim\,$100~Mm 	
over CME expansions \cite{LiuW.cavity-oscil.2012ApJ...753...52L}.

The {\it impulsiveness} of the CME lateral expansion, characterized by its {\it acceleration}, 
is key to EIT wave generation \cite{Patsourakos.Vourlidas.EIT-wave-review.2012SoPh..281..187P}. 
This is backed up on theoretical grounds that pistons with greater and/or more prolonged accelerations
tend to produce stronger nonlinear waves or shocks 
(see \opencite{VrsnakCliver.corona-shock-review.2008SoPh..253..215V} for a review). 
EIT waves often first appear at considerable distances of $\approx$100~Mm from the epicenter, 	
delayed by a few minutes from the initiation of the rapid CME expansion
({\it e.g.} \opencite{LiuW.cavity-oscil.2012ApJ...753...52L}).
Such space--time delays allow the lateral expansion to
reach speeds of several hundred $\kmps$ with accelerations up to a few hundred $\m \pss$,
and thus to build up sufficient compression to the ambient corona to produce detectable EUV signals,
as predicted for 	
shock formation ({\it e.g.} 	
\opencite{Zic.piston-driver-shock.2008SoPh..253..237Z}; see \sect{subsect_model}).

There is a noticeable trend that, regardless of the associated CME final speeds or flare sizes,
eruptions without appreciable lateral expansion (or with but non-impulsive)
tend to produce no or weak on-disk EIT-wave signatures.	
Examples include quiescent filament eruptions moving mainly radially with gentle acceleration profiles.
Future statistical analysis of the quantitative relationship between 
the CME lateral expansion and EIT-wave formation is required.	

\subsection{Association With CMEs and Flares}
\label{subsect_relat_flare-CME}



\begin{figure}[thbp]      
 \begin{center} 
 \includegraphics[width=1.0\textwidth]{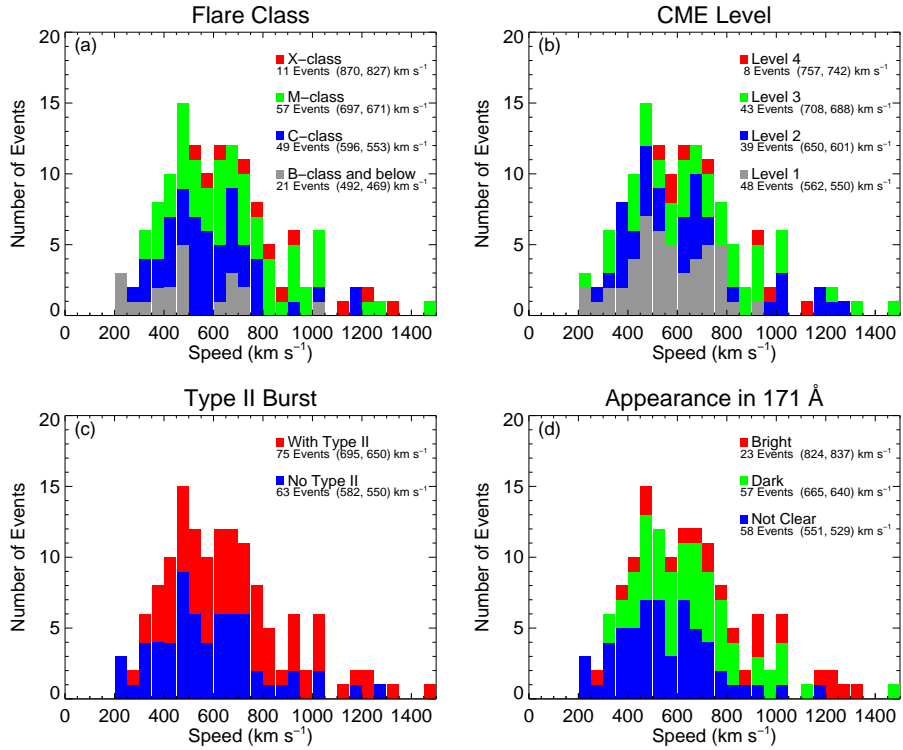}
 \end{center}
 \caption[]{
 Histograms of speeds of 138 on-disk EIT waves observed by AIA,
 with breakdowns in flare class, CME level, radio Type-II burst association, and 171~\AA\ wave signal.
 The mean and median speeds of each category are listed
 (from \opencite{NittaN.AIA.wave.stat.2013ApJ...776...58N}).
 } \label{Nitta_stat_hist.eps}
 \end{figure}

It has been widely recognized that {CMEs}, especially fast and wide ones,
are closely related to EIT waves and serve as a necessary but not sufficient condition
\cite{Biesecker.EIT-wave-association.2002ApJ...569.1009B}.	
For example, there were five times more front-side CMEs than EIT waves during a 13-month period
\cite{CliverE.EIT-wave-origin.2005ApJ...631..604C}. 
Recent AIA observations indicated that their association is {\it not as strong
as previously thought}	
\cite{NittaN.AIA.wave.stat.2013ApJ...776...58N}.
As shown in \fig{Nitta_stat_hist.eps}b, although faster EIT waves still tend to be associated
with stronger CMEs, more than $1/3$ of the 138 waves are associated with weak outflows 
reaching heliocentric distances $\leq\,$$5 \Rsun$ (labeled CME level~1)
that are not typical, large-scale CMEs with three-part structures
\cite{HundhausenA.SMM.CME.def.1984JGR....89.2639H,SchwennR.CME.terminology.1996Ap&SS.243..187S}.
Meanwhile, the correlation between on-disk EIT wave speeds 	
and radial CME speeds,		
although positive, is very weak
(see Figure~8 of \opencite{NittaN.AIA.wave.stat.2013ApJ...776...58N}).
These results are not surprising because lateral and radial expansions of CMEs
are not necessarily correlated and, as stressed above, it is the early-phase lateral expansion 
in the low corona (often below the FOVs of CME-detecting coronagraphs)
that is more relevant to EIT wave generation. 
%
%

The association of EIT waves with {flares} is generally considered weak.
For example, no EIT wave was detected in 14 major flares without CMEs
\cite{ChenPF.EIT-wave-not-flare-driven.2006ApJ...641L.153C}, 
and about 50\,\% of 176 waves were associated with minor flares of B-class or lower
\cite{CliverE.EIT-wave-origin.2005ApJ...631..604C}. 
At times, flare energy release is too late, too little to account for
EIT waves ({\it e.g.} \opencite{Veronig.STEREO-EUVI-wave.2008ApJ...681L.113V}).
These results contradict the interpretation of EIT waves as blast waves
generated by impulsive flare heating.
On the other hand, as shown in \figs{Nitta_stat_hist.eps}a and \ref{Nitta_stat_hist.eps}b,  
there are similar ``normal" type speed distributions of AIA-detected waves
in association with flares and CMEs. 	
That is, the mid-class flares (C and M) and CMEs (levels~2 and 3)
correspond to the majority of the EIT waves of intermediate speeds.
There is a tendency of association of faster EIT waves with stronger flares and CMEs,
which implies an {\it energetic correlation} between these three phenomena that
are integral components of typical solar eruptions.


\subsection{Association With Moreton Waves, Type-II Bursts, and SEPs}
\label{subsect_relat_typeII}


{\bf Chromospheric \Ha Moreton waves} are strongly associated with a subset of EIT waves, {\it i.e.} those sharp, 
well-defined ``S-waves" (\opencite{Wills-Davey.EIT-wave-review.2009SSRv..149..325W} and references therein)
and Type-II radio bursts \cite{VrsnakCliver.corona-shock-review.2008SoPh..253..215V}.
Their overall degree of correlation with EIT waves, considering both sharp and diffuse types,
is low, with only 1\,\% in 173 events \cite{Biesecker.EIT-wave-association.2002ApJ...569.1009B} 
and 21\,\% in 14 events of simultaneous EIT and \Ha observations 
associated with $\geq\,$M-class flares \cite{Okamoto.EIT-Moreton-wave-filament-oscil.2004ApJ...608.1124O}.
Moreton waves have much {\it narrower angular widths} than their EIT-wave counterparts,
{\it e.g.} with a mean of $92 \degree$ {\it vs.}~$193 \degree$ in
13 events \cite{ZhangYZ.Moreton.wave.2011PASJ...63..685Z}.	
Early case studies ({\it e.g.} \opencite{Warmuth.EIT.wave.debate.2010AdSpR..45..527W})
indicated that a fast-mode shock undergoing deceleration and decay can produce both
a Moreton wave of typically 500\,--\,$1000 \kmps$ and a fractionally slower
EIT wave in the late stage. The high median wave speed $>\,$$600 \kmps$ revealed by AIA
\cite{NittaN.AIA.wave.stat.2013ApJ...776...58N} suggests that EIT waves can be
detected in early stages as well.	
During the 9 August 2011 X6.9 flare, cospatial wave fronts at initial speeds
of 800\,--\,$1000 \kmps$ were detected from the upper photosphere to the corona
in \Ha and AIA's UV (1700 and 1600~\AA) and EUV channels
\cite{AsaiA.2010AugX6.9.Moreton.AIA.wave.2012ApJ...745L..18A,ShenYD.LiuY.Moreton.wave.photosphere2012ApJ...752L..23S}.
On the other hand, the forward inclination of an EIT wave front in the low corona
implies a {\it delay} of up to minutes of a Moreton wave from its coronal counterpart
\cite{Afanasyev.Uralov.shock-EUV-wave-I.2011SoPh..273..479A,LiuW.cavity-oscil.2012ApJ...753...52L},
consistent with the analysis of the Moreton wave during the 14 February 2011 M2.2 flare
\cite{WhiteS.2011Feb.Moreton.AIA.waves.2011SPD....42.1307W}.
	

It presents a challenge to draw a general conclusion on the Moreton--EIT wave relation,	
because Moreton waves have been {\it rarely observed}, {\it e.g.}
only two reported to date in the SDO era, in contrast to more than 210 EIT waves.	
Aside from noncontinuous ground-based \Ha coverage, this two orders of magnitude
difference in detection rate suggests that some {\it more stringent conditions} 
are required for Moreton waves than ordinary EIT waves.
Such conditions, {\it e.g.} a strong shock with Mach number $>\,$2 \cite{Balas.Moreton.wave.coronal.2007ApJ...658.1372B}, 
must be able to produce a compression strong enough to penetrate deep down to the chromosphere. 
It was proposed that \ion{He}{1}~10830~\AA\
waves formed in the upper chromosphere could be a link between coronal EIT waves
and chromospheric Moreton waves \cite{Vrsnak.HeI.wave.2002A&A...394..299V}, 
but the rarity of such observations and the complexity of line formation present another challenge 
({\it e.g.} \opencite{GilbertH.HeI.EIT.wave.cospatial.2004ApJ...607..540G}).	
Moreton waves should not be confused with sequential chromospheric brightenings 
along narrow channels without a well-defined wave front,	
ascribed to electron precipitation from high-corona magnetic reconnection
\cite{Balas.seq.chrom.bright.2005ApJ...630.1160B}.

{\bf Metric Type-II radio bursts} are signatures of coronal shocks that are more often considered to
be driven by CMEs than by flares (\opencite{VrsnakCliver.corona-shock-review.2008SoPh..253..215V}).	
A Type-II burst is generally accepted as a sufficient but not necessary condition
for an EIT wave. 90\,\% of the type IIs during 1997 were associated with EIT waves 	
\cite{KlassenA.1997.catalog.EIT.wave.TypeII.2000A&AS..141..357K},
while only 29\,\% of the 173 EIT waves from 1997-Mar to 1998-Jun were associated with Type~IIs,
with somewhat higher correlation rates at disk center and the limb than at intermediate longitudes
\cite{Biesecker.EIT-wave-association.2002ApJ...569.1009B}.
As shown in \fig{Nitta_stat_hist.eps}c, 54\,\% of the 138 AIA-detected waves 
(\opencite{NittaN.AIA.wave.stat.2013ApJ...776...58N}, \citeyear{2014.Nitta.typeII.EIT.wave.SolPhy})
were accompanied by Type~IIs and had a slightly higher median speed 
(650 {\it vs.}~$550 \,\kmps$) than Type-II free waves.
Faster EIT waves with speeds $\gtrsim\,$$800 \kmps$ have stronger associations with Type~IIs,
implying a common shock origin.
Meanwhile, there are Type-II free, but fast EIT waves, 
which, like radio quiet, but fast and wide CMEs 
\cite{GopalswamyN.radio.quiet.fast.CMEs.2008ApJ...674..560G},
may be due to large fast-mode speeds of the ambient corona,
making shock formation difficult, or due to large angles
of the wave-driving eruptions from the Sun--observer line.	
Recently, \inlinecite{KouloumvakosA.typeII.in.sheath.CME.EUV.wave.2014SoPh..289.2123K}
reported a Type-II burst originating from the sheath 
between a CME bubble and a preceding shock in both radial and lateral directions.

{\bf Solar energetic particles (SEPs)} are generally considered to be accelerated
by CME shocks and/or flares (see \opencite{ReamesD.SEP.2source.2013SSRv..175...53R} for a review). 
In the last two decades, there has been increasing evidence of 	
correlation between SEPs and Moreton and/or EIT waves 
({\it e.g.} \opencite{KocharovL.SEP.Moreton.wave.1994SoPh..155..149K};
\opencite{TorstiJ.10MeV-proton.EIT.wave.1999ApJ...510..460T};
\opencite{KruckerS.EIT-wave-SEP.1999ApJ...519..864K}),
suggesting an association of such waves with shocks.
Recent efforts were made to identify a possible spatio--temporal correlation between 
the releases of SEPs and the arrivals of EIT waves at their source regions 
connected by open magnetic-field lines to observers at multiple viewpoints 	
({\it e.g.} \opencite{Kozarev.AIA.EUVwv.2010Jun12.2011ApJ...733L..25K};
\opencite{RouillardA.SEP.STEREO.EUV.wave.IM.shock.2011ApJ...735....7R},
\citeyear{RouillardA.SEP.STEREO.EUV.wave.Longitude.2012ApJ...752...44R};
\opencite{NittaN.SEP.EIT.wave.2012AIPC.1436..259N}; \opencite{ParkJ.SEP.STEREO.AIA.EUV.wave.2013ApJ...779..184P};
\opencite{MitevaR.SEP.EIT.wave.cycle23.2014SoPh..tmp...37M}).
%
However, their physical relation has yet to be established.
Open questions include whether the flank of a CME-driven shock in a quasi-perpendicular orientation,
likely manifested as a low-corona EIT wave, or the high-corona nose of the shock in a quasi-parallel orientation
is more important in accelerating and/or transporting SEPs.
It was predicted that very fast ($\gtrsim\,$$1000 \kmps$) shocked EIT waves, missed by previous instruments,
can be responsible for rapid onsets of some SEPs including ground level events
\cite{NittaN.GEL.EIT.wave.2012SSRv..171...61N}.
This can be potentially verified with high-cadence AIA data.
 \begin{figure}[thbp]      
 \begin{center} 
 \includegraphics[width=5.5cm, angle=90]{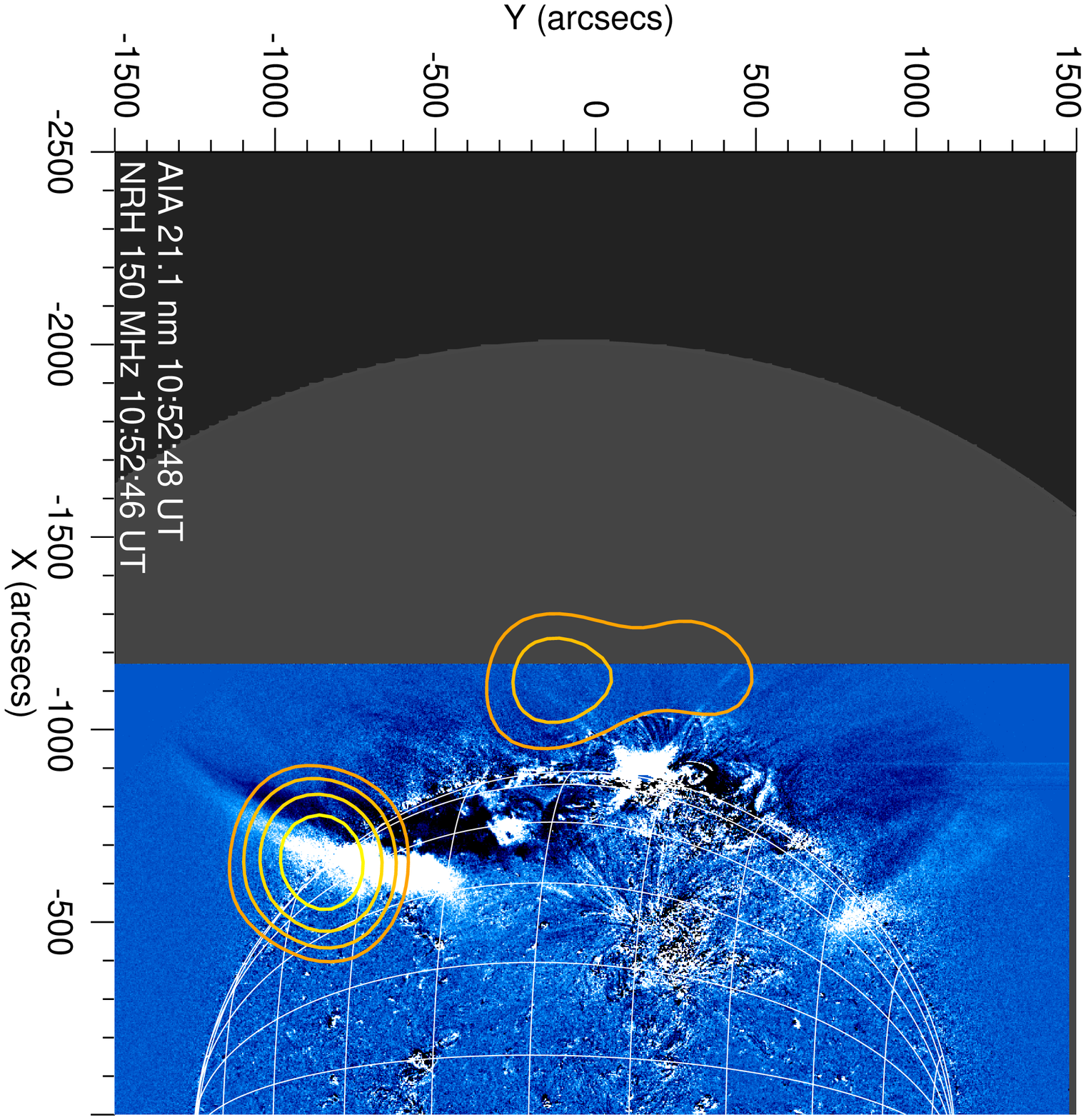}			
 \includegraphics[width=5.8cm, angle=90]{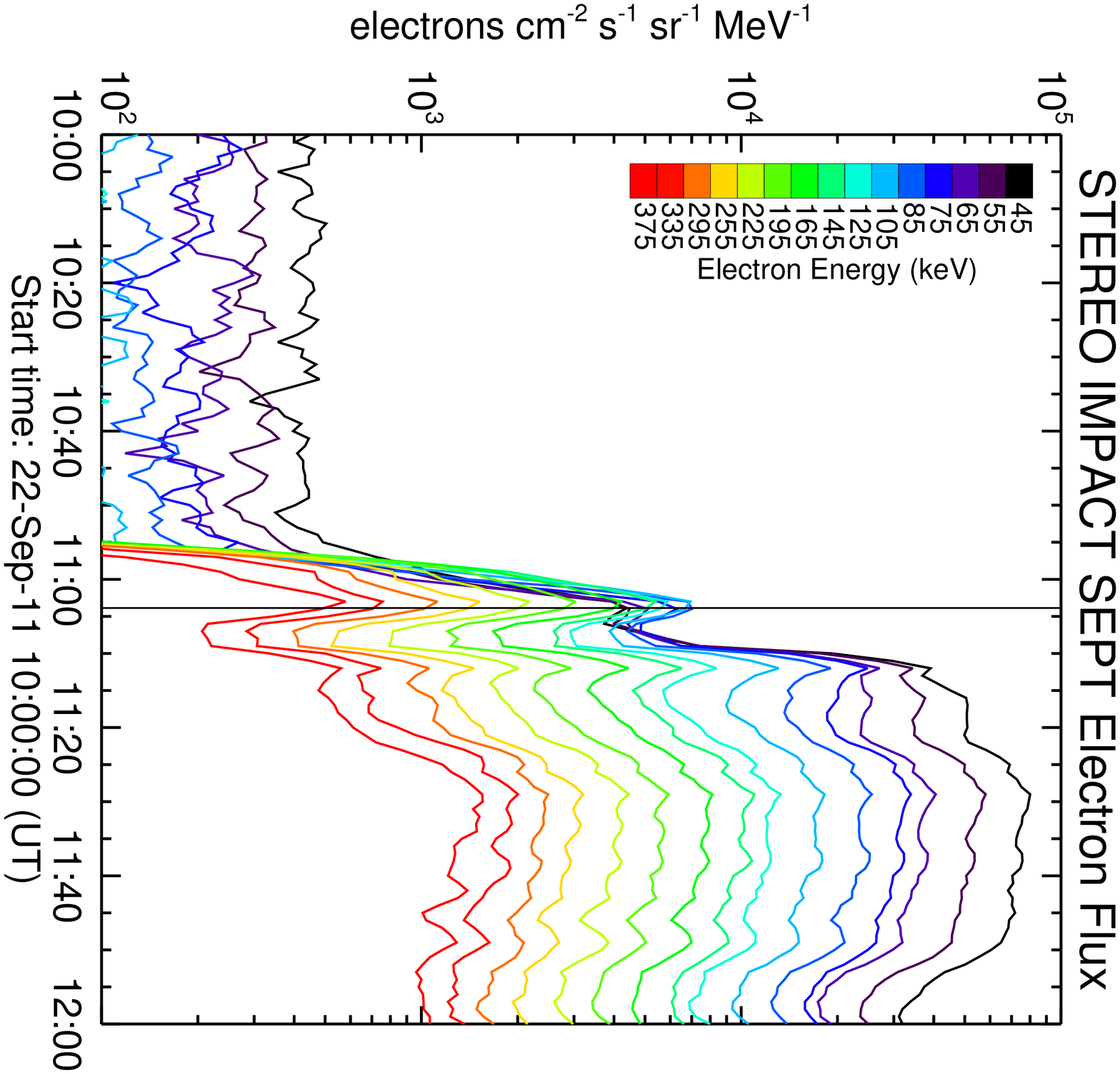}
 \end{center}
 \caption[]{
   {Left}: Cospatial EIT wave front in an AIA 211~\AA\ image and 
 150~MHz contours from the Nan\c{c}ay radioheliograph (NRH) traveling along the solar limb on 22 September 2011.
   {Right}: Energetic electron fluxes at 45\,--\,375~keV detected by 	
 the {\it Solar Electron Proton Telescope} on STEREO-B (from \opencite{CarleyE.AIA.EUV.wave.shock.e-acc.2011-09-22X1.4.2013NatPh...9..811C};
 reproduced by permission of the Nature Publishing Group).
 } \label{SEP.eps}
 \end{figure}
%

\fig{SEP.eps} shows an example of a joint study of an EIT wave, radio bursts, and SEPs
\cite{CarleyE.AIA.EUV.wave.shock.e-acc.2011-09-22X1.4.2013NatPh...9..811C}. 
In this event, a cospatial EIT wave front and 150~MHz source propagate
parallel to the solar surface, interpreted as a CME-flank-driven, 
quasi-perpendicular shock 	
that efficiently accelerate electrons 	
which produce the observed plasma emission.%
 \footnote{ 	
 A similar metric radio source associated with a Moreton--EIT wave was reported by 
 \inlinecite{Vrsnak2005.radio.EIT.wave.ApJ...625L..67V} and alternatively interpreted
 as optically thin gyrosynchrotron emission.
 }
Energetic electrons producing a Type-III burst are detected {\it in-situ} 	
at their expected arrival time (marked by the vertical line, right panel).
\referee{A followup study further indicated that the Type-II-burst associated shock forms
at the CME flanks where the inferred \AlfvenA-speed map shows a local minimum
\cite{ZuccaP.2011Sep22.X1.4.shock.form.height.2014arXiv1402.4051Z}.
}


\section{Quasi-periodic Fast Propagating (QFP) Wave Trains}
\label{sect_QFP}


\subsection{General Properties: Morphology and Kinematics}		
\label{subsect_QFPgen}

\begin{figure}[thbp]      
 \begin{center}
 \includegraphics[height=5.1cm]{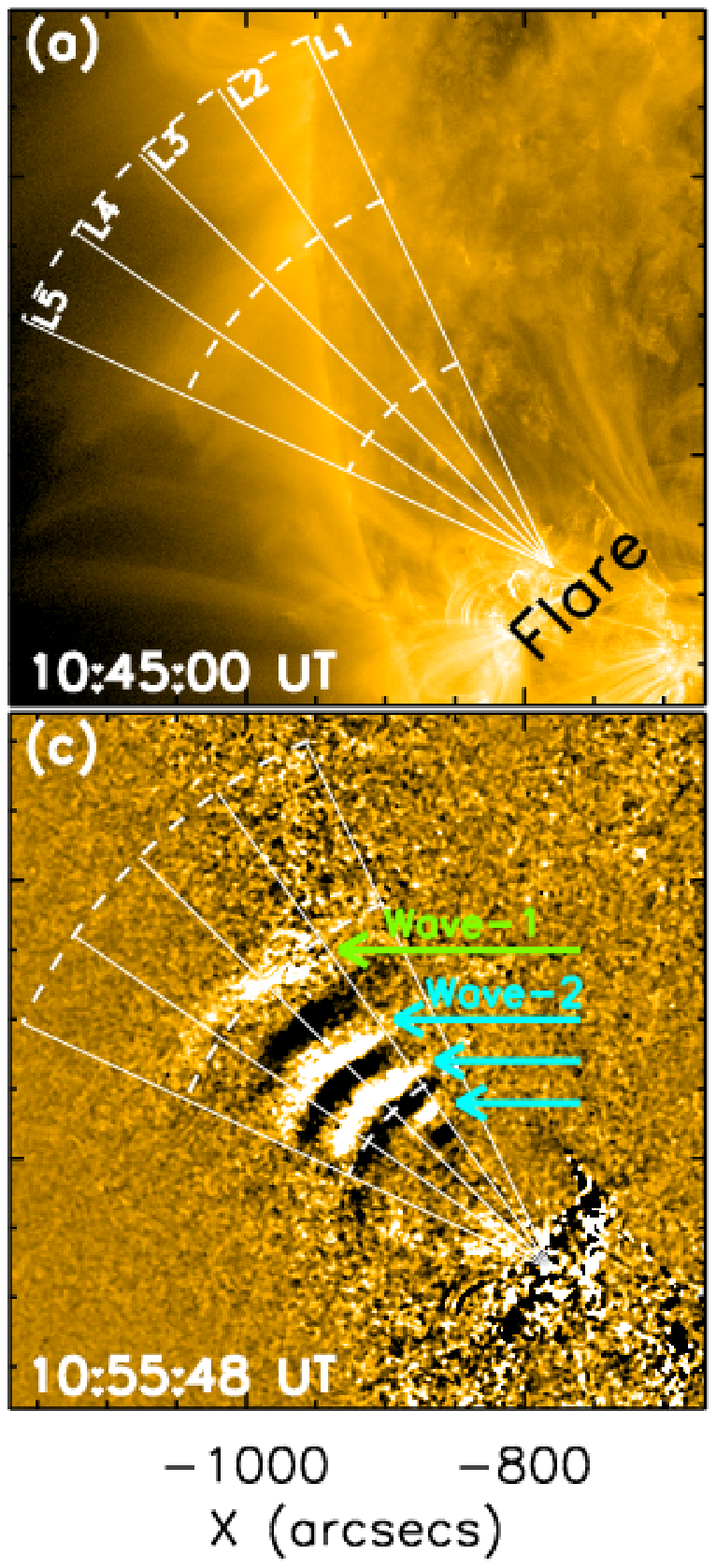}	
 \includegraphics[height=5.3cm]{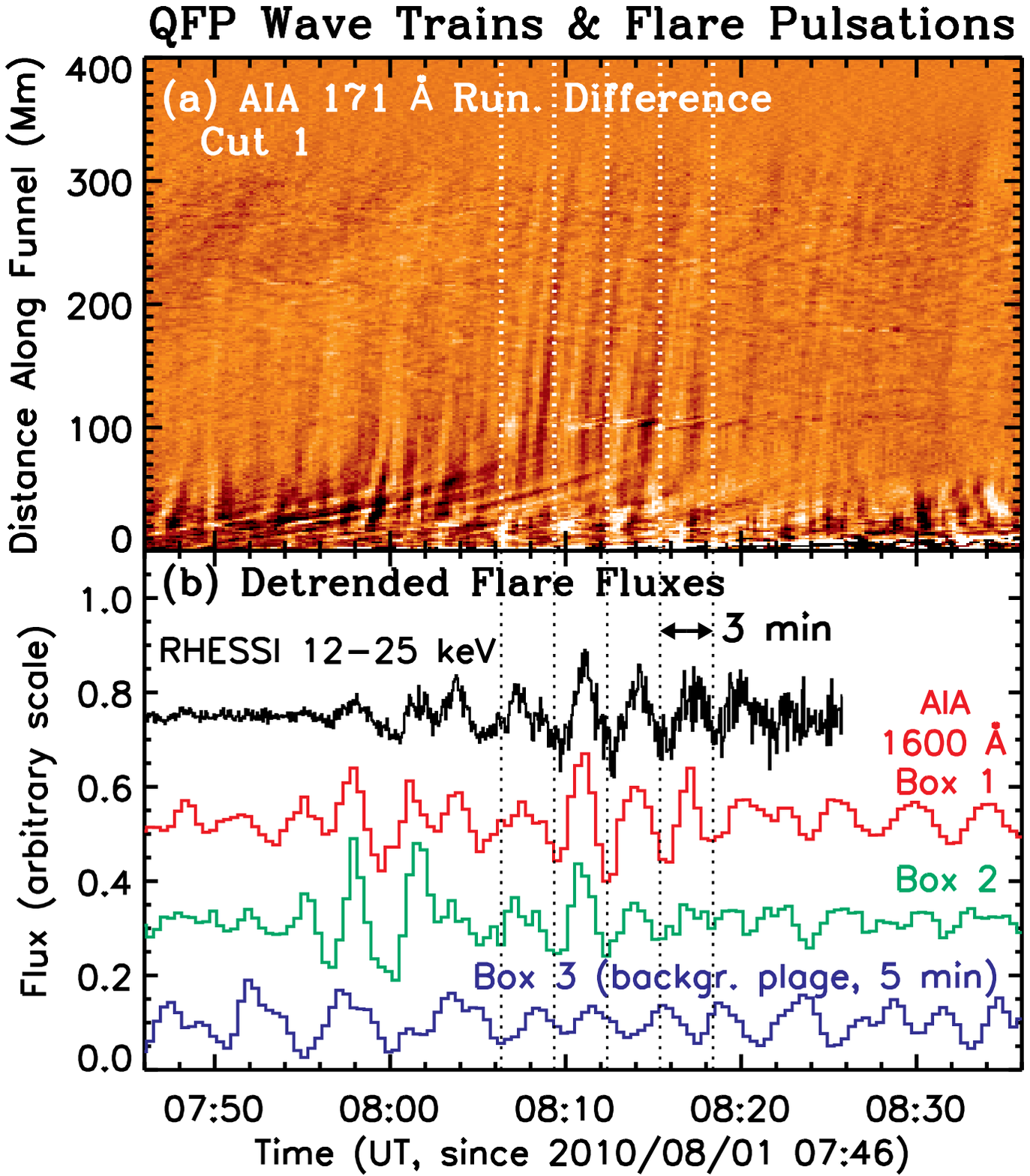}	
 \includegraphics[height=5.3cm]{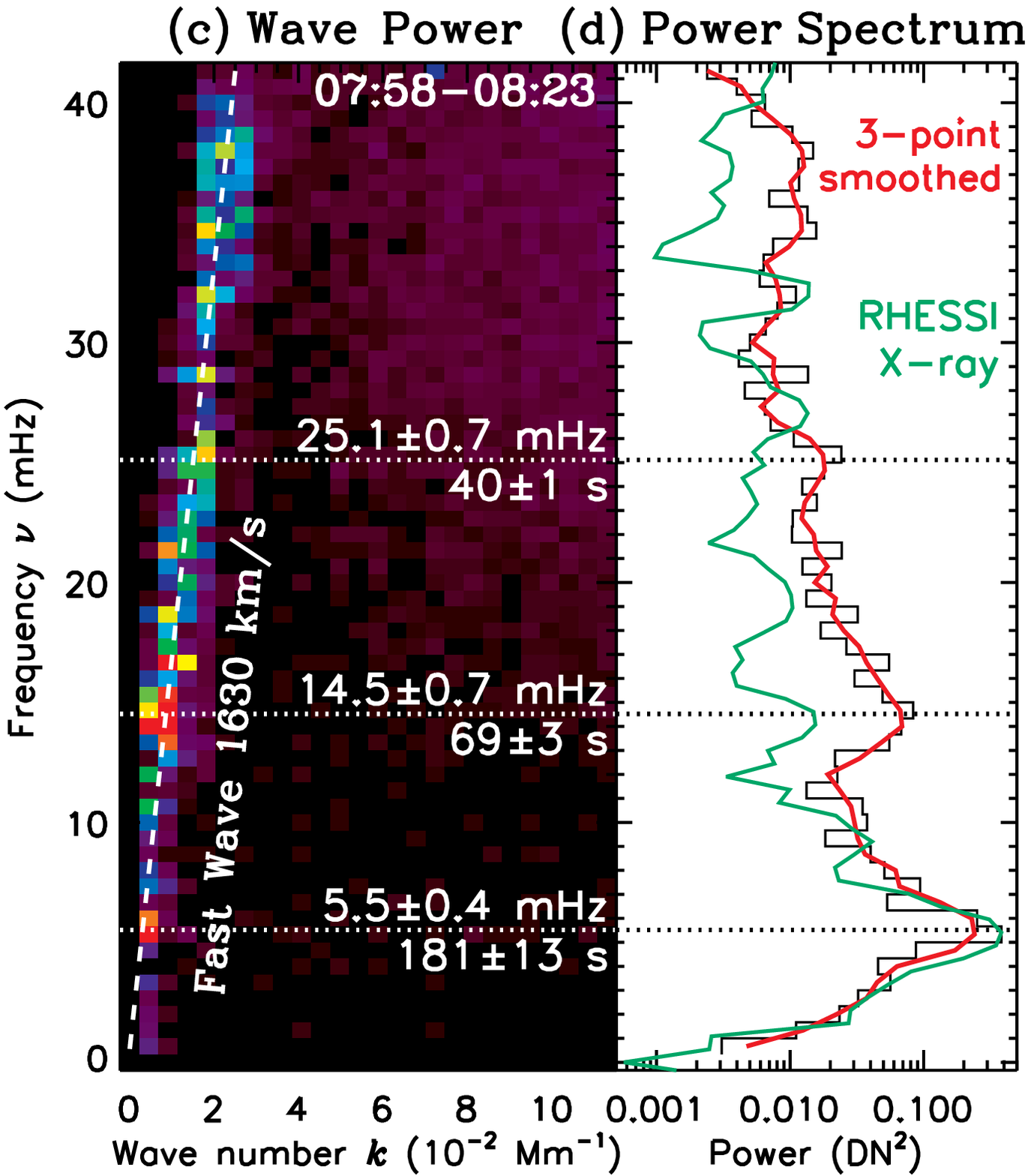}						
 \end{center}
 \caption[]{
 Examples of QFP wave trains observed by AIA and their frequency distributions.
   {Left}: Distinct QFP trains from a flare on 30 May 2011
 shown in 171~\AA\ running-difference (bottom) and direct (top) images 
 (from \opencite{YuanD.QFP.distinct.trains.2013A&A...554A.144Y}).	
   {Middle}: QFP waves  shown in a 171~\AA\ space--time plot (top) and their correlated
 flare pulsations (bottom) displaying a dominant three-minute period.
   {Right}: Fourier power or $k$--$\omega$ diagram showing a bright ridge for
 the QFP waves and wave-number averaged power as a function 
 of frequency on the right.
 The middle and right panels are from \inlinecite{LiuW.AIA-1st-EITwave.2010ApJ...723L..53L}
 for the 01 August 2010 event.
 } \label{QFP-overview.eps}
 \end{figure}

Quasi-periodic fast propagating (QFP) wave trains were one of the serendipitous discoveries of SDO/AIA
\cite{LiuW.AIA-1st-EITwave.2010ApJ...723L..53L,LiuW.FastWave.2011ApJ...736L..13L}
and are strong evidence of quasi-periodic fast-mode magnetosonic waves.
As shown in \fig{QFP-overview.eps} (left), they usually appear as a series of arc-shaped, 
1\,--\,5\,\% intensity variations emanating from flare kernels in active regions
and traveling upward along {\it funnel- or conic-shaped paths} within narrow angles (say, $10 \degree$\,--\,$60 \degree$)
often outlined by coronal loops. 
They propagate at {\it high speeds} of 500\,--\,$2200 \kmps$ up to 200\,--\,$400 \Mm$ from their origins,
and can last episodically from about ten~minutes to more than one~hour with numerous pulses following one another.
These characteristics distinguish QFP wave trains from comparably slower EIT waves that propagate 
in wider angular extents to greater distances across the solar disk or Moreton waves that travel on the chromosphere.
%
%

To date, more than 15 QFP events have been identified in the AIA database, and 
far more may have been detected. A fraction of these events were analyzed in detail, 
whose characteristics are summarized in \tab{table_QFPs}.
Prior to AIA, periodic disturbances traveling at such high speeds and evidencing
fast-magnetosonic wave trains were only imaged in a solar eclipse \cite{WilliamsDR.fast-mode-eclipse.2002MNRAS.336..747W}
and in post-flare supra-arcades 	
\cite{Verwichte.fast-mode-Apr21-supra-arcade.2005A&A...430L..65V}.
More common, non-imaging evidence was found in radio observations, such as
decimetric bursts of 10\,--\,80~second periods in a coronal fan above a null point 
\cite{MeszarosovaH.fast-mode.in.fan.above.null.2013SoPh..283..473M}
and fiber bursts of short 1\,--\,2~second periods \cite{KarlickyM.fiber.burst.fast.sausage.wave.train.2013A&A...550A...1K}.
\begin{table}[bthp]	
\scriptsize			
\caption{\small QFP wave trains observed in AIA's 171~\AA\ channel.}
\tabcolsep 0.01in	
\begin{tabular}{llcllllll}
\hline 
\multicolumn{3}{c}{Associated Flare} && 
\multicolumn{4}{c}{QFP Wave Characteristics}  &  References   \\
\cline{1-3}                \cline{5-8}
Date             & Start & GOES  &&  Init.~speed  & Deceler.     & Range & Period     &  \\
{\tiny ddmmmyy} & Time  & Class &&  {\tiny [$\kmps$]}   & {\tiny [$\km \pss$]} & {\tiny [Mm]}  & Peaks      &  \\
                  &       &       &&              &              &       & {[seconds]}  &  \\
\hline 
08Apr10 & 02:30 & B3.8  && 450\,--\,1200  & 0.2\,--\,5.8 & 330 & 40\,--\,240       & \inlinecite{LiuW.AIA-1st-EITwave.2010ApJ...723L..53L}%
\tabnote{These results were not presented there, but QFPs are evident in their Figure~3h.}   \\
01Aug10 & 07:25 & C3.2  && 2200       &         & 400 & 40,\,69,\,181   & \inlinecite{LiuW.FastWave.2011ApJ...736L..13L}   \\
08Sep10 & 23:00 & C3.3  && 1000\,--\,1200 & 3\,--\,4    & 320 & 30\,--\,240       & \inlinecite{LiuW.cavity-oscil.2012ApJ...753...52L}  \\
25Mar11 & 23:08 & M1.0  && 1000\,--\,1300  &         & 250 & $\approx$180     & \inlinecite{KumarP.AIA.plasma.blob.EUV.wave.2013A&A...553A.109K}%
\tabnote{Only two pulses were present and associated with an ejecta impact rather than a flare.} \\  
30May11 & 10:48 & C2.8  && 830        &         & 220 & 25\,--\,400       & \inlinecite{ShenYD.LiuY.QFP.wave.2012ApJ...753...53S}   \\  
 \multicolumn{3}{c}{}	
                         && 740\,--\,850   & 1.3\,--\,2.3 &    & 38,\,40,\,58   & \inlinecite{YuanD.QFP.distinct.trains.2013A&A...554A.144Y} \\
23Apr12 & 17:38 & C2.0  && 690        & 1.0     & 300 & 80     &  \inlinecite{ShenYD.LiuYu.QFP.171.193.2013SoPh..288..585S} \\
\hline
\multicolumn{2}{l}{Typical Range}
                & B\,--\,C   && 500\,--\,2200  & 1\,--\,4    & 200\,--\,400 &  25\,--\,400 & Total: 6 events, 7 articles  \\

\hline \end{tabular}
\label{table_QFPs} \end{table}

Although at times QFPs can be identified at their source flares
\cite{LiuW.FastWave.2011ApJ...736L..13L}, 
their initial appearances are often some distance away ($\gtrsim\,$$100 \Mm$).	
Such a distance is required for amplitude growth or a preferential LOS to be satisfied
for detecting them \cite{CooperF.LOS.detect.kink.sausage.2003A&A...397..765C}.
In fact, their amplitudes usually increase with distance and then decrease,
likely because of the interplay between amplification due to
density stratification and attenuation due to geometric expansion of the funnel
\cite{YuanD.QFP.distinct.trains.2013A&A...554A.144Y}. 
QFPs often exhibit {\it strong decelerations} on the order 
of 1\,--\,$4\km \pss$, likely due to the decrease of the fast-magnetosonic speed with distance 	
from the active region core.	
In an extreme case, they rapidly decelerate and terminate
while approaching a CME front from behind, which was attributed to 
enhanced damping or dispersion in such a turbulent environment	
\cite{LiuW.cavity-oscil.2012ApJ...753...52L}.
Expected dispersive evolution with decreasing periods was also identified
\cite{YuanD.QFP.distinct.trains.2013A&A...554A.144Y}.
Many of these characters have been reproduced in MHD simulations 	
\cite{Ofman.Liu.fast-wave.2011ApJ...740L..33O,Pascoe.wing.QFPs.funnel.2D.MHD.2013A&A...560A..97P}.

QFPs often travel upward in {\it open funnels}, but occasionally in opposite directions
along {\it closed loops} between conjugate flare ribbons \cite{LiuW.FastWave.2011ApJ...736L..13L} 
and even two sympathetic flares across the Equator \cite{LiuW.Ofman.counter.wave.2014}.
These counter-propagating waves seem to be generated individually in association with their source flares,
but it is possible that some of them are reflected repeatedly between the footpoints of closed loops.

\subsection{Periodicities, Correlation with Flares, and Physical Origin}
\label{subsect_freq}

QFP waves are observed in a wide range of periods from 25
to $\approx$400~seconds, with the lower end limited by the Nyquist frequency 
of $42 \mHz$ given by AIA's 12-second cadence. In Fourier power spectra or $k$--$\omega$ diagrams
({\it e.g.} \fig{QFP-overview.eps}, right),
QFPs appear as bright, nearly straight ridges passing through the origin,
which describe their {\it dispersion relations} and indicate temporally averaged
phase [$\nu/k$] and group [$d\nu/dk$] speeds indistinguishable 
given AIA's spatio--temporal resolution. 
Individual peaks of power on the ridge are often concentrated within a period range of 40\,--\,240~seconds. 
Some QFP periods, {\it e.g.} two to three~minutes or shorter, as shown in \fig{QFP-overview.eps} (middle),		
are correlated with {\it quasi-periodic pulsations} (QPPs) of accompanying flare 
emissions commonly seen from radio to hard X-rays 
({\it e.g.} \opencite{Fleishman.QPP-periodic-acc.2008ApJ...684.1433F};
\opencite{InglisA.DennisB.QPP.not.slow.modes.2012ApJ...748..139I};
\opencite{DollaL.QPP.2011Feb15.X2flare.2012ApJ...749L..16D};
\opencite{SuJT.AIA.QPP.sausage.and.slow-mode.2012ApJ...755..113S}). 	
Such correlations suggest a common physical origin.

Proposed mechanisms for flare QPPs (see \opencite{Nakariakov.Melnikov.QPP.2009SSRv..149..119N} for a review)
fall into {\it two categories}: pulsed energy release intrinsic to magnetic reconnection
and MHD oscillations, both relevant to the generation of QFP wave trains 
\cite{LiuW.FastWave.2011ApJ...736L..13L,ShenYD.LiuYu.QFP.171.193.2013SoPh..288..585S}.
For example, oscillatory reconnection at X-type null points
\cite{McLaughlin.periodic-waves_by-oscil-reconn.2012ApJ...749...30M},
repetitive generation and coalescence of plasmoids 	
\cite{KliemB.puls.recon.2000A&A...360..715K}		
and their fast ejections ({\it e.g.} with a period of about two~minutes,  
\opencite{LiuW.cusp.flare.20120719M7.2013ApJ...767..168L}),
or current sheet fluctuations induced by super-\Alfvenic beams 
and associated Kelvin--Helmholtz instability nonlinear oscillations 
\cite{OfmanSui.HXR-oscil.2006ApJ...644L.149O}
can cause {\it episodic energy release} in forms of plasma heating and particle acceleration 
that can result in emission pulsations and excite magnetosonic waves.

Meanwhile, {\it MHD oscillations} with periods determined by resonance or dispersion
can modulate flare energy release or emission
\cite{Foullon.QPP.kink-mode.2005A&A...440L..59F}	
and trigger magnetosonic waves. In particular, the strong three-minute QFP wave signal 
(see \fig{QFP-overview.eps}, middle; \opencite{LiuW.FastWave.2011ApJ...736L..13L})
may be related to slow-mode magnetosonic waves		
leaking from the three-minute chromospheric sunspot oscillations
\cite{JessD.3min.slow-mode.coronal.fan.2012ApJ...757..160J}
that can trigger flare QPPs \cite{Sych.3min-QPP-slow-mode.2009A&A...505..791S}.
Long period ($\gtrsim\,$$300$~seconds) QFP waves have been ascribed to
the leakage into the corona of pressure-driven $p$-modes 	
\cite{ShenYD.LiuY.QFP.wave.2012ApJ...753...53S}.		
In this case, it implies that during QFP-associated eruptions, 
some yet to be determined special mechanisms or conditions 
({\it e.g.} cone-shaped CME wakes seen in white-light eclipse images, \opencite{HabbalS.cone.wake.2011ApJ...734..120H}) 
must amplify these waves, which are otherwise expected to be seen anytime and anywhere on the Sun.

The funnel-shaped QFP paths often along large-scale coronal loops (open or closed)
indicate the presence of {\it waveguides}, because in an otherwise homogeneous low-$\beta$ plasma, 
fast-modes would propagate more isotropically both along and across magnetic-field
lines like EIT waves. A waveguide can form in a channel of lower fast-magnetosonic speed [$v_{\rm f}$], 
say, due to density enhancements, than the surrounding medium and thus trap waves by internal reflection.
Another possibility is a leaky waveguide where a high $v_{\rm f}$ region 
is surrounded by a low $v_{\rm f}$ background ({\it e.g.} \opencite{OfmanL.Alfven.heat.CH.leaky.wave.guide.1995JGR...10023413O}).
Such waveguides of finite widths results in dispersion with components of different frequencies 
propagating at different speeds. This can produce additional periods 
\cite{Pascoe.wing.QFPs.funnel.2D.MHD.2013A&A...560A..97P}, independent of a periodic driver.
An impulsively generated fast-mode wave can dispersively evolve
into a quasi-periodic wave train \cite{Roberts.coronal-seismology.1984ApJ...279..857R},
as manifested in AIA detection of distinct wave trains each 	
associated with a radio-burst episode \cite{YuanD.QFP.distinct.trains.2013A&A...554A.144Y}.




\subsection{Wave Trains Inside and Outside CME Bubbles}
\label{subsect_2trains}

\begin{figure}[thbp]      
 \begin{center}
 \includegraphics[width=0.35\textwidth]{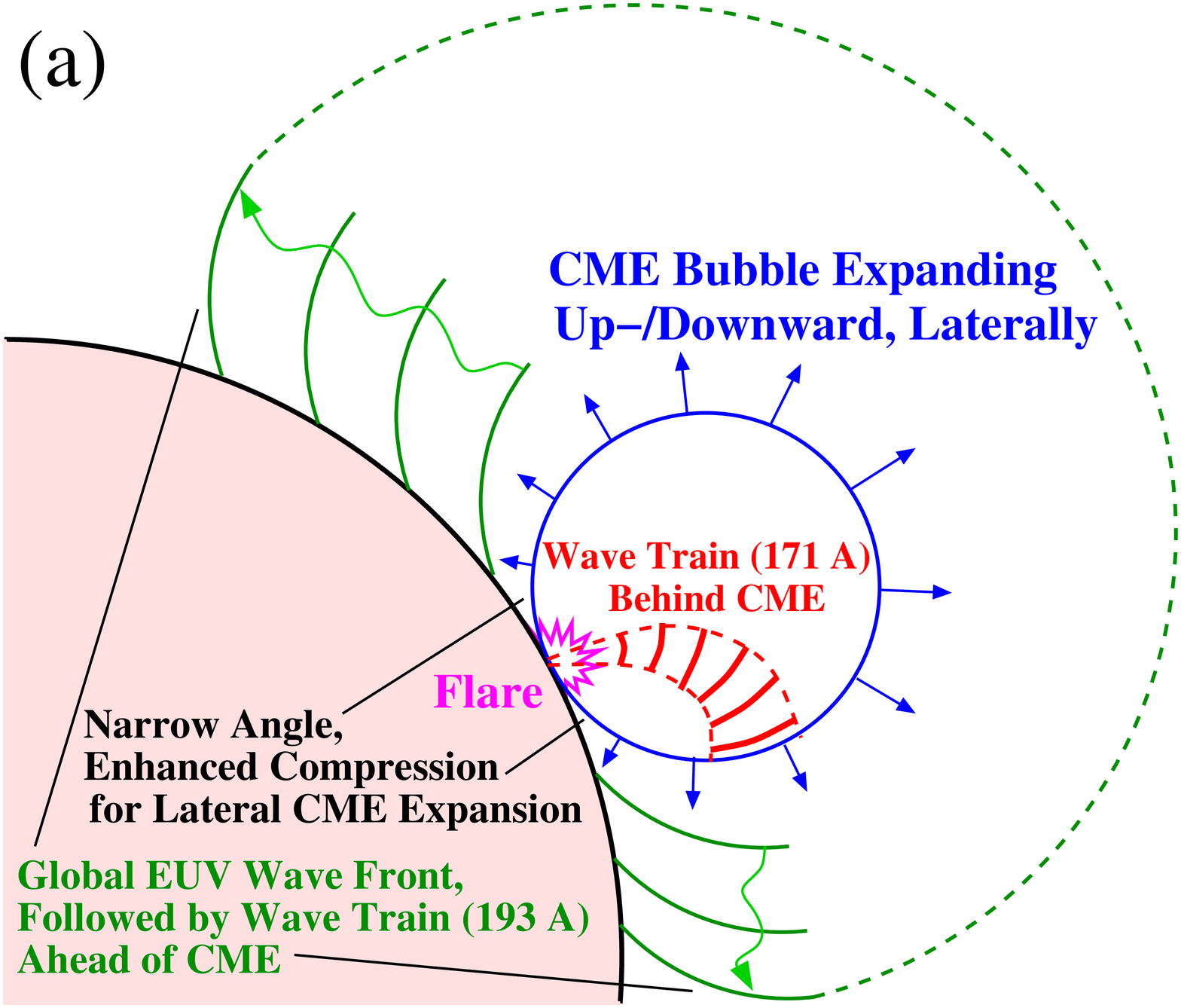}
 \includegraphics[width=0.6\textwidth]{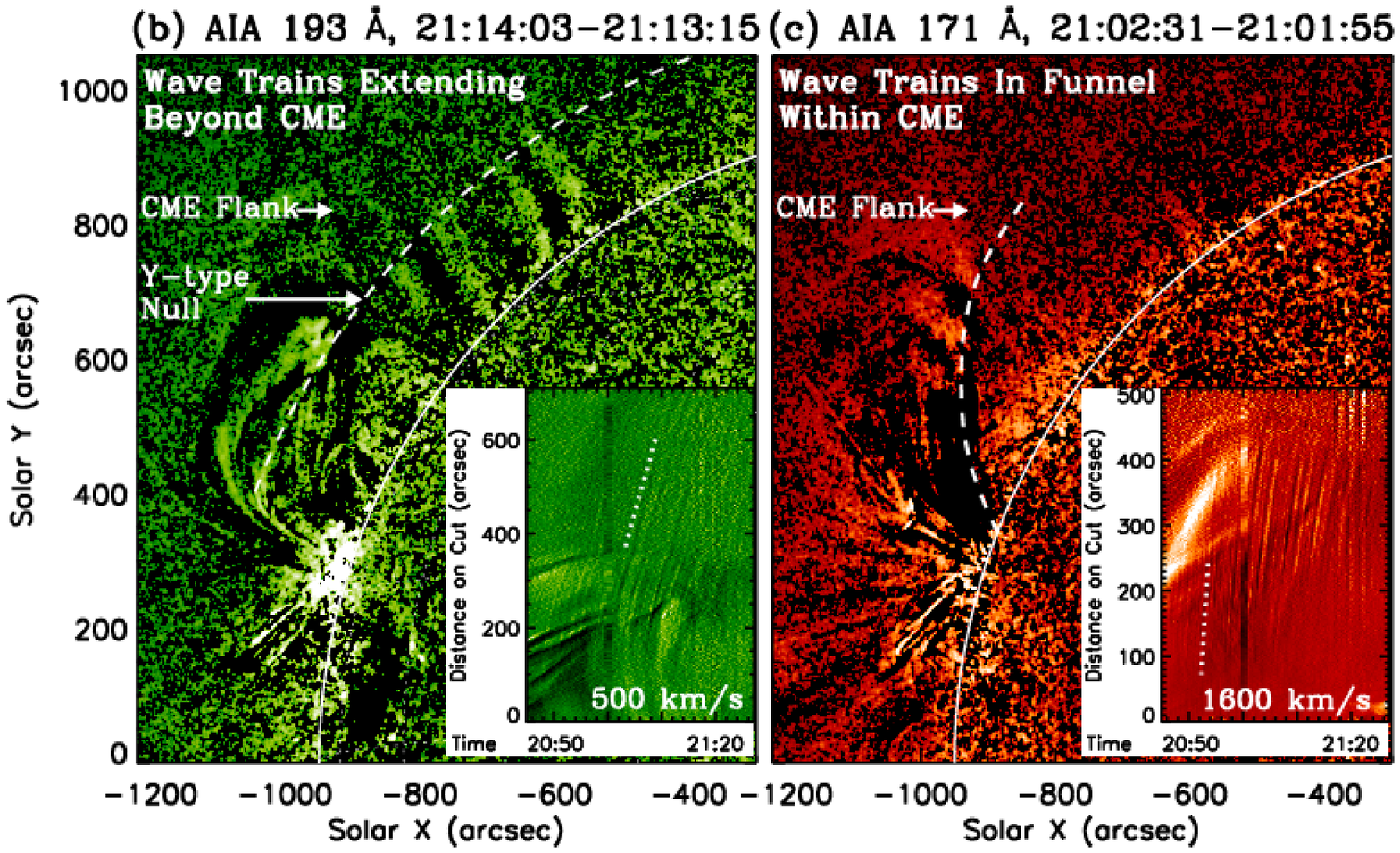}		
 \end{center}
 \caption[]{
  (a) Schematic of quasi-periodic wave trains outside (green) and inside (red) 
a CME bubble (blue) (from \opencite{LiuW.cavity-oscil.2012ApJ...753...52L}). 
The former propagates along the solar surface and follows the leading front of the broad EIT wave pulse
that is driven by the lateral and downward 	
CME expansions initiated at an elevated height.
The narrow angle formed between the CME bubble and the underlying chromosphere
can cause enhanced compression due to the lateral CME expansion (see \sect{subsect_gener}). 
The QFP wave trains behind the CME originate at the flare and propagate along a narrow funnel.
  (b) and (c) Example of two wave trains in the same event that appear in running-difference AIA images
preferentially outside (193~\AA, green) and inside (171~\AA, red) 
the CME (from \opencite{LiuW.Ofman.Downs.in-outside-QFPs.2014}).	
The insets are space--time plots obtained from cuts indicated by the white dashed lines in the images.
The wave trains exhibit as zebra-like steep stripes at typical speeds of 500 and $1600 \kmps$.
Note in (b) an abrupt change in wave speed at the CME flank shown as the outermost shallow stripe
near $s=350 \arcsec$.
 } \label{QFP-2trains.eps}
 \end{figure}

QFPs in coronal funnels are apparently different from quasi-periodic wave trains 
within broad EIT wave pulses (see \sect{subsect_bimod}) in temperature, speed, 
and spatial domain with respect to CMEs. As shown in \fig{QFP-2trains.eps},
QFPs are best (and often only) seen at 171~\AA\ (suggestive of lower temperatures)
within CME bubbles, while EIT wave trains are most evident at 193 and 211~\AA\ ahead of CME flanks.
	%
	%
When these wave trains are both detected in the same event off the limb
\cite{LiuW.cavity-oscil.2012ApJ...753...52L,LiuW.Ofman.Downs.in-outside-QFPs.2014}
or possibly on the disk \cite{LiuW.AIA-1st-EITwave.2010ApJ...723L..53L,%
ShenYD.LiuY.QFP.wave.2012ApJ...753...53S,ShenYD.LiuYu.QFP.171.193.2013SoPh..288..585S},
the funnel QFPs are usually two to three times faster than their EIT wave counterparts.
This is consistent with the rapid decrease of the fast-magnetosonic speed away from active regions.
Another possible contribution to the higher speeds of QFPs is the faster fall-off of density
with height due to gravitational stratification of the cooler 171~\AA\ emitting plasma and thus higher \Alfven speeds 
\cite{Ofman.Liu.fast-wave.2011ApJ...740L..33O}.
%
%

Aside from their differences, such wave trains sometimes have considerable spatial overlap
({\it e.g.} \opencite{ShenYD.LiuYu.QFP.171.193.2013SoPh..288..585S})
and even share some similar periods with the accompanying flare pulsations
\cite{LiuW.cavity-oscil.2012ApJ...753...52L}.
EIT wave trains detected thus far are always accompanied by funnel QFPs, but not {\it vice versa}.
In off-limb events	
the QFP funnel at times gradually turns from vertical to horizontal,
and the EIT wave trains travel horizontally along the solar surface.

Because of these similarities, we posit that that EIT wave trains beyond CME bubbles
are the {\it continuation} of QFP wave trains within funnels. 
In other words, both types of wave trains can share the same excitation agent but
exhibit different propagation behaviors, such as temperatures, speeds, and periods or wavelengths, 
determined or modified by their respective media, for example, due to dispersion.
This continuation can happen under favorable conditions, {\it e.g.} when a funnel waveguide turns
horizontal with distance or when QFPs are refracted back toward the solar surface.
Such conditions, somewhat stringent, may explain the relative infrequency
of EIT wave trains compared with funnel QFPs.
In fact, the increase of \Alfven speed with height on
the quiet Sun due to the faster fall-off of density than magnetic field can create
a horizontal waveguide aligned above the solar surface
({\it e.g.} \opencite{Afanasyev.Uralov.shock-EUV-wave-I.2011SoPh..273..479A}; 
\opencite{Kwon_EUV.wave.seismology.2013ApJ...776...55K}). This waveguide
can form an extension of an asymptotically horizontal funnel, 
confine EIT wave trains to their observed heights $\lesssim\,$100~Mm or one to two density
scale heights, and produce their forward inclinations
({\it cf.} \opencite{LiuW.cavity-oscil.2012ApJ...753...52L}).
As QFPs evolve into EIT wave trains on the quiet Sun, abrupt changes in speed can occur  
at topological interfaces or CME flanks between  	
media of different fast-magnetosonic speeds
({\it e.g.} \figs{QFP-2trains.eps}b and \ref{QFP-2trains.eps}c; \opencite{ShenYD.LiuYu.QFP.171.193.2013SoPh..288..585S}).

Another possible origin of EIT wave trains is {\it leakage}
from funnel waveguides, which can happen for sausage modes, for example, with wavelengths greater than the
cutoff value and incident angles at the funnel boundary being smaller than 
the total internal reflection angle \cite{Pascoe.wing.QFPs.funnel.2D.MHD.2013A&A...560A..97P}.
These possibilities remain to be validated with more detailed observations and 
more sophisticated MHD models.


\section{Small-scale EUV Waves and Wave-like Phenomena}
\label{sect_mini-wave}



In addition to large-scale waves, AIA has observed various small-scale waves, 
including mini-EUV waves and nonlinear waves or vortices associated with magnetic
Kelvin--Helmholtz instabilities, which add to the multitude of EUV wave and wave-like phenomena.
An exhaustive list of such events detected by AIA and their characteristics are given in \tab{table_mini-wave}.
\begin{table}[bthp]	
\scriptsize			
\caption{\small AIA observed small-scale EUV waves.}
\tabcolsep 0.02in	
\begin{tabular}{llcccll}
\hline 
\multicolumn{7}{c}{\bf (1) mini-EUV waves}  \\
\hline
Date     & Start & Duration   & Speed\tabnote{Initial or average speed.}     & Range & Associated     &  References  \\	
{\tiny ddmmmyy}
         & Time  & [minutes] & {[$\kmps$]} & {[Mm]}  & Phenomena       \\	
\hline 	
\multicolumn{7}{c}{{\bf (1.1)} Ephemeral-region mini EIT waves} \vspace{0.03in} \\		
21Oct10 & 20:37 & 13     & 270\,--\,350  & 250   & micro-sigmoid/B1.7 flare & \inlinecite{ZhengRS.EUV.wave.micro-sigmoid.2012ApJ...753L..29Z} \\
11Nov10 & 18:10 & 10\,--\,20 & 280\,--\,500  & 360   & surge-driven, homologous  & \inlinecite{ZhengRS.homologous.EUV.wave.EFR.2012ApJ...747...67Z} \\	
01Dec10 & 02:54 & 17     & 220\,--\,250  & 180   & mini-CME       & \inlinecite{ZhengRS.AIA.EUVwv.miniCME.2011ApJ...739L..39Z}  \\
01Mar11 & 13:00 & 13     & 260\,--\,350  & 260   & failed filament eruption & \inlinecite{ZhengRS.EUV.wave.failed.eruption.2012A&A...541A..49Z} \\
04Oct12 & 06:16 & 12     & 300\,--\,360  & 300   & micro-sigmoid/CME/flare  & \inlinecite{ZhengRS.loop.oscil.EUV.wave.micro-sigmoid.2013MNRAS.431.1359Z}  \\
\hline	
\multicolumn{2}{l}{Typical Range}  
                 & 10-20  & 200\,--\,500  & \multicolumn{2}{l}{180-360}  & Total: 5 events \vspace{0.1in}  \\
\hline 
\multicolumn{7}{c}{{\bf (1.2)} Quiet-Sun mini-EUV waves} \vspace{0.03in} \\
20Jul10 & 11:00 & 8      & 35\,--\,85    & 20    & quiet-Sun cyclones & \inlinecite{ZhangJun.LiuYang.AIA.cyclone.2011ApJ...741L...7Z} \vspace{0.15in} \\
\hline 

\multicolumn{7}{c}{\bf (2) Kelvin--Helmholtz instability vortices associated with CMEs/ejecta}  \\
\hline 
Date     & Start & GOES   & Speed\tabnote{Phase speed.}   & $\lambda$     &  \multicolumn{2}{l}{References}  \\
{\tiny ddmmmyy}
         & Time  & Class  & [$\kmps$] & [Mm]             \\
\hline 
08Apr10 & 02:30 & B3.7   & 6\,--\,14     & $\leq\,$7      & \multicolumn{2}{l}{\inlinecite{Ofman.Thompson.AIA.KH.instab.2011ApJ...734L..11O}}  \\
03Nov10 & 12:07 & C4.9   & 420       & 18            & \multicolumn{2}{l}{\citeauthor{Foullon.AIA.KH.instab.2011ApJ...729L...8F} 
   (\citeyear{Foullon.AIA.KH.instab.2011ApJ...729L...8F}, \citeyear{Foullon.AIA.KH.instab-2.detail.2013ApJ...767..170F});
   \inlinecite{BainH.typeII.plasmoid.AIA.2010-11-03.2012ApJ...750...44B}\tabnote{On the associated Type-II burst and AIA observations of the plasmoid ejection.} }  \\
24Feb11 & 07:23 & M3.5   & 310       & 14            & \multicolumn{2}{l}{\inlinecite{MostlU.KH.instab.AIA.MHD.model.2013ApJ...766L..12M}}  \\
\hline	
\multicolumn{2}{l}{Typical Range}  
                 & B\,--\,M   & 10\,--\,400   & \multicolumn{1}{l}{7\,--\,18}   & \multicolumn{2}{l}{Total: 3 events}  \\		

\hline \end{tabular}
\label{table_mini-wave} \end{table}

\subsection{Mini-EUV Waves}
\label{subsect_mini-euv}

Mini-EUV waves associated with small-scale eruptions are miniature versions
of EIT waves. They were first detected by STEREO/EUVI 
\cite{InnesD.mini.CME.EUV.wave.2009A&A...495..319I,Podladchikova.EUVI.mini.wave.2010ApJ...709..369P}
and then commonly seen by SDO/AIA 
(\opencite{ZhangJun.LiuYang.AIA.cyclone.2011ApJ...741L...7Z}; \opencite{ZhengRS.AIA.EUVwv.miniCME.2011ApJ...739L..39Z},
\citeyear{ZhengRS.EUV.wave.failed.eruption.2012A&A...541A..49Z,ZhengRS.EUV.wave.micro-sigmoid.2012ApJ...753L..29Z,%
ZhengRS.homologous.EUV.wave.EFR.2012ApJ...747...67Z,ZhengRS.loop.oscil.EUV.wave.micro-sigmoid.2013MNRAS.431.1359Z}).
They share similar characteristics with EIT waves, including quasi-circular shapes, 
often uniform speeds sometimes with decelerations, instigating loop oscillations,
being driven or triggered by energy release events
such as mini-CMEs and micro-flares, as well as spatial-decoupling from their drivers.
The key difference between EIT waves and mini-EUV waves is their {\it distinct source regions}
that define their energy budgets and physical characters.
Global EIT waves are associated with release of large amounts of free energy
on the order of $10^{32} \erg$ via CMEs or flares in {\it active regions}. Mini-EUV waves originate
{\it away from active regions} with orders of magnitude less free energy available, giving rise
to their somewhat smaller ranges of 20\,--\,$400 \Mm$,	
shorter lifetimes of 10\,--\,30~minutes, 	
and weaker wave intensities and dimmings.

Accompanying a variety of small-scale eruptions, including mini-CMEs, 
micro-flares, mini-filament and micro-sigmoid eruptions, jets, and surges, 
mini-EUV waves fall into two categories (see \tab{table_mini-wave}):
i) Those from {\it ephemeral regions} involving flux emergence or cancellation
tend to be larger in size and faster at typically 200\,--\,$500 \kmps$,
and are thus interpreted as fast-mode waves
({\it e.g.} \opencite{ZhengRS.AIA.EUVwv.miniCME.2011ApJ...739L..39Z}).
They closely resemble EIT waves and hence we call them ``mini-EIT waves".
ii) Those from the {\it quiet Sun}, for example, 
triggered by supergranular flows \cite{InnesD.mini.CME.EUV.wave.2009A&A...495..319I}
or by coronal cyclones associated with rotating network magnetic fields 
\cite{ZhangJun.LiuYang.AIA.cyclone.2011ApJ...741L...7Z}, tend to be smaller and slower at typically 10\,--\,$100 \kmps$,
and thus possibly evidence slow-mode waves \cite{Podladchikova.EUVI.mini.wave.2010ApJ...709..369P} 
or non-wave coronal reconfigurations.
\inlinecite{InnesD.mini.CME.EUV.wave.2009A&A...495..319I} estimated that 
quiet-Sun mini-eruptions can be numerous	
at a high rate of 1400 per day, with one third producing mini-EUV waves.
Considering the ensemble of EUV waves of such hierarchic sizes and speeds
\cite{Patsourakos.Vourlidas.EIT-wave-review.2012SoPh..281..187P}, 
mini-EUV waves can constitute an extension of the size distribution of large-scale EIT waves,
like the extension of their drivers, {\it i.e.} CMEs, toward smaller scales
\cite{SchrijverC.miniCME.size.distr.2010ApJ...710.1480S,InnesD.quiet.sun.explosions.2013SoPh..282..453I},
which involve (impulsive) mass motions of some kind.

\subsection{Magnetic Kelvin--Helmholtz Instability Waves}
\label{subsect_KHI}

 \begin{figure}[thbp]      
 \begin{center}
 \includegraphics[height=3.8cm]{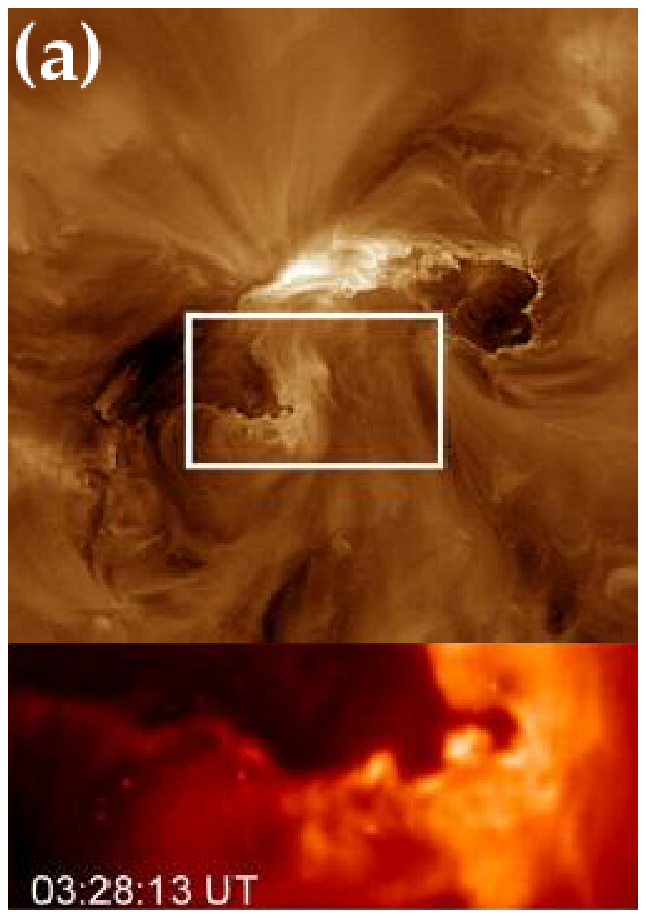} 	
 \includegraphics[height=4.0cm]{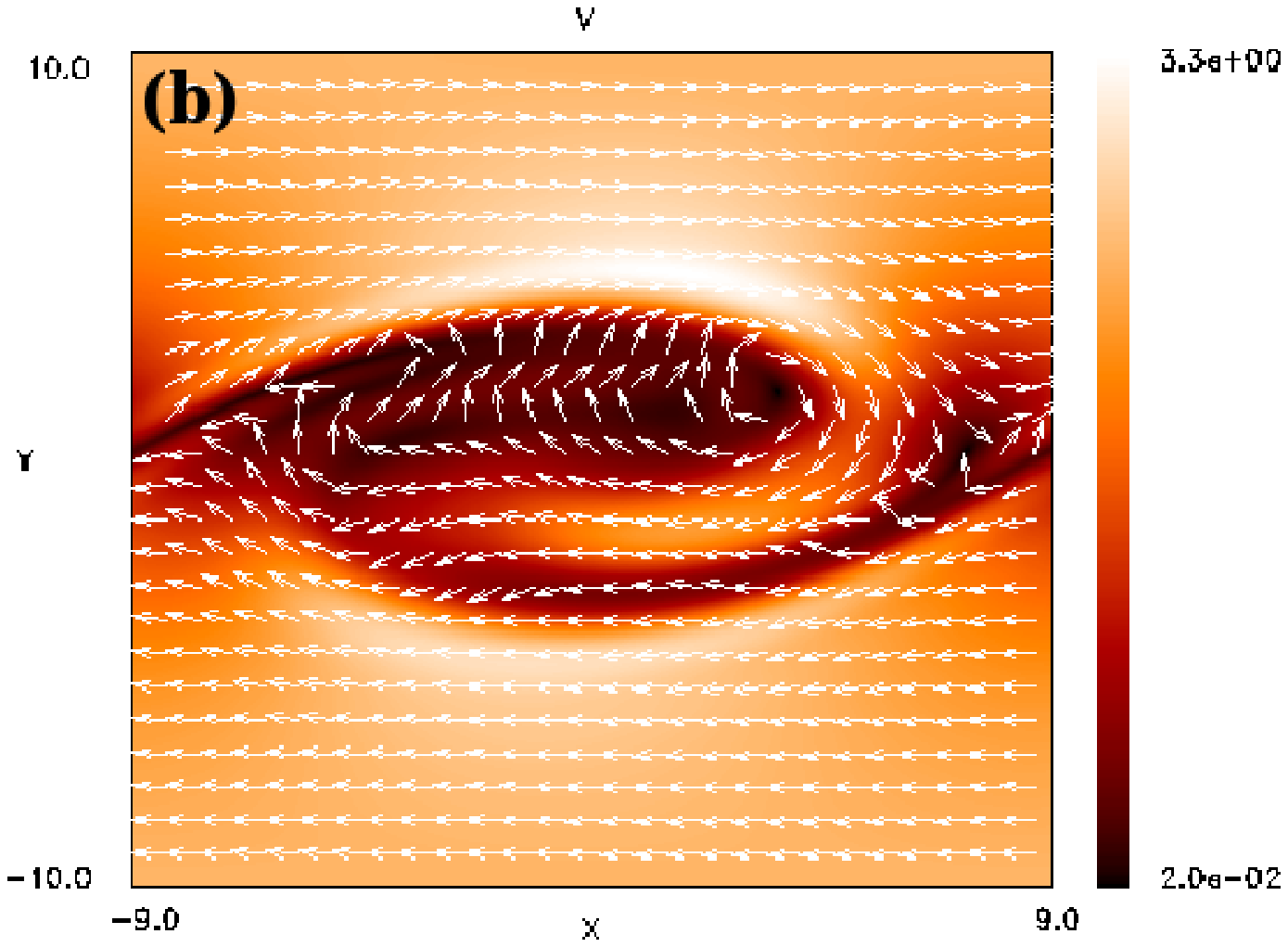}
 \includegraphics[height=3.9cm]{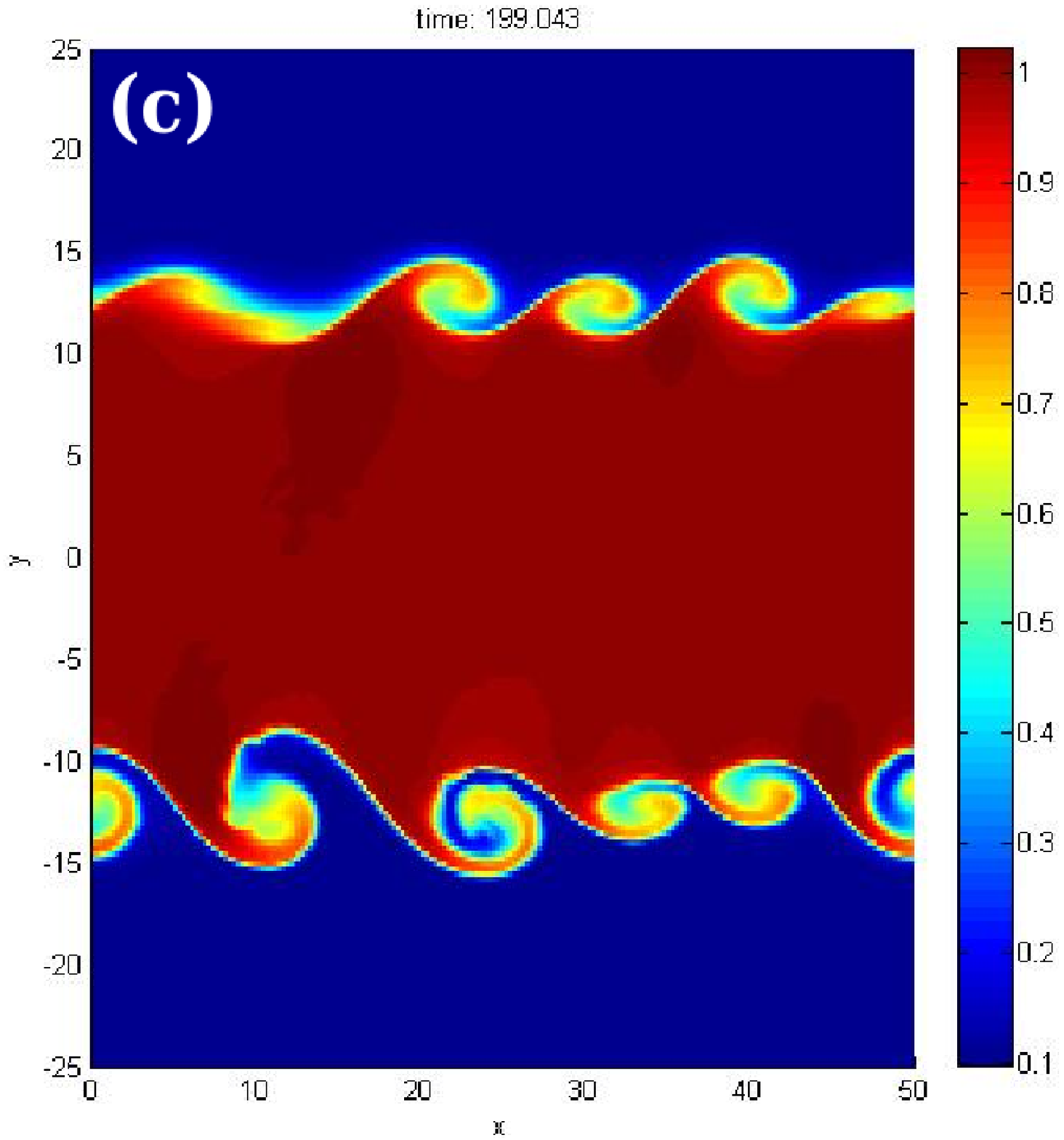}
 \end{center}
 \caption[]{	
 Small-scale,	
 nonlinear waves associated with magnetic Kelvin--Helmholtz instabilities
 observed by AIA on 08 April 2010 at 193~\AA\ (a)		
 and simulated plasma velocity (b) (both from \opencite{Ofman.Thompson.AIA.KH.instab.2011ApJ...734L..11O}).
  (c) Simulated plasma density for the KHI event associated with the 24 February 2011 M3.5 flare
 (from \opencite{MostlU.KH.instab.AIA.MHD.model.2013ApJ...766L..12M}). 
 } \label{mini_wave.eps}
 \end{figure}
%
Another serendipitous discovery of AIA was the detection of Kelvin--Helmholtz Instabilities (KHIs)
in the magnetized corona, which can play important roles in mass and energy transfer during solar eruptions
(\opencite{Ofman.Thompson.AIA.KH.instab.2010AGUFMSH14A..02O}, \citeyear{Ofman.Thompson.AIA.KH.instab.2011ApJ...734L..11O};
\opencite{Foullon.AIA.KH.instab.2011ApJ...729L...8F}, \citeyear{Foullon.AIA.KH.instab-2.detail.2013ApJ...767..170F}; 
\opencite{MostlU.KH.instab.AIA.MHD.model.2013ApJ...766L..12M}).
Observed at velocity-shear boundaries, {\it e.g.} flanks of high-speed CMEs or ejecta,
KHIs appear as a series of vortices of wavelengths on the order of 10~Mm traveling along the boundaries
at phase speeds of 10\,--\,$400 \kmps$ (see \fig{mini_wave.eps}a and \tab{table_mini-wave}). 	
KHIs were also detected in prominences 
\cite{BergerT.bubble-RT-instable.2010ApJ...716.1288B,RyutovaM.RT.KH.instab.promin.2010SoPh..267...75R} 
and high-corona streamers 	
\cite{FengL.streamer.KH.instab.2013ApJ...774..141F}.
Numerical models ({\it e.g.} \figs{mini_wave.eps}b and \ref{mini_wave.eps}c) played a crucial role in identifying KHIs by comparing observations 
with simulated formation and growth of the associated nonlinear waves 
\cite{Ofman.Thompson.AIA.KH.instab.2011ApJ...734L..11O,SolerR.KH.instab.ion-neutral.coupling.2012ApJ...749..163S,%
MostlU.KH.instab.AIA.MHD.model.2013ApJ...766L..12M,NykyriK.FoullonC.KH.seismology.2013GeoRL..40.4154N,%
ZaqarashviliT.KH-instab.twisted.tube.solar.wind.2014A&A...561A..62Z}.
Such models also allow seismological inference of the strengths and relative orientations
of the magnetic fields across the velocity shear boundary.

\section{Coronal Seismology Using EUV Waves}
\label{sect_seism}



Like any propagating disturbances, EUV waves can provide potential diagnostics for their medium, 
{\it i.e.} the (large-scale) solar corona \cite{Ballai.global-corona-seism.2007SoPh..246..177B,Ballai.EIT-wave-seismology.2008IAUS..247..243B}. 
Compared with the rapid development of local coronal seismology using loop oscillations
in the last two decades \cite{Nakariakov.wave-review.2005LRSP....2....3N}, 
advances in global coronal seismology using large-scale EUV waves 
have been hindered mainly because of the long debate on the nature of EIT waves.
Now that this controversy is coming to an end and it is clear that at least their leading fronts 
are true fast-mode waves, global seismology is becoming a reality.
We summarize below recent advances in this area
using EIT waves and QFP wave trains to infer the {\it magnetic, thermal, and energy}
properties of the corona. 
Note that because of the density stratification and the $\propto n^2$ dependence of EUV emission,
EUV wave seismology is usually limited to the inner corona, {\it e.g.} up to heights of $\approx$100\,--\,200~Mm.
Seismology in the outer corona resorts to alternative observables, such as 
white-light streamer waves \cite{ChenYao.streamer.seismology.2011ApJ...728..147C,FengSW.streamer.wave.cycle23.2011SoPh..272..119F}
and quasi-periodic radio bursts \cite{ZaqarashviliT.radio.seismology.2013A&A...555A..55Z}.

\subsection{Magneto-Seismology}
\label{subsect_Bseism}

Inferring the elusive coronal magnetic field is one of the primary goals of coronal seismology,
which complements other approaches including polarimetric measurements using 
Zeeman and Hanle effects \cite{LinH.coronal-B.2000ApJ...541L..83L},		
extrapolations of surface magnetograms 
\cite{ZhaoXP.Hoeksema.source-surface-model.1994SoPh..151...91Z,DeRosaM.NLFFF.compare.2009ApJ...696.1780D},
and radio observations of gyro-resonance emission 
\cite{GaryD.HurfordG.radio.Bfield.1994ApJ...420..903G,WhiteS.KunduM.radio.Bfield.1997SoPh..174...31W}.
The fast-magnetosonic speed weakly depends on the direction of the wave vector with respect
to the magnetic field, which has a strong vertical component in the quiet Sun	
and is thus practically assumed to be perpendicular to the
predominantly horizontal EIT wave vectors. The fast-magnetosonic speed therefore reduces to a simple form 
 \beq
 v_{\rm f}= \sqrt{v_{\rm A}^2 + c_{\rm s}^2},
 \label{vf_eq} \eeq 
where $v_{\rm A}= B / \sqrt{ 4 \pi \rho}$ is the \Alfven speed, which depends on the magnetic-field strength [$B$]
and mass density [$\rho$], and $c_{\rm s} \propto \sqrt{T}$ is the sound speed, which depends on
the temperature [$T$]. A single temperature is usually assumed for simplicity, although the corona
is multithermal.
With the knowledge of $v_{\rm f}$, $T$, and $\rho$, the magnetic-field strength can be calculated in cgs units as
 \beq
   B = \sqrt{ 4 \pi \rho (v_{\rm f}^2 - c_{\rm s}^2(T)) }.
  \label{B_eq} \eeq

\begin{figure}[thbp]      
 \begin{center}
 \includegraphics[height=3.8cm]{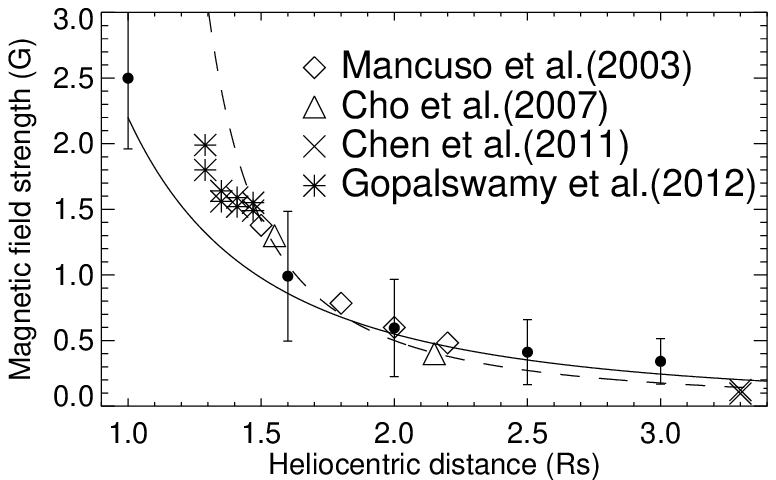}
 \includegraphics[height=4.4cm]{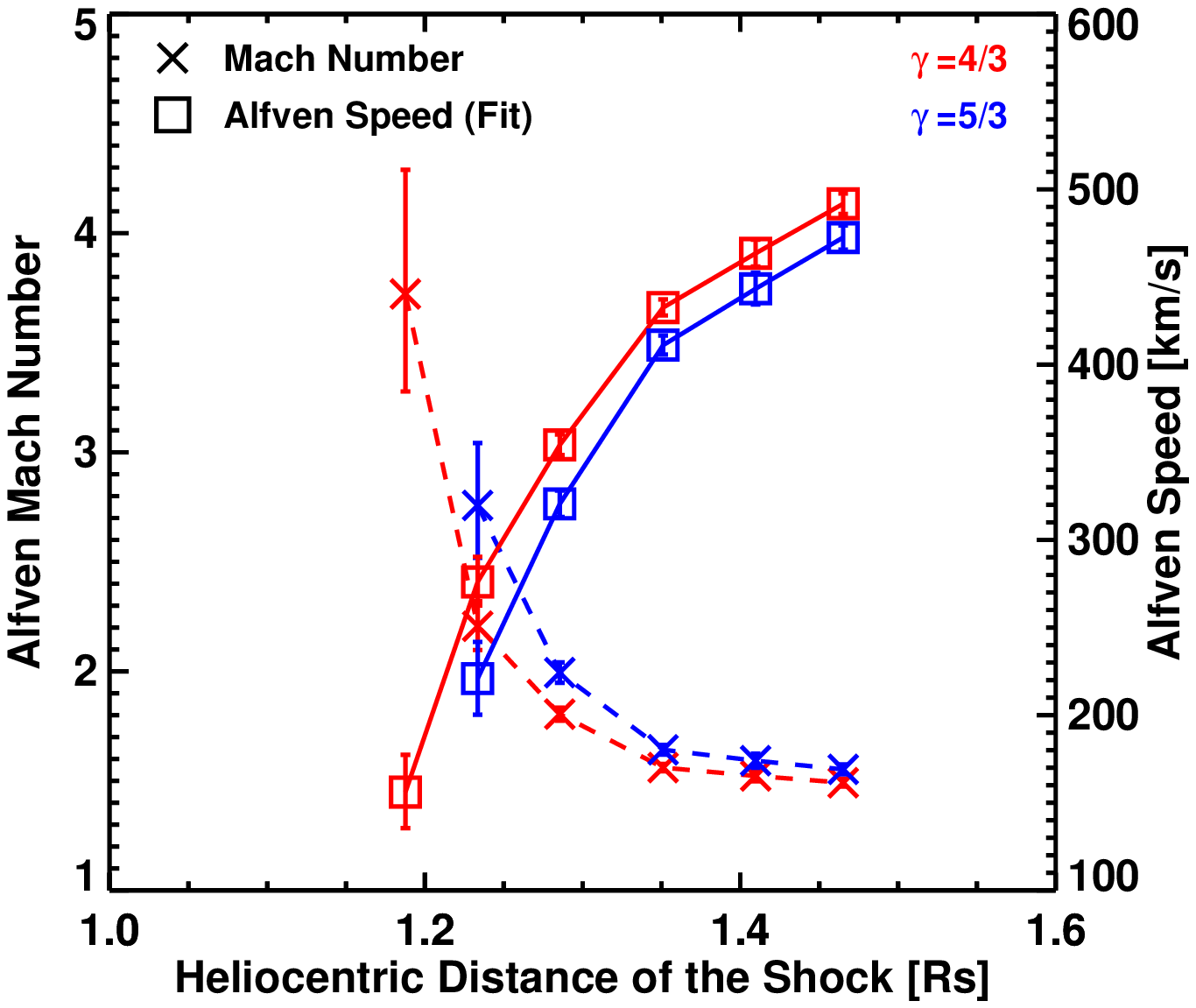}
 \end{center}
 \caption[]{	
 Examples of coronal-seismology results.
  {Left}: magnetic-field strength inferred from the lateral propagation speed  
 of a coronal wave on 04 August 2011 as a function of heliocentric distance
 (from \opencite{Kwon_STEREO-EUV-wave-high-corona.2013ApJ...766...55K}).
  {Right}: Height distribution of the derived \Alfven speed and Mach number 
 from the 13 June 2010 EIT wave and shock
 (from \opencite{GopalswamyN.2010Jun13.AIA.shock.Bseismology.2012ApJ...744...72G}).
 } \label{seism.eps}
 \end{figure}

For EIT waves that are {\it linear fast magnetosonic}, one can equate their measured speeds and $v_{\rm f}$
with uncertainties subject to projection effects among others. 
In the quiet-Sun corona, the temperature is on the order of 1\,--\,2~MK giving
$c_{\rm s} = 150$\,--\,$210 \kmps$. 	
Density measurements are more challenging because of LOS integration and unknown filling factors.
Fortunately, the uncertainty in magnetic field due to density is small because of 
the $\sqrt{\rho}$ dependence ({\it e.g.} \opencite{NakariakovOfman.B-from-oscil.2001A&A...372L..53N}).
Early studies used {\it atmosphere models} to estimate the density 
\cite{WarmuthA.MannG.Alfen.speed.model.EIT.wave.seismology.2005A&A...435.1123W,%
Ballai.global-corona-seism.2007SoPh..246..177B}.
More accurate estimates were made with {\it density-sensitive line ratios}, such as \ion{Si}{10} 258~\AA/261~\AA,
from spectroscopic measurements. Applying this to {\it Hinode}/EIS data,	
\inlinecite{West.EIT-wave-seismology.2011ApJ...730..122W}
inferred a weak quiet-Sun magnetic field of $\approx$$1 \G$ in 2009 during the deep solar minimum,
while \inlinecite{LongD.EIT.wave.seismology.2013SoPh..288..567L}	
found 2\,--\,6~G at heights of 70\,--\,130~Mm in 2010\,--\,2011 during the rise phase of this solar cycle.
Alternatively, using {\it tomographic density reconstruction} \cite{KramarM.tomograph.3D.STEREO.2009SoPh..259..109K},
\citeauthor{Kwon_STEREO-EUV-wave-high-corona.2013ApJ...766...55K}
(\citeyear{Kwon_STEREO-EUV-wave-high-corona.2013ApJ...766...55K,Kwon_EUV.wave.seismology.2013ApJ...776...55K})
found fields of 0.4\,--\,2.5~G at heliocentric distances of 1\,--\,$3 \Rsun$
in the extended corona (see \fig{seism.eps}, left) and higher field strengths and plasma $\beta$ in streamers than in coronal holes.
%
%

For {\it fast-mode shock} cases, $v_{\rm f}$ can be obtained from the measured EIT wave speed [$v_{\rm EIT}$]
and the fast-mode Mach number [$M_{\rm f}= v_{\rm EIT}/v_{\rm f}$] ($\approx$ the \Alfven Mach number [$M_{\rm A}$]
for low-$\beta$ plasmas) inferred from the Rankine­--Hugoniot jump condition 
given the {\it compression ratio} [$n/n_0$].
The latter can be estimated with {\it Type-II burst band-splitting} [$n/n_0= (f_{\rm U}/f_{\rm L})^2$]
where $f_{\rm U}$ and $f_{\rm L}$ are the upper and lower branch frequencies, respectively
({\it e.g.} \opencite{MaSL.AIA.shock.2011ApJ...738..160M}; 
\opencite{KouloumvakosA.typeII.in.sheath.CME.EUV.wave.2014SoPh..289.2123K}),
or with the {\it EUV intensity ratio} [$n/n_0= \sqrt{I/I_0}$] by
ignoring heating and thus obtaining upper limits 	
\cite{Veronig.dome-wave.2010ApJ...716L..57V,KienreichI.STB.homologous.EUV.waves.2011ApJ...727L..43K,
MuhrN.STEREO.EUVwave.2011ApJ...739...89M,ShenYD.LiuY.true.wave.2012ApJ...754....7S}.
Such estimates have yielded $n/n_0$ and $M_{\rm f}$ values of 1\,--\,1.5
and magnetic fields of 1.5\,--\,3.5~G.		
Alternatively, \inlinecite{GopalswamyN.2010Jun13.AIA.shock.Bseismology.2012ApJ...744...72G}
obtained the Mach number (see \fig{seism.eps}, right) from 	
the {\it shock-standoff distance} and the curvature radius of the shock-driving flux rope,
and the density from the Type-II burst frequency according to an atmosphere model.
They obtained magnetic fields of 1.3\,--\,1.5~G at heliocentric distances of 1.2\,--\,$1.5 \Rsun$.

For coronal funnels rooted in active regions in which QFP wave trains are observed to propagate,
their typical speed is $1000 \kmps \gg c_{\rm s}$, and thus $v_{\rm f} \approx v_{\rm A}$ and 
$B \approx v_{\rm f} \sqrt{ 4 \pi \rho} $.
For QFPs in the linear fast-mode regime,
AIA passband responses were used to estimate lower limits of $\rho$
and thus lower limits of $B$ in the range of 2\,--\,8~G 
\cite{LiuW.FastWave.2011ApJ...736L..13L,ShenYD.LiuY.QFP.wave.2012ApJ...753...53S,ShenYD.LiuYu.QFP.171.193.2013SoPh..288..585S}.

\subsection{Thermal-Seismology}
\label{subsect_Tseism}

Thermal-seismology with EUV waves, proposed by \inlinecite{DownsC.MHD.2010-06-13-AIA-wave.2012ApJ...750..134D}, 
is a relatively new approach to probe the thermal properties of the local corona.
The temperature distribution or differential emission measure (DEM) of 
the coronal plasma has been traditionally inferred from {\it snapshots} of multiple EUV passband data 
using forward modeling ({\it e.g.} \opencite{Aschwanden.auto.AIA-T-EM.2013SoPh..283....5A}) 
or inversion ({\it e.g.} \opencite{WeberM.DEM.inversion.2004IAUS..223..321W};
\opencite{SchmelzJ.AIA.DEM.2011ApJ...731...49S}),	
each with its own advantages and disadvantages (\opencite{GuennouC.AIA.DEM1.isothermal.2012ApJS..203...25G};
\citeyear{GuennouC.AIA.DEM2.multithermal.2012ApJS..203...26G}). Large-scale EUV waves, by perturbing the corona
and producing {\it temporal} variations, can provide additional constraints	
not only on the wave itself but also on its medium. 

For example, as discussed in \sect{subsect_therm}, the general trend of 193/211~\AA\ brightening
and 171~\AA\ darkening at EIT wave fronts indicates heating of the perturbed corona
from an initial temperature range of 0.8\,--\,1.6~MK determined by the peak responses of these channels,
while the exception of 193~\AA\ darkening in some cases indicates higher initial temperatures $\gtrsim\,$1.6~MK.
The temperature resolution of all AIA passbands combined is estimated at $0.03 \log T$
\cite{GuennouC.AIA.DEM1.isothermal.2012ApJS..203...25G}.
Such a sensitivity is manifested in those so-called tri-color composite emission maps
(see \figs{bimod_MaSL.eps} and \ref{thermal.eps}, right),	
which show that as little as a 0.3\,\% temperature rise with or without adiabatic compression can account for
typical EUV intensity changes of a few percent \cite{DownsC.MHD.2010-06-13-AIA-wave.2012ApJ...750..134D}.

Note that, because of their sensitive dependence on local plasma and magnetic conditions, 
any change in time or space of kinematic or thermal properties of EUV waves
serves as a probe to the temporal evolution or spatial inhomogeneity
of the large-scale corona. For example, wave reflections or abrupt speed changes
indicate interfaces between drastically different plasmas or magnetic fields.
Such a potential using large-scale EUV waves in magneto--thermal seismology is yet to be exploited.

\subsection{Energy-Seismology}
\label{subsect_Mseism}

Aside from local coronal seismology applications, 	
forced transverse oscillations discussed in \sect{subsect_oscil}, with their kinetic energies, can be used to
estimate the energy contents of their triggering agents, {\it i.e.} EIT waves.
The estimate depends on proper evaluation of the coupling between the EIT wave 
and the oscillating loop, and the fractional reflection, trapping, and transmission 
of the wave energy, {\it e.g.} by using realistic MHD models of such events 
({\it e.g.} \opencite{Ofman.wave.2007ApJ...655.1134O}; \opencite{DownsC.MHD.2010-06-13-AIA-wave.2012ApJ...750..134D}).
Such estimates were obtained at $10^{23}$\,--\,$10^{26} \erg$
as lower limits from loop oscillations
\cite{Ballai.EIT-wave-trigger-loop-oscil.2005ApJ...633L.145B,Ballai.global-corona-seism.2007SoPh..246..177B}
and at $10^{28} \erg$ from oscillations of a flux-rope coronal cavity
\cite{LiuW.cavity-oscil.2012ApJ...753...52L}. The latter is close to 
a more complete estimate of $10^{29} \erg$ by including kinetic, radiative loss,
and conductive energies \cite{Patsourakos.Vourlidas.EIT-wave-review.2012SoPh..281..187P},
and comparable to the typical energy of $10^{28} \erg$
of a micro-flare \cite{HannahI.hsi.microflare.stat-II.2008ApJ...677..704H}.

For QFP waves, one can assume the observed EUV intensity variations due to density perturbations alone,
which is reasonable because unlike EIT waves, QFPs analyzed so far show no sign of heating or cooling.
The kinetic energy of the perturbed plasma can thus be obtained and used as a lower limit to estimate
the wave-energy flux which was found in the range of (0.1\,--\,$2.6) \times 10^5 \erg \pcms \ps$
\cite{LiuW.FastWave.2011ApJ...736L..13L,ShenYD.LiuYu.QFP.171.193.2013SoPh..288..585S}.
Such fluxes are sufficient for {\it coronal heating} in active regions
\cite{Withbroe.Noyes.coronal-heating-flux.1977ARA&A..15..363W},
although the overall contribution by QFPs could be small 
due to their low occurrence rate (only during flares/CMEs).

\section{Methodologies for EUV Wave Research}
\label{sect_method}


Methodologies adopted in EUV wave research have direct impact on our capabilities to
achieve the best science from observations and to probe their
underlying physics. To this end, advanced data-analysis techniques and
numerical or analytical models have played important roles in recent years.

\subsection{Data Analysis Techniques}
\label{subsect_analysis}


A key task for EUV wave research is to identify the wave front
and track it in space and time.
Many early studies were based on observer's visual identification
in running- or base-difference images and on point-and-click measurements, 
which introduce an element of subjective interpretation of the data.	
The immense data volume with the advent of SDO/AIA made this traditional technique 
impractically labor intensive and necessitated {\it automatic wave detection}.
The first two such algorithms were
i) the Novel EIT wave Machine Observing ({\sf NEMO}: \opencite{Podladchikova.NEMO.2005SoPh..228..265P})
package utilizing image profiles, recently updated with a clustering
technique for dimming extraction \cite{Podladchikova.NEMO.update.2012SoPh..276..479P},
and ii) the mapping technique of \inlinecite{Wills-Davey.auto-detect-EIT-wave.2006ApJ...645..757W},
using the Huygens principle to track wave trajectories.

%
 \begin{figure}[thbp]      
 \begin{center}
 \includegraphics[height=4.5cm]{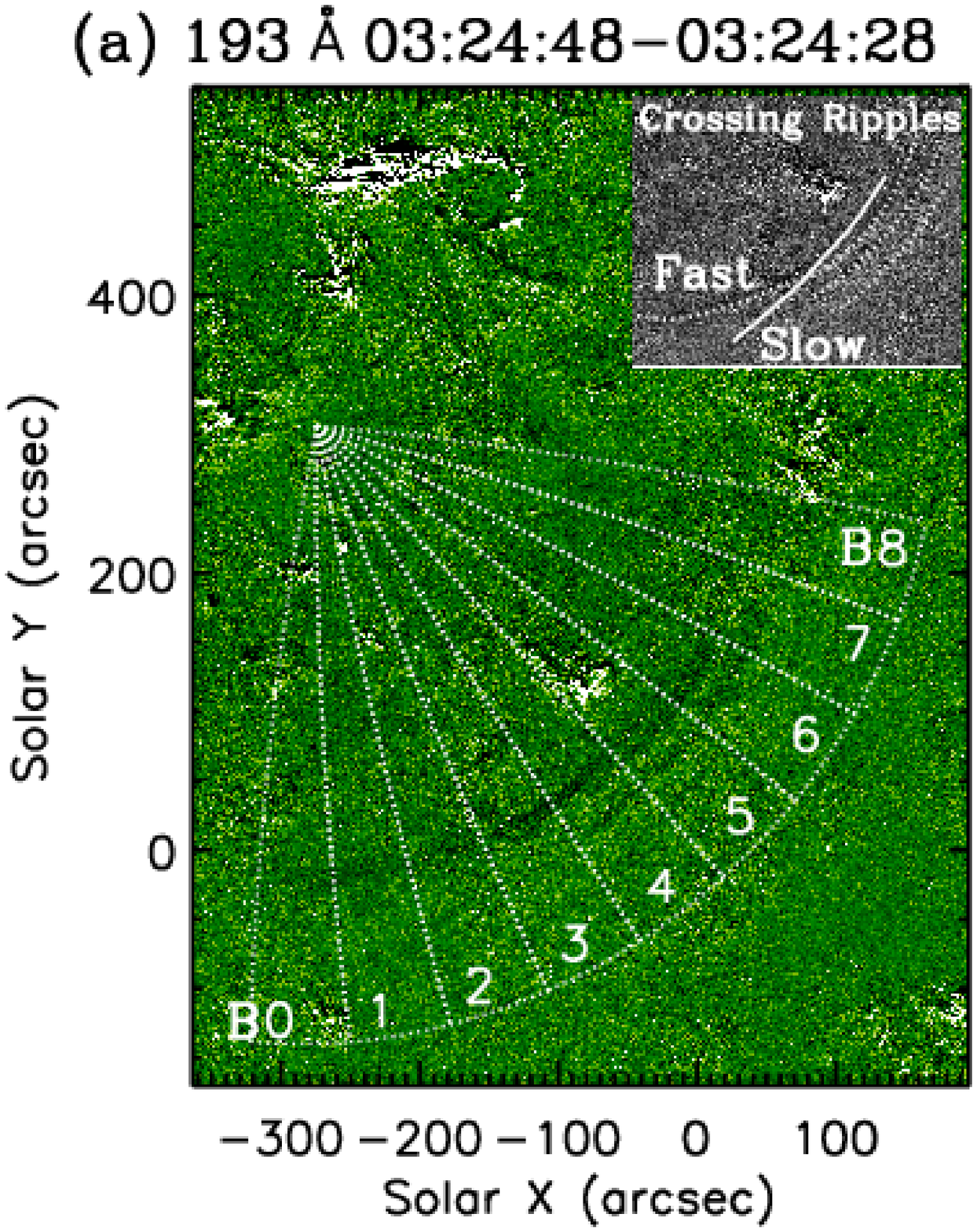}	
 \includegraphics[height=4.5cm]{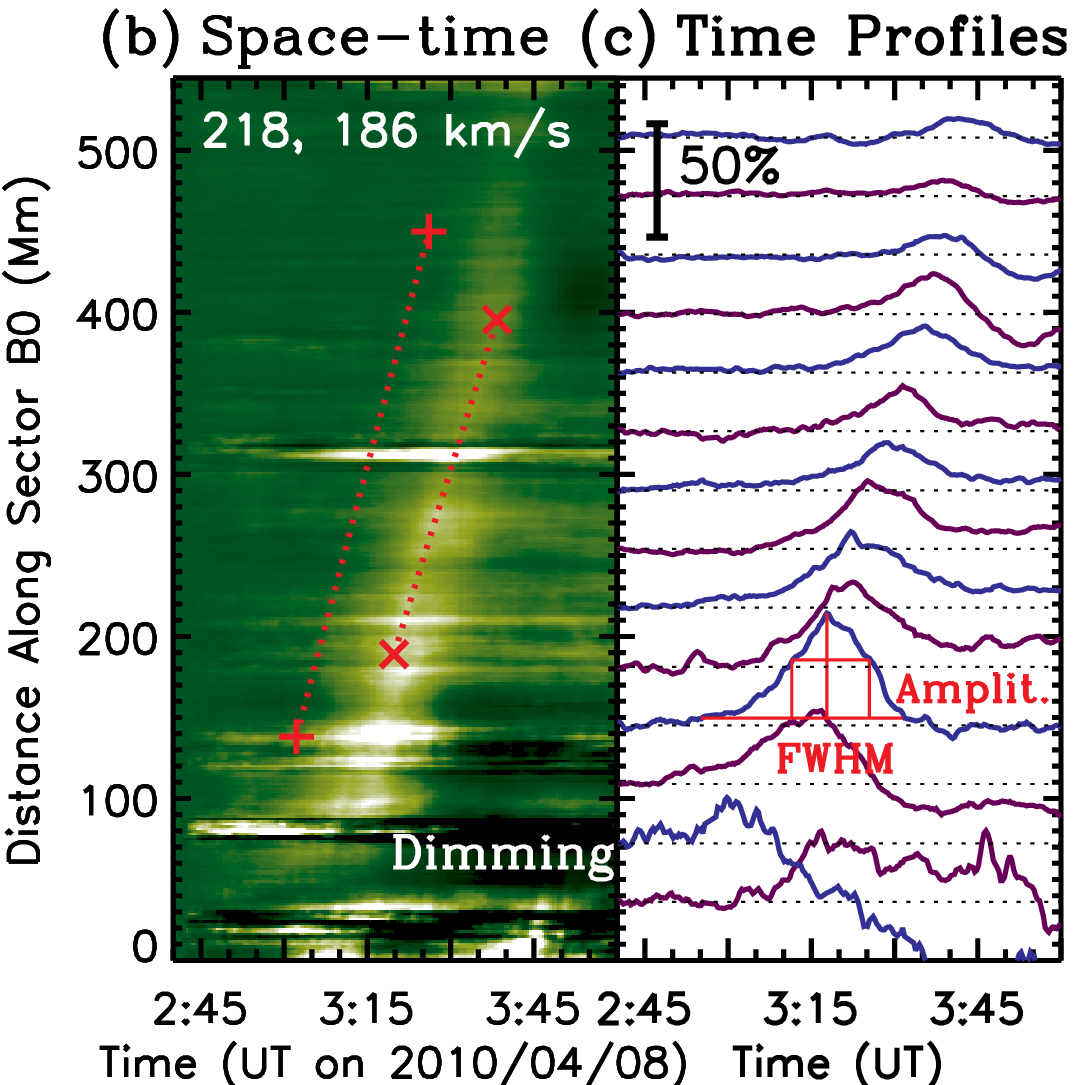}
 \includegraphics[height=3.85cm]{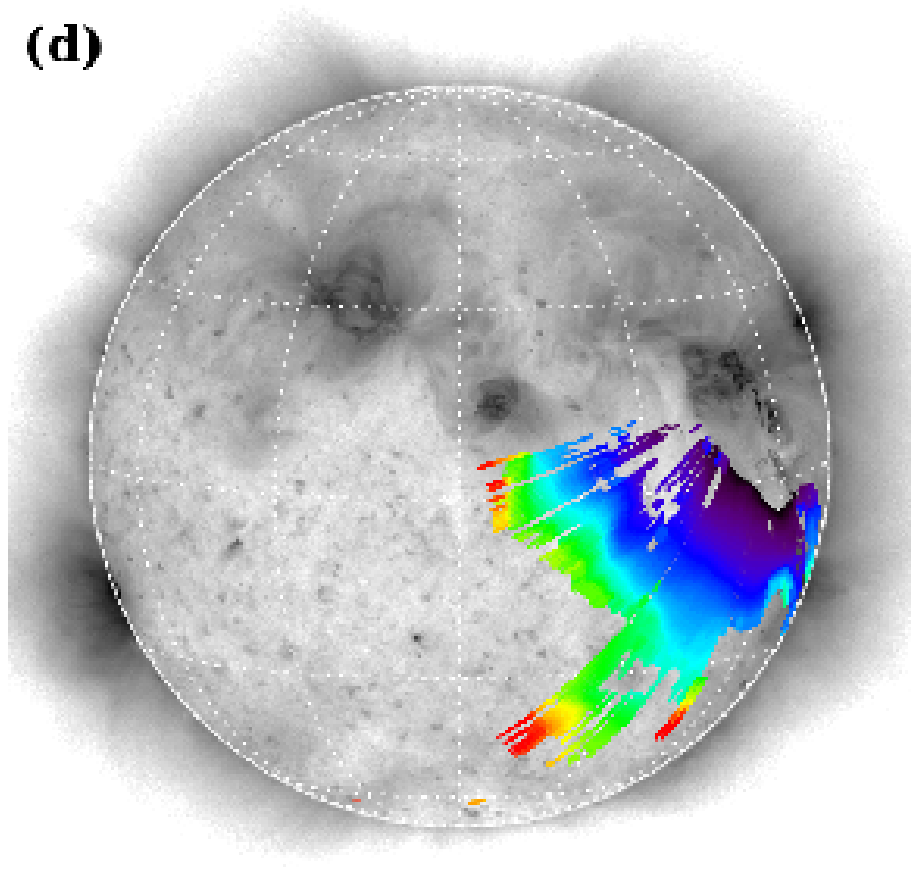}		
 \end{center}
 \caption[]{	
 Examples of EIT wave analysis techniques.
  (a) Spherical sectors overlaid on a running-difference image of the 08 April 2010 EIT wave for obtaining
 (b) a base-difference space--time plot and (c) intensity profiles at fixed locations,
 from which pulse amplitude, FWHM, and onset and peak positions/speeds can be measured
 (from \opencite{LiuW.AIA-1st-EITwave.2010ApJ...723L..53L}).
  (d) Wave ground tracks identified by the {\sf CorPITA} algorithm
 \cite{LongD.EIT.wave.CorPITA.package.2014arXiv1403.6722L} with \editor{colors from black to red} indicating
 time elapsed in 20~minutes
 (from \opencite{ByrneJ.auto.kinem.CME.wave.2013A&A...557A..96B}).	
 } \label{data-tech.eps}
 \end{figure}
%
As the cornerstone of various recently developed algorithms 
({\it e.g.} \opencite{Veronig.dome-wave.2010ApJ...716L..57V}), {\sf NEMO} projects
EUV images onto the solar surface and obtains intensity profiles 
along spherical sectors	
originating from the epicenter ({\it e.g.} \fig{data-tech.eps}, left), thus recovering ``ground tracks" of EIT waves.
Next, a common practice is to compose such 1D profiles at different times 
into a 2D space--time diagram or stack plot (\fig{data-tech.eps}, middle), from which wave properties
such as front kinematics and pulse amplitudes and widths
can be retrieved automatically ({\it e.g.} \opencite{LiuW.AIA-1st-EITwave.2010ApJ...723L..53L})
or by repeated visual measurements.
A similar approach adopted in the Coronal Pulse Identification and Tracking Algorithm
({\sf CorPITA}; \opencite{LongD.EIT.wave.CorPITA.package.2014arXiv1403.6722L}; \fig{data-tech.eps}, right),
which has been implemented in 
the SDO computer vision project \cite{MartensP.SDO-feature-finding.2012SoPh..275...79M},
involves fitting 1D profiles with a Gaussian to characterize kinematics
and wave dispersion.
Derivatives of these approaches employ cuts in various geometries to accommodate
specific situations, such as cuts at constant heights above the limb to track off-limb waves.
Recently, \inlinecite{ByrneJ.auto.kinem.CME.wave.2013A&A...557A..96B}
adopted robust statistical techniques to remedy undesired numerical effects 
arising from deriving EUV wave kinematics from scattered space--time data.

The above ground-track techniques neglect heights of EIT waves by assuming that all emission
originates from the spherical solar surface. Near disk center, this inconsistency causes only a negligibly small error
in travel distance at a constant height, {\it e.g.} $\lesssim\,$1\,\% for $h=100 \Mm$.	
Near the limb, however, any change in height ({\it e.g.} vertical propagation)
can result in significant overestimates of velocities projected onto the spherical surface, 
as demonstrated in 3D MHD simulations \cite{Hoilijoki.EUV-wave.view.angle.2013SoPh..286..493H}.
3D reconstruction of EIT waves from STEREO observations has been recently
attempted with triangulation techniques \cite{DelanneeC.EUV.wave.height.drop.2013arXiv1310.5623D}.
In doing so, extra care must be taken to account for optically thin EUV wave emission,
whose integration along different lines of sight never corresponds to the same feature,
unlike optically thick emission ({\it e.g} some line emission from prominences).

For tracking waves and oscillations in general, 	
a range of algorithms were or are currently being developed, such as a coherence/travel-time based approach 
\cite{McIntoshS.coherence.wave.tracking.2008SoPh..252..321M},
an interactive web system \cite{Sych.web.wave.detection.2010SoPh..266..349S},
a Bayesian-based approach \cite{IrelandJ.Bayesian.oscil.detect.2010SoPh..264..403I},
phase-speed measurements with 	
various fitting techniques \cite{YuanD.auto.measure.phase.speed.2012A&A...543A...9Y},
and EIT wave detection using advanced image processing and the {\sf SunPy} python library
\cite{IrelandJ.auto.detect.EIT.wave.2012AAS...22020117I}.

Fourier analysis of wave power and dispersion relation with $k$--$\omega$ diagrams 
is a common technique in helioseismology, but was previously underused in coronal seismology
mainly because of the scarcity of high-cadence data
\cite{DeForestC.TRACE-1600-fast-mode.2004ApJ...617L..89D,TomczykMcIntosh.coronal-time-distance-seism.2009ApJ...697.1384T}.
AIA filled this gap and offered a new venue for this technique,
which has been widely used for analysis of QFP wave trains as noted earlier
\cite{LiuW.FastWave.2011ApJ...736L..13L,ShenYD.LiuY.QFP.wave.2012ApJ...753...53S}.

\subsection{Numerical and Analytical Models}
\label{subsect_model}

%


Recent modeling efforts pertinent to large-scale EUV waves fall into two categories:
2D/3D MHD simulations for EIT waves and QFP wave trains and 
analytical or numerical models for shock formation and propagation.

Early 2D MHD models of EIT waves in Cartesian geometries, although simple, 
were able to capture some essential features,
such as the bimodal composition of a fast-mode (shock) wave ahead of a CME driven compression
\cite{ChenPF.EIT-wave-MHD.2002ApJ...572L..99C,ChenPF.EIT-wave.2005ApJ...622.1202C,%
PomoellJ.MHD-EIT-wave-lateral-expansion.2008SoPh..253..249P}.
More recent 2D models revealed the presence of a fast-mode shock and its reflection from the chromosphere
together with a slow-mode shock and velocity vortices surrounding a CME 
\cite{WangHongjuan.EIT-slow-mode-wave.2009ApJ...700.1716W,MeiZX.LinJ.2012ScChG..55.1316M}.

3D MHD models capture some crucial physics missing in 2D models that assume 
an invariant third dimension. 
For example, the linear and nonlinear coupling between the 
fast mode and other MHD modes in inhomogeneous plasma requires 3D modeling.
3D models in Cartesian geometries focus on {\it local} behaviors of EUV waves,
such as reflections or transmissions and flare-triggered loop oscillations
in bipolar active regions \cite{OfmanThompson.EIT-wave-fast-mode.2002ApJ...574..440O,Ofman.wave.2007ApJ...655.1134O},
non-wave current shells surrounding CMEs 
\cite{Delannee.EIT-wave-current-shell.2008SoPh..247..123D,Schrijver.Aulanier.2011Feb15.X2.AIAwv.2011ApJ...738..167S},
and dome-shaped EIT waves from rotating active regions
(\opencite{SelwaM.dome-EUV-wave-MHD-model.2012ApJ...747L..21S}, 
\citeyear{SelwaM.dome-EUV-wave-MHD-AR-models.2013SoPh..284..515S}).
Recently, \inlinecite{Hoilijoki.EUV-wave.view.angle.2013SoPh..286..493H} demonstrated
that, because of LOS integration, observed EIT wave speeds are highly dependent on viewing angles 
and can vary by nearly a factor of two.
3D models in spherical geometries using more realistic
magnetic fields from extrapolations of magnetograms allow for tracking
{\it global} propagation of EIT waves. For example, \inlinecite{Wu.EIT-fast-MHD-wave.2001JGR...10625089W}
found that a fast-mode wave in a coronal region of $\beta \approx 1$ can explain an observed EIT wave.	
Using the coronal module of the Block-Adaptive Tree Solar-wind Roe Upwind Scheme ({\sf BATSRUS}) code
\cite{Roussev.BATS-R-US_code-coronal-module.2003ApJ...595L..57R},
\inlinecite{CohenO.EITwave.non-wave.both.2009ApJ...705..587C} confirmed the wave/non-wave bimodality,
and \inlinecite{Schmidt.Ofman.3D-MHD-eitwv.2010ApJ...713.1008S} reproduced the observed 
EIT wave reflection from a coronal hole.

 \begin{figure}[thbp]      
 \begin{center}
 \includegraphics[height=3.8cm]{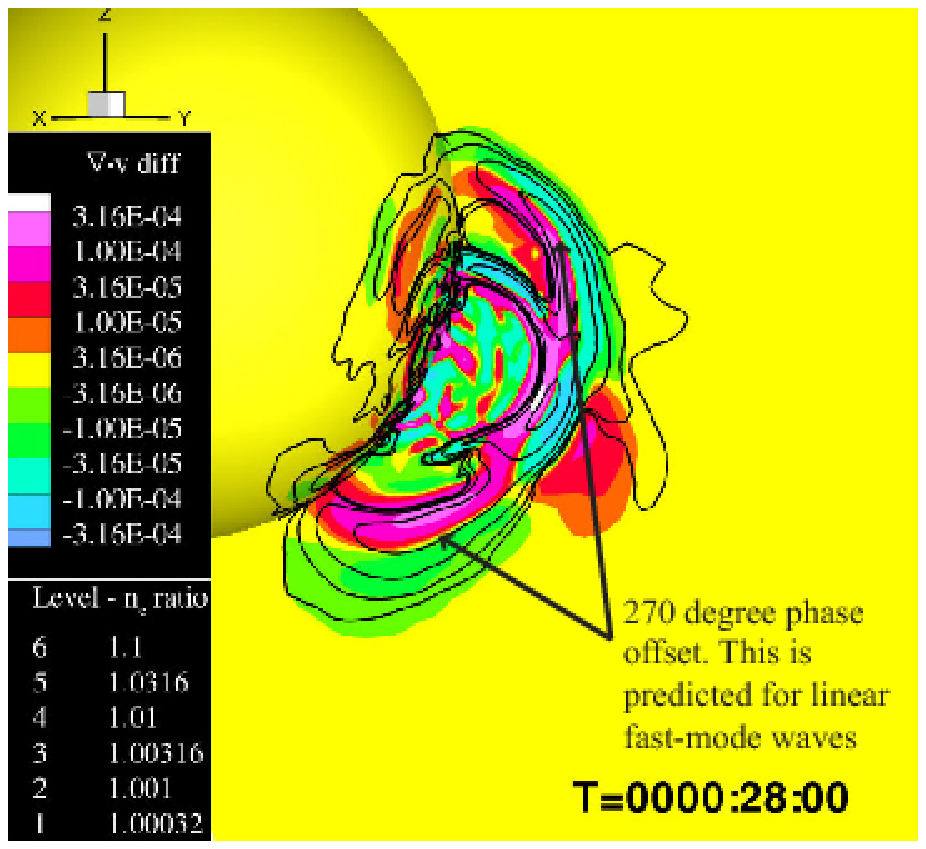}		
 \includegraphics[height=3.8cm]{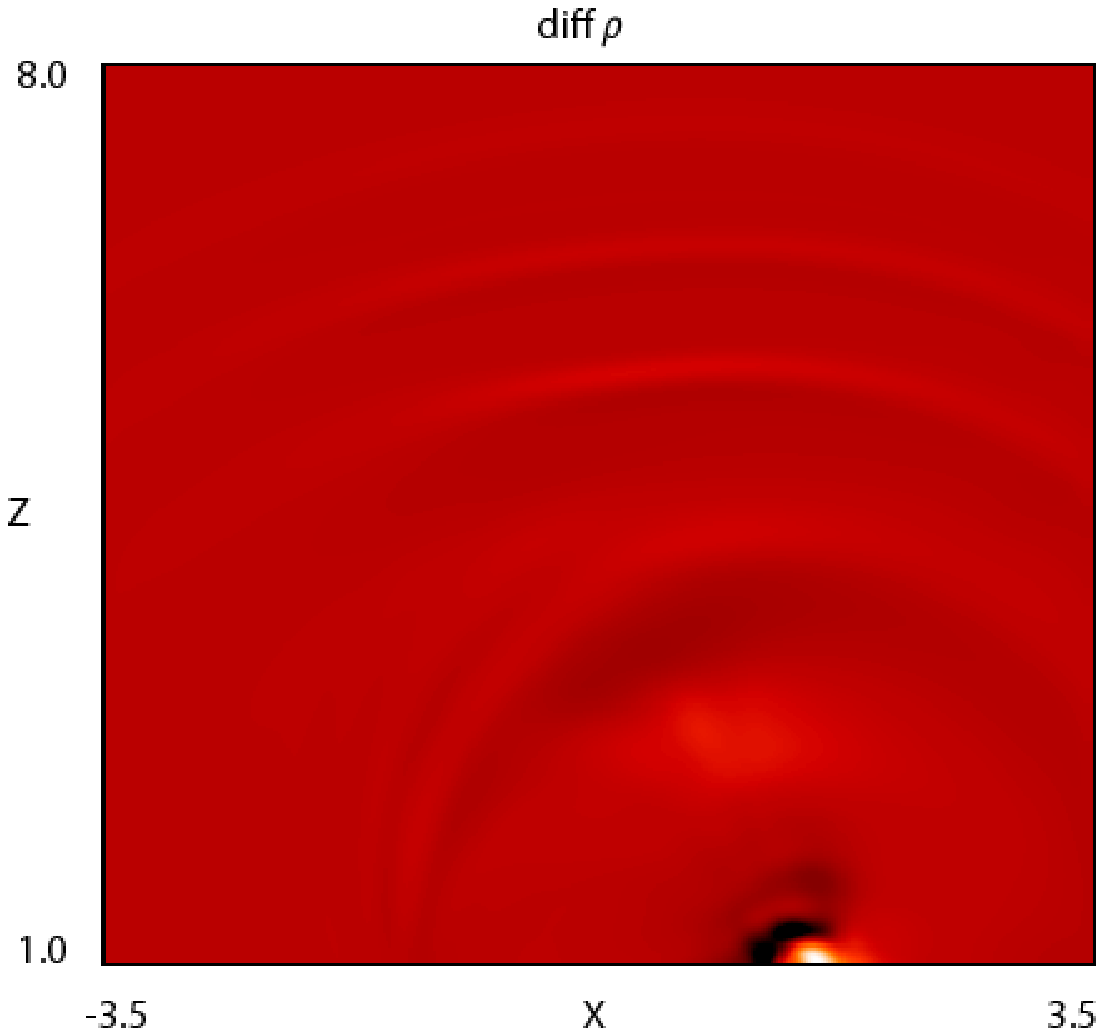}		
 \includegraphics[height=3.8cm, bb=73 36 195 161, clip=]{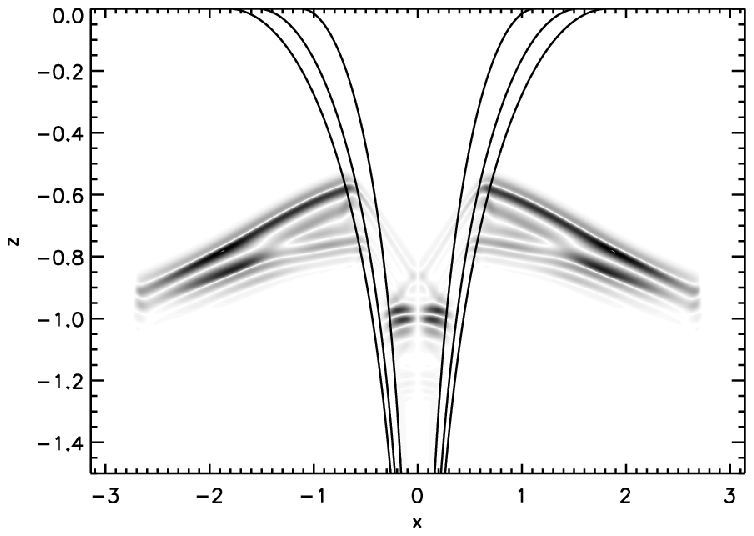}		
 \end{center}
 \caption[]{	
 Examples of numerical simulations for EUV waves.
  {Left}: Divergence of perturbed velocity in color
 showing a sign reversal from negative to positive (green to pink; {\it i.e.}
 from compression to expansion)	
 that coincides with the maximum density enhancement (black contours), 
 as expected for a compressible fast-mode wave component
 of an EIT wave (from \opencite{DownsC.MHD.EUVIwave.2011ApJ...728....2D}).
  {Middle}: Density perturbations of a modeled QFP train 	
 (from \opencite{Ofman.Liu.fast-wave.2011ApJ...740L..33O}).
  {Right}: Simulated velocity perturbations for confined and leaked ``wing"
 QFP waves from a funnel-shaped waveguide (from \opencite{Pascoe.wing.QFPs.funnel.2D.MHD.2013A&A...560A..97P}). 
 } \label{MHD-models.eps}
 \end{figure}
%
A recent advance in 3D global MHD models is to {\it synthesize EUV emission} by convolving
simulated density and temperature distributions with instrument response functions
(\opencite{DownsC.MHD.EUVIwave.2011ApJ...728....2D}, \citeyear{DownsC.MHD.2010-06-13-AIA-wave.2012ApJ...750..134D}),
as previously done for 2D models \cite{ChenPF.synthesis-EIT-Moreton-wave.2005SSRv..121..201C}.
This allows for direct comparison with observables, {\it e.g.} from EUVI or AIA, 
to constrain the models		
and perform thermal-seismology. 
\inlinecite{DownsC.MHD.2010-06-13-AIA-wave.2012ApJ...750..134D} established a unified picture 
supporting the hybrid wave/non-wave model of EIT waves.
Specifically, as shown in \fig{MHD-models.eps} (left), they found through phase analysis that 
the highest density enhancement is cospatial with the maximum compression
located in the outer EUV front, as expected for a compressible fast-mode wave.
However, the bulk CME motion and an associated current shell and reconnection front
are located further inside, as part of the non-wave, inner EUV component.

QFP wave trains were modeled as fast-mode magnetosonic waves with a 3D MHD code in a Cartesian geometry 
\cite{Ofman.Liu.fast-wave.2011ApJ...740L..33O}. They are excited by periodic velocity pulsations
at one pole of a bipolar active region and produce density fluctuations by compressibility
(see \fig{MHD-models.eps}, middle) that match the observed QFP signatures \cite{LiuW.FastWave.2011ApJ...736L..13L}, including
reflections within closed loops. A funnel-shaped {\it leaky waveguide} is achieved by 
the fast-magnetosonic speed distribution determined by the diverging magnetic field and density stratification. 
A similar waveguide formed with density enhancement within a funnel was constructed
with a 2D MHD model by \inlinecite{Pascoe.wing.QFPs.funnel.2D.MHD.2013A&A...560A..97P}. 
They reproduced observed QFP amplitude growth and deceleration 
\cite{YuanD.QFP.distinct.trains.2013A&A...554A.144Y} and 
found additional ``wing" wave trains leaking from the funnel (see \fig{MHD-models.eps}, right).

Shock formation and propagation have been extensively investigated both analytically and numerically
(see \opencite{VrsnakCliver.corona-shock-review.2008SoPh..253..215V} for a review),
which is relevant to a subset of strong EIT waves that are shocked.
Analytical solutions were obtained for piston-driven shock formation in 1D planar geometries
\cite{MannG.shock-formation.1995JPlPh..53..109M,VrsnakB.Lulic.coronal-shock-1.2000SoPh..196..157V}, 
in 2D cylindrical and 3D spherical geometries \cite{Zic.piston-driver-shock.2008SoPh..253..237Z}, 
and applied to an expanding sphere with a rising center in comparison 
with Moreton-wave observations \cite{TemmerM.analytic-model-Moreton-wave.2009ApJ...702.1343T}. 
A recent development involved 2.5D MHD simulations
of shock formation driven by a cylindrical piston \cite{LulicS.coronal.shock.2.5D.MHD.2013SoPh..286..509L}. 
A general conclusion is the positive dependence of the (shock) wave amplitude and phase speed 
on the {\it acceleration of the piston}.
This bears strong implications for the early (lateral) CME expansion discussed in \sect{subsect_gener}.

Another line of research focuses on the propagation of a shock or linear MHD wave
after its generation.	
The pioneering work of \citeauthor{Uchida.Moreton-wave-sweeping-skirt.1968SoPh....4...30U}
(\citeyear{Uchida.Moreton-wave-sweeping-skirt.1968SoPh....4...30U},
\citeyear{UchidaY.flare.wave.typeII.1974SoPh...39..431U})
applied {\it linear geometric acoustics} or ray tracing
with the Wentzel--Kramers--Brillouin (WKB) approximation to follow fast-mode (weak shock) 		
wave propagation governed by the spatial (height) distribution of the fast-magnetosonic speed
and to explain Moreton waves and Type-II bursts.
The same approach was adopted in the modern observational context of EIT waves
\cite{WangYM.EIT-fastMHDwave.2000ApJ...543L..89W,Patsourakos.EUVI-wave.2009SoPh..259...49P}.
Recently, {\it nonlinear} effects such as energy dissipation and the dependence of wave speed on amplitude were included 
\cite{Afanasyev.Uralov.shock-EUV-wave-I.2011SoPh..273..479A,AfanasyevA.AR.magnetosphere.fast.shock.2013ARep...57..594A}.
The results were applied to a coherent picture unifying observations of EIT waves, Type-II bursts, and CMEs 
(\opencite{GrechnevV.shock-EUV-wave-I.2011SoPh..273..433G},
\citeyear{GrechnevV.shock-EUV-wave-III.2011SoPh..273..461G}).

\section{Conclusions and Prospects}
\label{sect_conclude}



We have presented a review of recent advances in EUV wave research focusing on 
new observations since the launch of SDO and related data-analysis techniques and models.
Thanks to its advanced capabilities, SDO/AIA not only played a critical role in ending the 15-year-long
debate on the nature of EIT waves, allowing them to be used for coronal seismology,
but also opened new research areas for newly discovered coronal phenomena,
such as QFP wave trains and magnetic KH instabilities with associated nonlinear waves.
We summarize below the current status and future prospects of these
topics. 	


Backed up with strong observational and numerical evidence, the {\it hybrid or bimodal nature
of EIT waves} has been established. In this general picture, an outer EUV front of a true fast-mode (shock) wave
travels ahead of an inner non-wave component of CME-driven compression.
Heating due to electric current dissipation or magnetic reconnection may contribute to the
EUV emission at the inner, CME front, but not the outer, true wave front.
AIA revealed an average EIT wave speed $>\,$$600 \kmps$ that is 
well expected for coronal fast-mode waves but much higher than
the typical speeds of 200\,--\,$400 \kmps$ from previous SOHO/EIT measurements.
A wide range of behaviors intrinsic to fast-mode waves are now commonly observed,
including quasi-periodic wave trains, reflections and transmissions, coherent periodicities, 
sequential structural oscillations, and heating-cooling cycles. 
The impulsive lateral and downward expansions 
of a CME are believed to be key in generating EIT waves (see \sect{sect_EIT}).
Outstanding questions regarding EIT waves include:
\begin{enumerate}
 \item quantitative relation between their generation and CME (lateral) expansion;
 \item their roles in transporting energy and triggering sympathetic eruptions;
 \item their physical relation with Type-II bursts, Moreton waves, and SEPs.
\end{enumerate}

As one of AIA's discoveries, QFP wave trains with typical speeds of 500\,--\,$2200 \kmps$ are 
evidence of fast-mode magnetosonic waves in funnel-shaped waveguides from active regions.
They are commonly associated with quasi-periodic flare pulsations (\sect{sect_QFP}). 
Open questions on QFPs include:
\begin{enumerate}
 \item the origin of periodicities, especially those not identified in flare pulsations with possible
 connections to three-minute sunspot and other (sub)surface oscillations;	
 \item their roles in energy transport and coronal heating;
 \item the relation between QFPs within funnels and quasi-periodic wave trains within EIT waves ahead of CME flanks.
\end{enumerate}

Small-scale EUV waves including mini-EUV waves and KHI waves are relatively new and
require further investigation to fully uncover the statistical distributions of their physical parameters.
Mini-EUV waves are less energetic but more numerous than their large-scale counterparts (\sect{sect_mini-wave}).
Thus their total energy budget could be significant for the quiet Sun.
As for nano- or micro-flares in the flare size distribution,
mini-EUV waves may play an important role in the full spectrum of EUV waves of hierarchic sizes.
Such possibilities could be topics of future research.

Seismological practice using EIT waves and QFPs to probe the coronal magnetic fields and
thermal states and wave-energy fluxes is being actively pursued. 
The currently inferred quiet-Sun magnetic fields are in the range of 1\,--\,10~G 
with uncertainties of about the same order (\sect{sect_seism}).
Improving this accuracy, {\it e.g.} with refined density and temperature estimates,
will be a critical future task. There are other potential diagnostic techniques 
to be explored, {\it e.g.} by including mini-EUV waves and using wave reflections and 
transmissions to probe topological interfaces.

A whole suite of data-analysis techniques is becoming mature and has started to
produce fruitful results (\sect{subsect_analysis}). However, automatic detection and tracking of EUV waves have not
been widely tested or used. Their performance in data-processing pipelines	
remains to be seen and will be critical to fully explore rich observations offered by AIA and other instruments.
We emphasize that detailed analysis of individual well-observed events
and statistical analysis of large samples are equally important.

Numerical models are crucial in lending credence to data interpretation
and in understanding 	
the underlying physics. Particularly useful are those 3D MHD models with realistic initial and boundary conditions	
that can produce synthesized observables to be directly compared with observations (\sect{subsect_model}).

In the years to come, the diagnostic power enabled by AIA's spatio--temporal and thermal coverage remains to be
fully exploited to answer the above open questions about coronal EUV waves. 
The future EUV imager and spectrometer onboard the {\it Solar Orbiter} mission, scheduled for launch in 2017,
and the EIS counterpart onboard the currently planned {\it Solar-C} mission will likely make further contributions.
Additional constraints can be obtained from complementary observations of the solar atmosphere beyond the corona
or at wavelengths outside the EUV regime.	
The {\it Interface Region Imaging Spectrograph} 
(IRIS: \opencite{DePontieuB.IRIS.mission.2014SoPh..tmp...25D}), launched in June 2013,
can detect potential UV signatures of coronal waves and Moreton waves in the transition region and chromosphere, 
such as Doppler, density, and temperature perturbations, and help identify the origin of QFP wave periodicities
in regard to flares and chromospheric oscillations.
The ground-based {\it Daniel K.~Inouye Solar Telescope} (DKIST), previously known as the {\it Advanced Technology Solar Telescope} (ATST), 
with first light expected in 2019, 	
will provide imaging and spectroscopic observations of the solar atmosphere from the photosphere
to the corona in visible and near-infrared regimes and offer critical plasma 
and magnetic field diagnostics simultaneously.





\begin{acks}
This work is supported by the NASA Living With a Star (LWS) Program (grant NNX11AO68G). 
Additional support to LO was provided by NASA grant 	
NNX12AB34G.		
\editor{	
Special thanks go to Barbara Thompson for inviting both authors to 	
the 2013 LWS SDO Science Workshop that led to this topical issue and this review.
We are grateful to the anonymous referee for constructive comments and suggestions that helped improve this article.
WL~thanks Nariaki Nitta, Cooper Downs, Barbara Thompson, Angelos Vourlidas, Peng-Fei Chen, Spiros Patsourakos, and Kyoung-Sun Lee
for critical comments on the manuscript and/or fruitful discussions.
We thank Suli Ma, Alexander Warmuth, Nariaki Nitta, Ding Yuan, Ute M\"{o}stl (now Ute Amerstorfer), 
Nat Gopalswamy, Jason Byrne, and David Pascoe for providing the original figures,		
and especially Cooper Downs, Liheng Yang, Ting Li, Eoin Carley, and Ryun-Young Kwon 
for customizing their figures to fit the layout of this article.
\figs{3mission.eps}c and \ref{3mission.eps}d, \ref{bimod_MaSL.eps}, \ref{0908_global-train.eps}, \ref{Nitta_stat_v-a.eps}~(right), 
\ref{reflect.eps}, \ref{908_oscil-fits.eps}a, \ref{908_oscil-fits.eps}c, and \ref{908_oscil-fits.eps}i, \ref{thermal.eps}, \ref{Nitta_stat_hist.eps}, \ref{QFP-overview.eps}~(middle and right), 
\ref{QFP-2trains.eps}a, \ref{mini_wave.eps}, \ref{seism.eps}, \ref{data-tech.eps}a\,--\,\ref{data-tech.eps}c, and \ref{MHD-models.eps}~(left and middle)
are reproduced by permission of the AAS.
\figs{Nitta_stat_v-a.eps}~(left), \ref{QFP-overview.eps}~(left), \ref{data-tech.eps}~(right), and \ref{MHD-models.eps}~(right) 
are reproduced with permission from Astronomy \& Astrophysics, \copyright~ESO.
}  
\end{acks}

%
%

\bibliographystyle{spr-mp-sola-cnd} 

{	
%
	
}

\end{article} 
\end{document}